\documentclass{article}
\usepackage{amsmath}
\usepackage{amssymb}
\usepackage{amsthm}
\usepackage{natbib}
\usepackage{graphicx}
\usepackage{algorithm}
\usepackage{algpseudocode}
\usepackage{url}
\usepackage{authblk}
\usepackage[a4paper,top=3cm, bottom=3.5cm, left=3cm, right=3cm]{geometry}

\usepackage{etoolbox}

\newtheorem{theorem}{Theorem}

\title{Improving Power by Conditioning on Less in Post-selection Inference for Changepoints}

\author{Rachel Carrington\thanks{r.j.carrington@lancaster.ac.uk} }
\author{Paul Fearnhead\thanks{p.fearnhead@lancaster.ac.uk}}
\affil{Department of Mathematics and Statistics, Lancaster University, UK}

\begin{document}

\maketitle

\begin{abstract}
Post-selection inference has recently been proposed as a way of quantifying uncertainty about detected changepoints. The idea is to run a changepoint detection algorithm, and then re-use the same data to perform a test for a change near each of the detected changes. By defining the $p$-value for the test appropriately, so that it is conditional on the information used to choose the test, this approach will produce valid $p$-values. We show how to improve the power of these procedures by conditioning on less information. This gives rise to an ideal post-selection $p$-value that is intractable but can be approximated by Monte Carlo. We show that for any Monte Carlo sample size, this procedure produces valid $p$-values, and empirically that noticeable increase in power is possible with only very modest Monte Carlo sample sizes. Our procedure is easy to implement given existing post-selection inference methods, as we just need to generate perturbations of the data set and re-apply the post-selection method to each of these. On genomic data consisting of human GC content, our procedure increases the number of significant changepoints that are detected 
when compared to the method of \cite{jewell-testing}.

\end{abstract}



\section{Introduction} \label{sec:intro}

Detecting abrupt changes in time-series, or other ordered, data has been one of the most active research areas of the past decade. It has applications in bioinformatics \cite[e.g.][]{braun2000multiple,olshen2004circular}, computer performance \cite[]{barrett2017virtual}, climate science \cite[]{reeves2007review,shi2022changepoint}, cyber security \cite[]{heard2014monitoring,fearnhead2019changepoint}, neuroscience \cite[]{aston2012evaluating,jewell2020fast}, and industrial process monitoring \cite[]{maleki2016development} amongst many others. There has been a wide range of methods that have been proposed, dealing with detecting different types of change, such as change in mean, variance or slope; different algorithms for searching for multiple changepoints, including binary segmentation and its variants \cite[]{olshen2004circular,fryzlewicz2014wild,baranowski2019narrowest}, moving window methods \cite[]{hao2013multiple,eichinger2018mosum,meier2021mosum}, $L_1$ penalised regression methods \cite[]{kim2009ell_1,tibshirani2014adaptive}, and dynamic programming approaches to maximising an $L_0$ penalised likelihood \cite[e.g.][]{killick2012optimal,maidstone2017optimal}; and for different types of data, such as high-dimensional data \cite[]{wang2018high}, network data \cite[]{wang2021optimal}, and general non-Euclidean data \cite[]{song2022asymptotic,dubey2020frechet}. See \cite{truong2020selective}, \cite{fearnhead2020relating} and \cite{shi2022comparison} for an overview of this area.

There has been much less work looking at quantifying the uncertainty of estimated changepoints. Whilst Bayesian methods \cite[e.g.][]{fearnhead2006exact} that sample from a posterior  over the number and location of the changepoints naturally give measures of uncertainty, assessing uncertainty for non-Bayesian methods is more challenging. Current work in this area includes the SMUCE method \cite[]{frick2014multiscale,li2016fdr,pein2017heterogeneous}, and global methods that try to give regions that produce sets of intervals, all of which must include a change at a pre-specified significance level \cite[]{fang2020segmentation,fryzlewicz2021robust,fryzlewicz2023narrowest}. These latter methods have advantages of being adaptive -- with the size of intervals reflecting the uncertainty in the location of possible changes. However they do not give measures of uncertainty for specific detected changes as the methods we consider below do. There are also methods such as the MOSUM procedure \cite[]{eichinger2018mosum} or tests based on self-normalisation \cite[]{zhao2022segmenting}, that can give asymptotic $p$-values against the null hypothesis of no change-point. For the MOSUM procedure, this enables an asymptotic $p$-value to be calculated for each window of data, that gives the probability under the null of some window having a larger test-statistic value than observed for that specific window.  We investigate the use of these $p$-values as measures of uncertainty for each detected change-point in Section \ref{sec:multiple-testing}. The asymptotics used to calculate these $p$-values Brownian bridge, are often plagued with slow convergence \cite[]{fearnhead2022detecting} which can lead to conservative tests and a slight loss of power. We see some evidence of this in our empirical results  (see Section \ref{sec:multiple-testing}).

A different approach is to first estimate the changepoints, and then assign a measure of significance to each detected change. The challenge here is to avoid so-called double peeking at the data \cite[]{zhao2021defense}, where you use the same data both to detect a change and then to test for the change, as a naive implementation of the test based on using the same data twice will be invalid. This is because, in the absence of any change, the detection process will bias you to performing tests that are more likely to have small $p$-values. This results in tests where the $p$-values are neither uniform, nor stochastically bounded below by a uniform distribution \cite[see e.g.][]{jewell-testing}.

One simple approach to circumvent this is sample splitting \cite[]{rinaldo2019bootstrapping}, where you use a proportion of the data to detect changes and the other other part to perform a test for each detected change. However using only part of the data for each of detection and testing is sub-optimal. Instead post-selection inference ideas for regression \cite[]{berk2013valid,fithian2014optimal,kuchibhotla2022post} have recently been applied to the changepoint setting. These allow the same data to be used for detection and testing, but with the $p$-values for each change being calculated conditional on information from the data that includes whatever information is used to choose the test that is being performed. These are called post-selection $p$-values.

Methods for calculating post-selection $p$-values have been developed for the change in mean problem with Gaussian noise and for a range of detection algorithms. \cite{hyun-post-selection} develop an approach for binary segmentation and its variants, and for the fused lasso; while \cite{jewell-testing} and \cite{duy-more} propose methods that work if changes are detected using an $L_0$ penalised likelihood.  Furthermore \cite{jewell-testing} show how to improve on the method of \cite{hyun-post-selection} by conditioning on less information when defining the post-selection $p$-value, and show that conditioning on less information can lead to a substantial increase in power. There has also been recent work on post-selection inference beyond the change in mean problem \cite[e.g][]{chen2023quantifying}.

Our work is motivated by further wanting to reduce the information that one conditions on when calculating the post-selection $p$-value. Current methods condition on the projection of the data that is orthogonal to the test statistic. If, as is common, our test has a null hypothesis where, say, the mean of the data does not change within a region about the tested changepoint, then we can reduce this to conditioning on the data outside the region and an appropriate sufficient statistic, such as the sample mean, within the region. This, together with whatever aspect of the detected changes is used to pick the test, is the minimum amount of information that we need to condition on to make the post-selection $p$-value well-defined. In Section \ref{sec:optimal} we show that, for a natural class of distributions for the data under the alternative, the resulting post-selection $p$-value is optimal. 

Unfortunately we cannot directly calculate this $p$-value. Instead we propose a simple Monte Carlo approximation. This is based on simulating new data within the region around the changepoint that is being tested, applying the existing post-selection inference methodology to each such data set, and then calculating a weighted average of the post-selection $p$-values for each data set. Importantly, we show that if one of the data sets we average over is the observed data, then this leads to a valid post-selection $p$-value, in that its distribution is uniform on $[0,1]$ under the null, regardless of the Monte Carlo sample size. Furthermore, it is simple to calculate provided we have a method that calculates a post-selection $p$-value based on conditioning on the projection of the data orthogonal to the test statistic. As such, our method applies to all changepoint scenarios considered in \cite{hyun-exact}, \cite{hyun-post-selection}, \cite{jewell-testing} and \cite{chen2023quantifying}. We present empirical results that show one can obtain a noticeable improvement in power even with modest Monte Carlo sample sizes, say of the order of 10. Whilst our method has been developed for the changepoint problem, the underlying ideas apply more widely, see Section \ref{sec:diss}. All proofs are given in the Appendix.

\section {Background}  \label{sec:back}

\subsection{Post-selection $p$-values for changepoints}

Suppose we have a data set $\boldsymbol{X} = \left( X_1, \ldots, X_T \right)$ and we fit a changepoint model which consists of $K$ changepoints $\mathcal{M}(\boldsymbol{X}) = \{ \hat{\tau}_1, \ldots, \hat{\tau}_K \}$. We are interested in quantifying the level of uncertainty associated with these changepoints: how confident can we be that the changepoints we have found correspond to real changes and not false discoveries? One approach is to compute $p$-values for each changepoint of interest.

One aspect in quantifying uncertainty in this way is deciding what we mean by $\tau$ being a changepoint. Or, more specifically, what null hypothesis do we want to test? In many applications we say that $\tau$ is a changepoint providing some aspect of the data (that we are interested in) changes at or close to $\tau$. That is, the null hypothesis would be that there is no change in some region centered on $\tau$.

Even once we have decided on the null hypothesis, naively applying a test for a change at each of $\hat{\tau}_1,\ldots,\hat{\tau}_K$ is not possible, as we have already used the data to detect the changepoints. If the data contains no changes, we would expect any detected changepoint to be where, by chance, the patterns of the data are similar to patterns produced by a change. This will bias the $p$-values \cite[see][for examples of this]{jewell-testing}.

To overcome this, we can correct the naive test to take account of the fact that we are using the data twice \cite[]{fithian2014optimal}. This can be done by calculating a post-selection $p$-value, which uses the distribution of the test statistic under the null but also conditional on any information used to choose the test that we are performing.

To make the idea concrete let $\mathcal{F}(\boldsymbol{X})$ denote the information we want to condition on. We have freedom over the choice of $\mathcal{F}(\boldsymbol{X})$, except that it must contain the information from the data that is used to choose the test we performing. So, for example, if we choose to test that $\tau$ is a changepoint based only on the property that $\tau$ is one of the estimated changepoints, then $\mathcal{F}(\boldsymbol{X})$ must include the information $\tau\in\mathcal{M}(\boldsymbol{X})$. The post-selection $p$-value is then 
\begin{equation*}
    \Pr ( \mathcal{T} \geq \mathcal{T}_{obs} ~ | ~ \mathcal{F} \left( \boldsymbol{X} \right) = \mathcal{F} \left(\boldsymbol{X}_{obs}\right) ),
\end{equation*}
for some test statistic $\mathcal{T}$, and where we use $\boldsymbol{X}_{obs}$ and $\mathcal{T}_{obs}$ to denote the observed data and test statistic respectively. The challenge is then how to calculate this post-selection $p$-value. This will require a careful choice of the information we condition on, both to make the post-selection $p$-value well defined, and tractable. 

\subsection{Post-selection $p$-value for change in mean}

For ease of presentation it is helpful to consider a specific example. We will consider the univariate change in mean model, for which methods for calculating post-selection $p$-values have been developed by \cite{hyun-exact}, \cite{hyun-post-selection}, \cite{jewell-testing} and \cite{duy-computing}. However, the ideas we introduce for increasing the power of post-selection inference apply more widely (see Section \ref{sec:other-H0}).

For the change in mean model, we assume the data is of the form
\begin{equation*}
    X_t = \mu_t + \epsilon_t, \hspace{1cm} t = 1, \ldots, T,
\end{equation*}
where $\mu_t$ is piecewise constant, with $\mu_{t+1} \neq \mu_t$ only at $K$ changepoints $\tau_1, \ldots, \tau_K$. We assume that $\epsilon_t \sim_{iid} N(0, \sigma^2)$, with $\sigma$ known.

We run a changepoint algorithm -- for example binary segmentation \cite[]{scott1974scott}, wild binary segmentation \cite[]{fryzlewicz2014wild}, narrowest-over-threshold \cite[]{baranowski2019narrowest}, fused lasso \cite[]{tibshirani2005sparsity}, or a penalised likelihood approach \cite[]{maidstone2017optimal} -- and detect a set of changepoints $\{\hat{\tau}_1, \ldots, \hat{\tau}_K\}$. We now want to test for a change at a particular estimated changepoint, which for simplicity we will denote $\hat{\tau}$.

As mentioned above, for many applications a natural null hypothesis is that there is no change in mean close to $\hat{\tau}$. There are various possible choices for what we mean by ``close to"; we will discuss this in more detail in Section \ref{sec:other-H0}, but for now we will assume that there is a pre-determined distance $h$ that is appropriate for our application. (We discuss other choices of null hypothesis in Section \ref{sec:other-H0}.)
Our null hypothesis is therefore
\begin{equation*}
    H_0: \text{ } \mu_{\hat{\tau}-h+1} = \cdots = \mu_{\hat{\tau}} = \mu_{\hat{\tau}+1} = \cdots = \mu_{\hat{\tau}+h},
\end{equation*}
with the alternative hypothesis being that there is at least one inequality.

Let $\boldsymbol{\nu}_{\hat{\tau}}$ be a $T$-dimensional vector whose $t$th entry is
\begin{equation*}
    (v_{\hat{\tau}})_t = \begin{cases}
        \frac{1}{h} & \text{if } \hat{\tau} - h < t \leq \hat{\tau} \\
        -\frac{1}{h} & \text{if } \hat{\tau} < t \leq \hat{\tau} + h \\
        0 & \text{if } t \leq \hat{\tau} - h \text{ or } t > \hat{\tau} + h.
    \end{cases}
\end{equation*}
Then, under $H_0$, and without conditioning on any information in the data, $\boldsymbol{\nu}_{\hat{\tau}}^T \boldsymbol{X} \sim N \left( 0, \frac{2\sigma^2}{h} \right)$. Under $H_1$, where there is a changepoint at or near $\hat{\tau}$, we would expect the mean of $\boldsymbol{\nu}_{\hat{\tau}}^T \boldsymbol{X}$ to be non-zero. We can therefore take the test statistic to be $\mathcal{T} = |\boldsymbol{\nu}_{\hat{\tau}}^T \boldsymbol{X}|$.

As above let $\mathcal{M}(\boldsymbol{X})=\{\hat{\tau}_1, \ldots, \hat{\tau}_K\}$. The information used to choose the null hypothesis to test is that $\hat{\tau}\in \mathcal{M}(\boldsymbol{X})$. Thus for our post-selection $p$-value we need a conditioning event that includes this information. Unfortunately it is not possible to just choose $\mathcal{F}(\boldsymbol{X})$ to be $\hat{\tau}\in \mathcal{M}(\boldsymbol{X})$, because the probability of this event depends on parameters that are unknown under the null hypothesis. 

To deal with this, current approaches \cite[]{jewell-testing,hyun-post-selection} condition also on the projection of the data that is orthogonal to $\boldsymbol{\nu}_{\hat{\tau}}$. Denote this orthogonal projection by $\boldsymbol{\Pi}_{\hat{\tau}}$, then this leads to the post-selection $p$-value
\begin{equation} \label{eq:p}
    \Pr(|\boldsymbol{\nu}_{\hat{\tau}}^T \boldsymbol{X}| > |\boldsymbol{\nu}_{\hat{\tau}}^T \boldsymbol{X}_{obs}| ~|~
        \hat{\tau}\in\mathcal{M}(\boldsymbol{X}), ~ \boldsymbol{\Pi}_{\hat{\tau}} \boldsymbol{X} = \boldsymbol{\Pi}_{\hat{\tau}} \boldsymbol{X}_{obs} ).
\end{equation}

While this is well-defined, calculating the required conditional distribution of $\boldsymbol{\nu}_{\hat{\tau}}^T \boldsymbol{X}$ is non-trivial. \cite{hyun-post-selection} show that for binary segmentation or the fused lasso, if you condition on further information, namely the order in which the changepoints are detected and the estimated sign of each changepoint, then the conditional distribution will be a truncated Gaussian. Furthermore the truncation region can be calculated by solving a series of linear equations.

Motivated by intuition that conditioning on less information will improve power \cite[]{fithian2014optimal,liu2018more}, 
\cite{jewell-testing} shows how to reduce the amount of information conditioned on, by avoiding having to condition on the order and signs of the changepoints. As we are conditioning on the projection of the data orthogonal to $\boldsymbol{\nu}_{\hat{\tau}}^T$,  $\boldsymbol{X}$ will be uniquely determined if, in addition, we know $\boldsymbol{\nu}_{\hat{\tau}}^T \boldsymbol{X}$.
Let $\phi = \boldsymbol{\nu}_{\hat{\tau}}^T \boldsymbol{X}$, and define the set of possible data sets that are possible as we vary $\phi$ by
\begin{equation}
  \label{eq:x-phi}
    \boldsymbol{X}'(\phi) = \boldsymbol{X}_{obs} - \frac{1}{||\boldsymbol{\nu}_{\hat{\tau}}||^2} \boldsymbol{\nu}_{\hat{\tau}} \boldsymbol{\nu}_{\hat{\tau}}^T \boldsymbol{X}_{obs} + \frac{1}{||\boldsymbol{\nu}_{\hat{\tau}}||^2} \boldsymbol{\nu}_{\hat{\tau}} \phi.
\end{equation}
If we define $\mathcal{S} = \{ \phi : \hat{\tau} \in \mathcal{M}(\boldsymbol{X}'(\phi)) \}$, then the $p$-value in  (\ref{eq:p}) is equal to
\begin{equation*}
     \Pr \left( |\phi| \geq |\boldsymbol{\nu}_{\hat{\tau}}^T \boldsymbol{X}_{obs}| ~ | ~ \hat{\tau} \in \mathcal{M}(\boldsymbol{X}'(\phi)) ~ \right) = \Pr \left( |\phi| \geq |\boldsymbol{\nu}_{\hat{\tau}}^T \boldsymbol{X}_{obs}| ~ | ~ \phi \in \mathcal{S} \right),
\end{equation*}
where, unconditionally,  $\phi \sim N ( 0, \sigma^2 ||\boldsymbol{\nu}_{\hat{\tau}}||^2 )$. \cite{jewell-testing} shows how the set $\mathcal{S}$ can be efficiently computed for changepoint methods including binary segmentation, $L_0$ segmentation and the fused lasso; in each case $\mathcal{S}$ is a union of intervals. Their methods can also be extended to other similar changepoint algorithms such as wild binary segmentation and narrowest-over-threshold.  The method of \cite{jewell-testing} leads to an increase in power compared to the approach of \cite{hyun-post-selection}, as it requires conditioning on less information. However, it still requires conditioning on $T - 1$ parameters that are orthogonal to $\boldsymbol{\nu}_{\hat{\tau}}$. The method we propose further reduces the amount of information we need to condition on, leading to greater power. 

\section{Conditioning on less information} \label{sec:method}

\subsection{The ideal post-selection $p$-value} \label{sec:ideal}

Instead of conditioning on $\boldsymbol{\Pi}_{\hat{\tau}} \boldsymbol{X}$, we could consider just conditioning on the minimum amount of information to make the post-selection $p$-value well-defined. Our null hypothesis fixes  $\mu_{\hat{\tau} - h + 1} = \cdots = \mu_{\hat{\tau} + h}$, so this contains no information about $\boldsymbol{\mu}$ for data points outside  of $\{ \hat{\tau} - h + 1, \ldots, \hat{\tau} + h \}$, nor does it specify the mean within this window.

Hence, as a minimum we need to condition on
\begin{equation}
  \label{eq:conditions}
    \begin{split}
        X_t & = X_{obs,t} \text{ for } t \in \{ 1, \ldots, \hat{\tau} - h, \hat{\tau} + h + 1, \ldots, T \} \\
        \frac{1}{2h} \sum_{i=\hat{\tau}-h+1}^{\hat{\tau}+h} X_t & = \frac{1}{2h} \sum_{i=\hat{\tau}-h+1}^{\hat{\tau}+h} X_{obs,t},
    \end{split}
\end{equation}
that is, the data outside of $\{ \hat{\tau} - h + 1, \ldots, \hat{\tau} + h \}$, and the sample mean of the data in this window, which is a sufficient statistic for the unknown, constant mean within the window. These have total dimension $T - 2h + 1$, so we gain an additional $2h - 2$ degrees of freedom compared to the method in \cite{jewell-testing}.

Let $\boldsymbol{B}$ be the $T \times \left( T - 2h \right)$ matrix obtained by removing  the columns corresponding to $\{ \hat{\tau} - h + 1, \ldots, \hat{\tau} + h \}$ from the $T \times T$ identity matrix, and let $\boldsymbol{a}$ be a $T$-dimensional vector such that
\begin{equation*}
    a_t = \begin{cases}
        \frac{1}{2h} & \text{ if } t \in \{ \hat{\tau} - h + 1, \ldots, \hat{\tau} + h \} \\
        0 & \text{ otherwise}.
    \end{cases}
\end{equation*}
Then the conditions in (\ref{eq:conditions}) are equivalent to:
\begin{equation*}
  \label{eq:cons-2}
    \begin{split}
        \boldsymbol{B}^T \boldsymbol{X} & = \boldsymbol{B}^T \boldsymbol{X}_{obs} \\
        \boldsymbol{a}^T \boldsymbol{X} & = \boldsymbol{a}^T \boldsymbol{X}_{obs}.
    \end{split}
\end{equation*}
Hence, the conditional $p$-value is
\begin{equation*}
    Pr_{H_0} (|\boldsymbol{\nu}_{\hat{\tau}}^T \boldsymbol{X}| > |\boldsymbol{\nu}^T \boldsymbol{X}_{obs}| ~ | ~ \hat{\tau} \in \mathcal{M}(\boldsymbol{X}), \boldsymbol{B}^T \boldsymbol{X} = \boldsymbol{B}^T \boldsymbol{X}_{obs}, \boldsymbol{a}^T \boldsymbol{X} = \boldsymbol{a}^T \boldsymbol{X}_{obs}).
\end{equation*}

To see how we can calculate this, we can rewrite $\boldsymbol{X}$ as
\begin{equation*}
  \begin{split}
    \boldsymbol{X} & = \boldsymbol{X} - \left( \boldsymbol{BB}^T + \frac{1}{||\boldsymbol{a}||_2^2} \boldsymbol{a a}^T + \frac{1}{||\boldsymbol{\nu}_{\hat{\tau}}||_2^2} \boldsymbol{\nu}_{\hat{\tau}} \boldsymbol{\nu}_{\hat{\tau}}^T \right) \boldsymbol{X} + \left( \boldsymbol{BB}^T + \frac{1}{||\boldsymbol{a}||_2^2} \boldsymbol{a a}^T + \frac{1}{||\boldsymbol{\nu}_{\hat{\tau}}||_2^2} \boldsymbol{\nu}_{\hat{\tau}} \boldsymbol{\nu}_{\hat{\tau}}^T \right) \boldsymbol{X} \\
        & = \left( \boldsymbol{I} - \left( \boldsymbol{BB}^T + \frac{1}{||\boldsymbol{a}||_2^2} \boldsymbol{a a}^T + \frac{1}{||\boldsymbol{\nu}_{\hat{\tau}}||_2^2} \boldsymbol{\nu}_{\hat{\tau}} \boldsymbol{\nu}_{\hat{\tau}}^T \right) \right) \boldsymbol{X} + \frac{1}{||\boldsymbol{\nu}_{\hat{\tau}}||_2^2} \boldsymbol{\nu}_{\hat{\tau}} \boldsymbol{\nu}_{\hat{\tau}}^T \boldsymbol{X} + \left( \boldsymbol{BB}^T + \frac{1}{||\boldsymbol{a}||_2^2} \boldsymbol{a a}^T \right) \boldsymbol{X} \\
      & = \boldsymbol{ZX} + \frac{1}{||\boldsymbol{\nu}_{\hat{\tau}}||_2^2} \boldsymbol{\nu}_{\hat{\tau}} \boldsymbol{\nu}_{\hat{\tau}}^T \boldsymbol{X} + \left( \boldsymbol{BB}^T + \frac{1}{||\boldsymbol{a}||_2^2} \boldsymbol{a a}^T \right) \boldsymbol{X},
  \end{split}
\end{equation*}
where $\boldsymbol{Z} = \boldsymbol{I} - \left( \boldsymbol{BB}^T + \frac{1}{||\boldsymbol{a}||_2^2} \boldsymbol{a a}^T + \frac{1}{||\boldsymbol{\nu}_{\hat{\tau}}||_2^2} \boldsymbol{\nu}_{\hat{\tau}} \boldsymbol{\nu}_{\hat{\tau}}^T \right)$. We include the term $\frac{1}{||\boldsymbol{\nu}_{\hat{\tau}||_2^2}} \boldsymbol{\nu}_{\hat{\tau}} \boldsymbol{\nu}_{\hat{\tau}}^T$ here to make explicit the dependence of $\boldsymbol{X}$ on the test statistic $\phi = \boldsymbol{\nu}_{\hat{\tau}}^T \boldsymbol{X}$.

$\boldsymbol{Z}$ is a $T \times T$ matrix with rank $2h-2$, so $\boldsymbol{ZX}$ follows a degenerate multivariate Gaussian distribution.  However, since $\boldsymbol{Z}$ is a symmetric matrix with all its non-zero eigenvalues equal to $1$, we can write $\boldsymbol{Z} = \boldsymbol{UU}^T$, where $\boldsymbol{U}$ is a $T \times (2h - 2)$ matrix with orthonormal columns. Under $H_0$, $\boldsymbol{U}^T \boldsymbol{X} \sim N( \boldsymbol{U}^T \boldsymbol{\mu}, \sigma^2 \boldsymbol{U}^T \boldsymbol{U} ) = N( \boldsymbol{0}, \sigma^2 \boldsymbol{I} )$. The matrix $\boldsymbol{U}$ is not uniquely defined, but the choice of basis is arbitrary. It can be found, for example, using the Singular Value Decomposition.

Let $\boldsymbol{\psi} = \boldsymbol{U}^T \boldsymbol{X}$ and, as before, let $\phi = \boldsymbol{\nu}_{\hat{\tau}}^T \boldsymbol{X}$. Then, given the information we are conditioning on, as we vary $\boldsymbol{\psi}$ and $\phi$ we get data
\begin{equation*}
    \boldsymbol{X} = \boldsymbol{X}'(\phi, \boldsymbol{\psi}) = \boldsymbol{U \psi} + \frac{1}{||\boldsymbol{\nu}_{\hat{\tau}}||_2^2} \boldsymbol{\nu}_{\hat{\tau}} \phi + \left( \frac{1}{||\boldsymbol{a}||_2^2} \boldsymbol{a a}^T + \boldsymbol{B B}^T \right) \boldsymbol{X}_{obs}.
\end{equation*}
Furthermore, under the null and without conditioning on further aspects of the data, such as the estimated changepoints, $\phi \sim N(0, \frac{2\sigma^2}{h})$ and $\psi_i \sim_{iid} N(0, \sigma^2)$. The resulting post-selection $p$-value is
\begin{equation} 
  \label{eq:ideal_p}
     P_{\phi,\boldsymbol{\psi}} \left( |\phi| \geq |\boldsymbol{\nu}_{\hat{\tau}}^T \boldsymbol{X}_{obs}| ~ | ~  \hat{\tau} \in \mathcal{M}(\boldsymbol{X}'(\phi, \boldsymbol{\psi})) ~ \right).
\end{equation}
As this $p$-value is obtained by conditioning on the least amount of information needed for it to be well-defined, we will call it the ideal $p$-value.

\subsection{Intuition behind new post-selection $p$-value} \label{sec:optimal}

\begin{figure}
    \centering
    \includegraphics{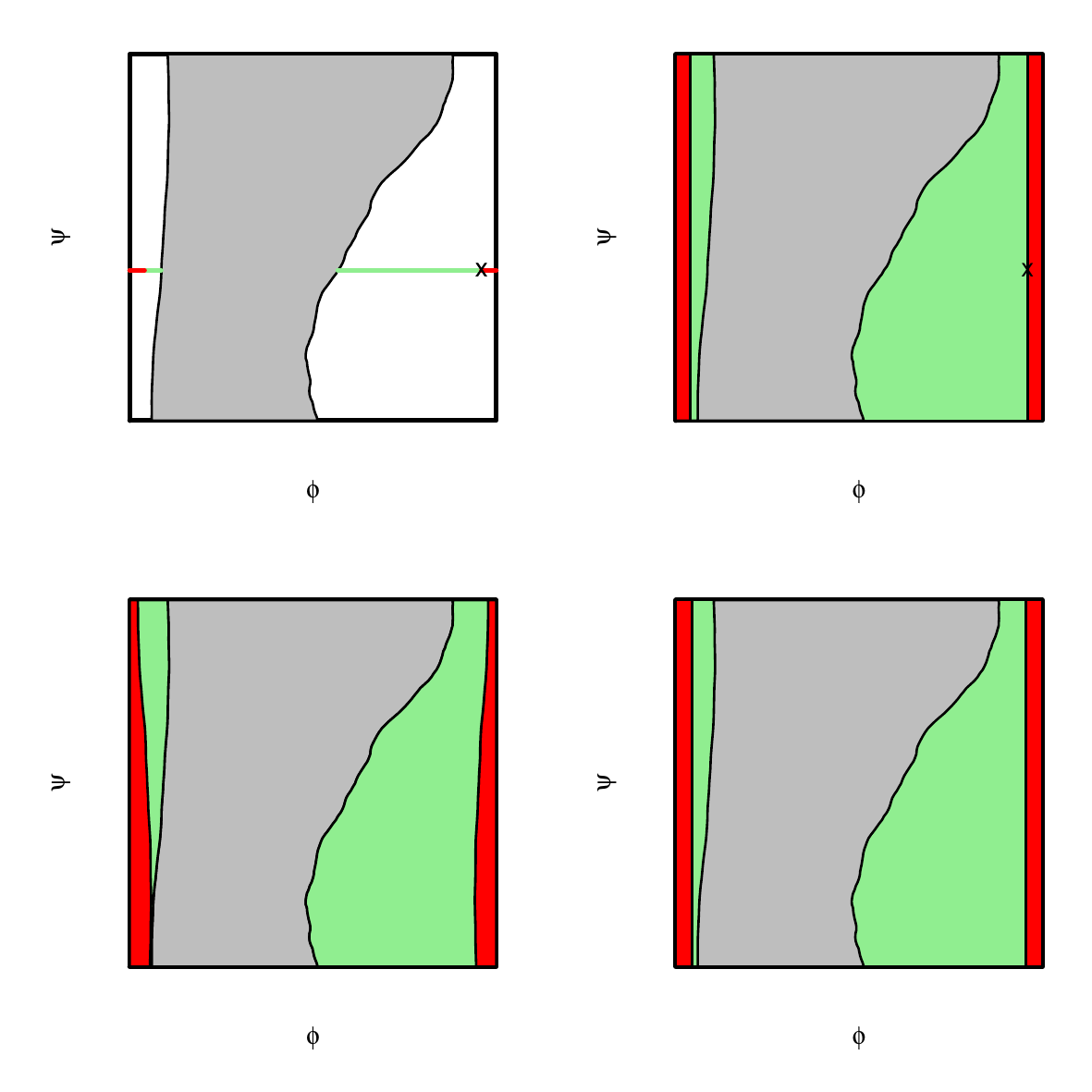}
    \caption{\small{Comparison of the $p$-value of \cite{jewell-testing} (left-hand column) and the ideal $p$-value (right-hand column) for the case of a univariate $\psi$ parameter. We have used the probability inverse mapping to transform $\phi$ and $\psi$ so that they are uniformly and independently distributed on $[0,1]$ under the prior. We view data sets as being a function of $(\phi,\psi)$, and the selection event -- which corresponds to the information in the data used to choose the test -- corresponds to a region of $(\phi,\psi)$ values (non-grey region in all plots). The observed data corresponds to a specific $(\phi,\psi)$ value  shown by a cross (top-row plots). For the method of \cite{jewell-testing}, the $p$-value is the probability of observing a more extreme value of $\phi$ conditional on the observed $\psi$-value. This is the proportion of the coloured line that is red in the top-right plot. For our method, the $p$-value is the (unconditional) probability of observing a more extreme value of $\phi$: the proportion of the non-grey area that is red (top-right plot). In the bottom row we show the data-sets, as represented by their $(\phi,\psi)$ value, that would give a post-selection $p$-value that is 0.2 or lower (red region in both plots).}}
    \label{fig:pvalueregions}
\end{figure}

To understand the difference between the ideal post-selection $p$-value (\ref{eq:ideal_p}) and the $p$-value of \cite{jewell-testing}, we give a schematic comparison in Figure \ref{fig:pvalueregions}. To enable us to present a plot we have supposed that $\psi$ is scalar -- we can see this as a special case where we fix all but one component of $\boldsymbol{\psi}$ -- and have also used the probability inverse mapping to transform $(\phi,\psi)$ from independent Gaussian to independent uniform on $[0,1]$. 

With this mapping, and given the conditioning in (\ref{eq:conditions}), data sets correspond to points in $(\phi,\psi)$-space, and under the null such points are uniform on the unit square. The conditioning on $\hat{\tau}$ being a detected changepoint corresponds to restricting the possible set of $(\phi,\psi)$ values -- to the non-grey area in Figure \ref{fig:pvalueregions}. We have plotted the $(\phi,\psi)$ value for the observed data by a cross in the top row of Figure \ref{fig:pvalueregions}. The $p$-value of \cite{jewell-testing} then fixes the $\psi$ value so the conditional distribution of $\phi$ is uniform on the coloured line -- i.e. all values that are consistent with detecting a change at $\hat{\tau}$ for that value of $\psi$. The $p$-value is the probability of observing a more extreme value than that for the data -- which is the proportion of the line that is red.

By comparison, the $p$-value of (\ref{eq:ideal_p}) allows $\psi$ to vary. It is thus the probability of observing a more extreme value of $\phi$ than that for the data over all possible $(\phi,\psi)$ values that are consistent with $\hat{\tau}$ being a detected change. This is the proportion of the non-grey area that is red in the top right plot of Figure \ref{fig:pvalueregions}. If we generate the data by simulating a $(\phi,\psi)$ point uniformly in the non-grey region, then it is simple to show that the distribution of either $p$-value will be uniform on $[0,1]$.

To see why the ideal $p$-value (\ref{eq:ideal_p}) is to be preferred, in the bottom row of Figure \ref{fig:pvalueregions} we plot the set of $(\phi,\psi)$ values that would correspond to data with a post-selection $p$-value of less than 0.2. For both $p$-values these give regions whose area is 0.2 of the non-grey area. The difference is the shape of the regions, with the ideal $p$-value consisting of requiring just $\phi$ greater than some constant, whereas the $p$-value of \cite{jewell-testing} has different regions for $\phi$ as we vary $\psi$. The former will have more power if we have alternative hypotheses that, compared to the null, place increasing probability on larger values of $|\phi|$.

To make this precise, let $\mathcal{A}$ be the projection of the data we condition on (\ref{eq:conditions}). Under the null, and conditional on $\mathcal{A}$ denote the density for $(\phi,\boldsymbol{\psi})$ as
\[
f(\phi)\prod_{i=1}^{2h-2}g(\psi_i).
\]
Consider alternative hypotheses that correspond to a density of $(\phi,\boldsymbol{\psi})$ of the form 
\begin{equation} \label{eq:alt}
    k(\phi) f(\phi)\prod_{i=1}^{2h-2}g(\psi_i),
\end{equation}
for some function $k(\phi)$. That is, under the alternative hypothesis the distribution of $\phi$ is altered, and $k(\phi)$ represents the ratio of density between the alternative and the null.

\begin{theorem} \label{thm:ideal}
Conditional on $\mathcal{A}$ and $(\phi,\boldsymbol{\psi})\in \mathcal{S}$, define $P_I$ to be the $p$-value given by (\ref{eq:ideal_p}), and $P^*$ be any other valid $p$-value, i.e. that satisfies that under the null
\[
    \Pr(P^* \leq \alpha) \leq \alpha, ~ \mbox{for all $\alpha \in [0,1]$.}
\]
Then under an alternative with density of the form (\ref{eq:alt}) for a function $k(\phi)=\tilde{k}(|\phi|)$ with $\tilde{k}$ increasing,
\[
    \Pr(P_I \leq \alpha) \geq \Pr(P^* \leq \alpha), ~ \mbox{for all $\alpha \in [0,1]$.}
\]
\end{theorem}
The condition that the distribution under the alternative is of form (\ref{eq:alt}) for appropriate $k(\phi)$, holds if under the alternative we have a common mean for $X_{\hat{\tau}-h+1},\ldots,X_{\hat{\tau}}$, and a different common mean for $X_{\hat{\tau}+1},\ldots,X_{\hat{\tau}+h}$, and that the density for the size of the change in mean is symmetric about 0 -- see Appendix \ref{sec:prooft1}  for details.

\subsection{Estimating $p$-values with sampling} \label{sec:p-value-ests}

Unfortunately it is not possible to analytically calculate the ideal post-selection $p$-value (\ref{eq:ideal_p}). Instead we will resort to using Monte Carlo to estimate it, under the assumption that we have a method for calculating the null distribution of $\phi$ given $\boldsymbol{\psi}$ -- this would refer to any combination of type of change, choice of null hypothesis and method for detecting the changepoints for which current post-selection inference methods exist.

Let 
\begin{equation*}
    \mathcal{S} = \{ (\phi, \boldsymbol{\psi}) : \hat{\tau} \in \mathcal{M} \left( \boldsymbol{X}'(\phi, \boldsymbol{\psi}) \right) \},
\end{equation*}
so the conditioning event for (\ref{eq:ideal_p}) corresponds to $(\phi,\boldsymbol{\psi})$. Furthermore, define
\[
\mathcal{S}_{\boldsymbol{\psi}} = \{ \phi : \hat{\tau} \in \mathcal{M} \left( \boldsymbol{X}'(\phi, \boldsymbol{\psi}) \right) \},
\]
the set of $\phi$ values corresponding to data where we estimate $\hat{\tau}$ as a changepoint, for a given value of $\boldsymbol{\psi}$. 
These regions $\mathcal{S}_{\boldsymbol{\psi}}$ will depend on the algorithm used to estimate the changepoints, but we can make use of existing methods to calculate them. For example, \cite{jewell-testing} show how to calculate these regions for the change in mean problem with changepoints estimated by binary segmentation (their method is easily extended to variants such as wild binary segmentation and narrowest over threshold) and $L_0$ penalised likelihood methods. In each case the set $\mathcal{S}_{\boldsymbol{\psi}}$ consists of a finite union of intervals in $\phi$.

Now that we have $2h - 2$ additional parameters in $\boldsymbol{\psi}$, the truncation region $\mathcal{S}$ becomes much more complicated to calculate explicitly, as the values of $\phi$ that yield $\hat{\tau} \in \mathcal{M}(\boldsymbol{X}'(\phi))$ depend on $\boldsymbol{\psi}$. However, for a given $\boldsymbol{\psi}^*$, by replacing $\boldsymbol{X}$ with $\boldsymbol{X}_{\psi^*} = \boldsymbol{U \Psi}^* + \frac{1}{||\boldsymbol{\nu}_{\hat{\tau}}||_2^2} \boldsymbol{\nu}_{\hat{\tau}} \boldsymbol{\nu}_{\hat{\tau}}^T \boldsymbol{X} + \frac{1}{||\boldsymbol{a}||_2^2} \boldsymbol{a a}^T \boldsymbol{X} + \boldsymbol{B B}^T \boldsymbol{X}$, we can calculate $\mathcal{S}_{\psi^*}$ using the method of \cite{jewell-testing}. We can then calculate a $p$-value conditional on $\boldsymbol{\psi}^*$:
\begin{equation*}
    p_{\psi^*} = \frac{\Pr(|\phi| \geq |\boldsymbol{\nu}_{\hat{\tau}}^T \boldsymbol{X}_{obs}| \cap \phi \in \mathcal{S}_{\psi^*})} {\Pr(\phi \in \mathcal{S}_{\psi^*})}.
\end{equation*}

To estimate the overall $p$-value, we take $N$ samples, $\{\boldsymbol{\psi}^{(1)}, \ldots, \boldsymbol{\psi}^{(N)}\}$, and calculate $\mathcal{S}_{\boldsymbol{\psi}^{(j)}}$ for each $\boldsymbol{\psi}^{(j)}$. We then estimate the $p$-value as
\begin{equation}
  \label{eq:pN}
  \frac{\Pr(|\phi| \geq |\boldsymbol{\nu}_{\hat{\tau}}^T \boldsymbol{X}_{obs}| \cap \phi \in \mathcal{S})}{\Pr(\phi \in \mathcal{S})} \approx
    \frac{\frac{1}{N} \sum_{j=1}^N \Pr(|\phi| \geq |\boldsymbol{\nu}_{\hat{\tau}}^T \boldsymbol{X}_{obs}| \cap \phi \in \mathcal{S}_{\boldsymbol{\psi}^{(j)}})}{\frac{1}{N} \sum_{j=1}^N \Pr(\phi \in \mathcal{S}_{\boldsymbol{\psi}^{(j)}})} = \hat{p}_N.
\end{equation}
This can also be written as a weighted average of individual $p$-value estimates
\begin{equation}
  \label{eq:weighted}
    \hat{p}_N = \frac{1}{\sum_{j=1}^N w_j} \sum_{j=1}^N w_j p_{\psi^{(j)}},
\end{equation}
where $w_j = \Pr(\phi \in \mathcal{S}_{\boldsymbol{\psi}^{(j)}})$. Since each $\mathcal{S}_{\boldsymbol{\psi}^{(j)}}$ consists of a union of intervals, and $\phi \sim N(0, \frac{2\sigma^2}{h})$, it is straightforward to calculate $w_j$ and $p_{\boldsymbol{\psi}^{(j)}}$.

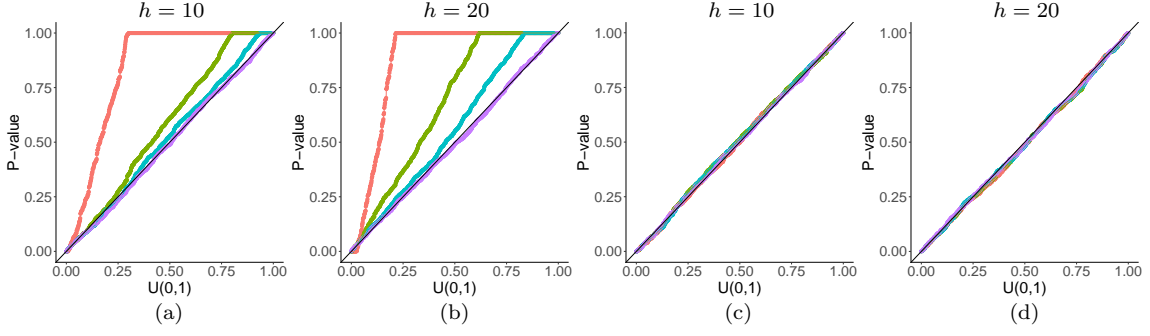
\begin{figure}
    \centering
    \includegraphics[width=0.25\linewidth]{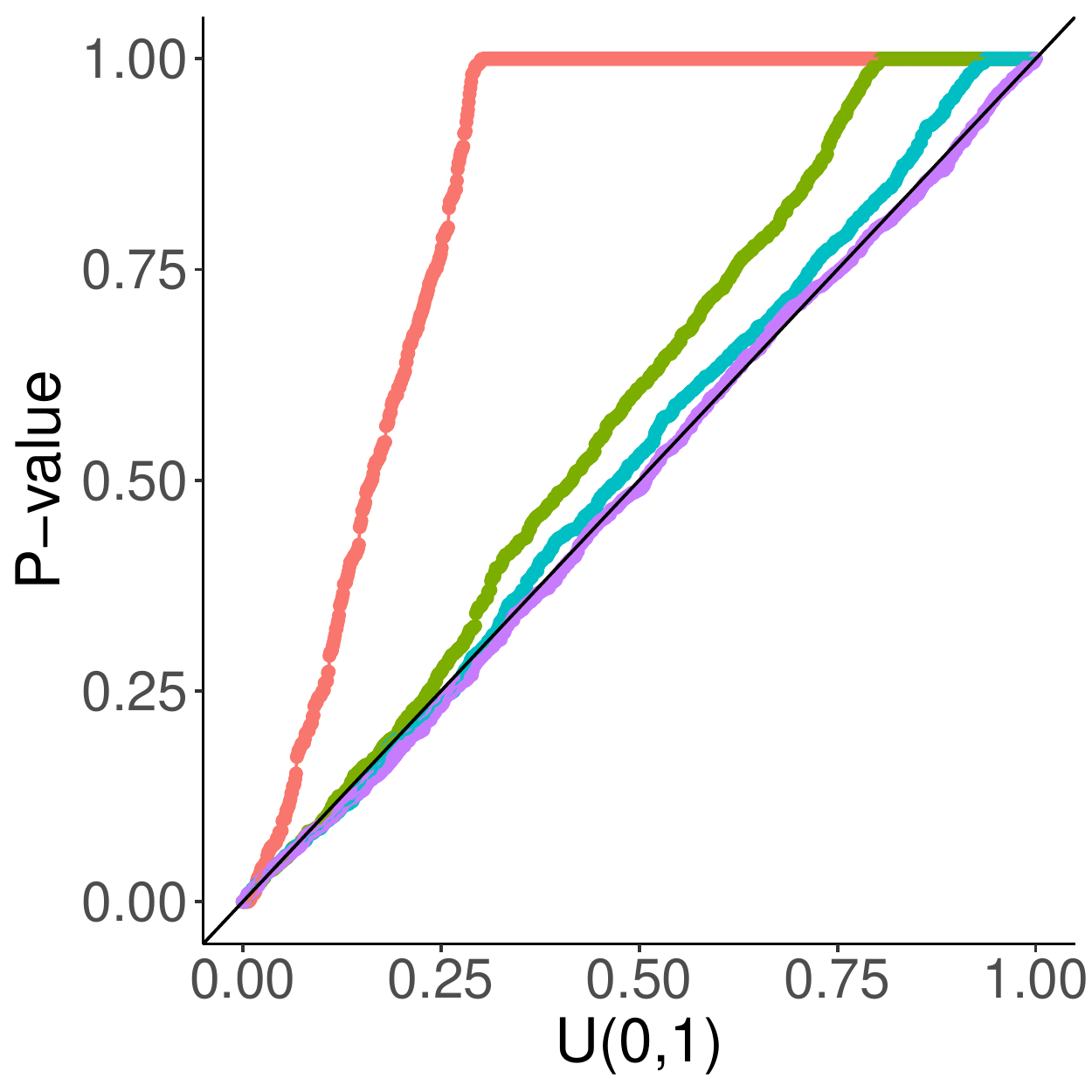}
        \put(-50,-7){\footnotesize{(a)}}
        \put(-56,107){\footnotesize{$h = 10$}}
    \includegraphics[width=0.25\linewidth]{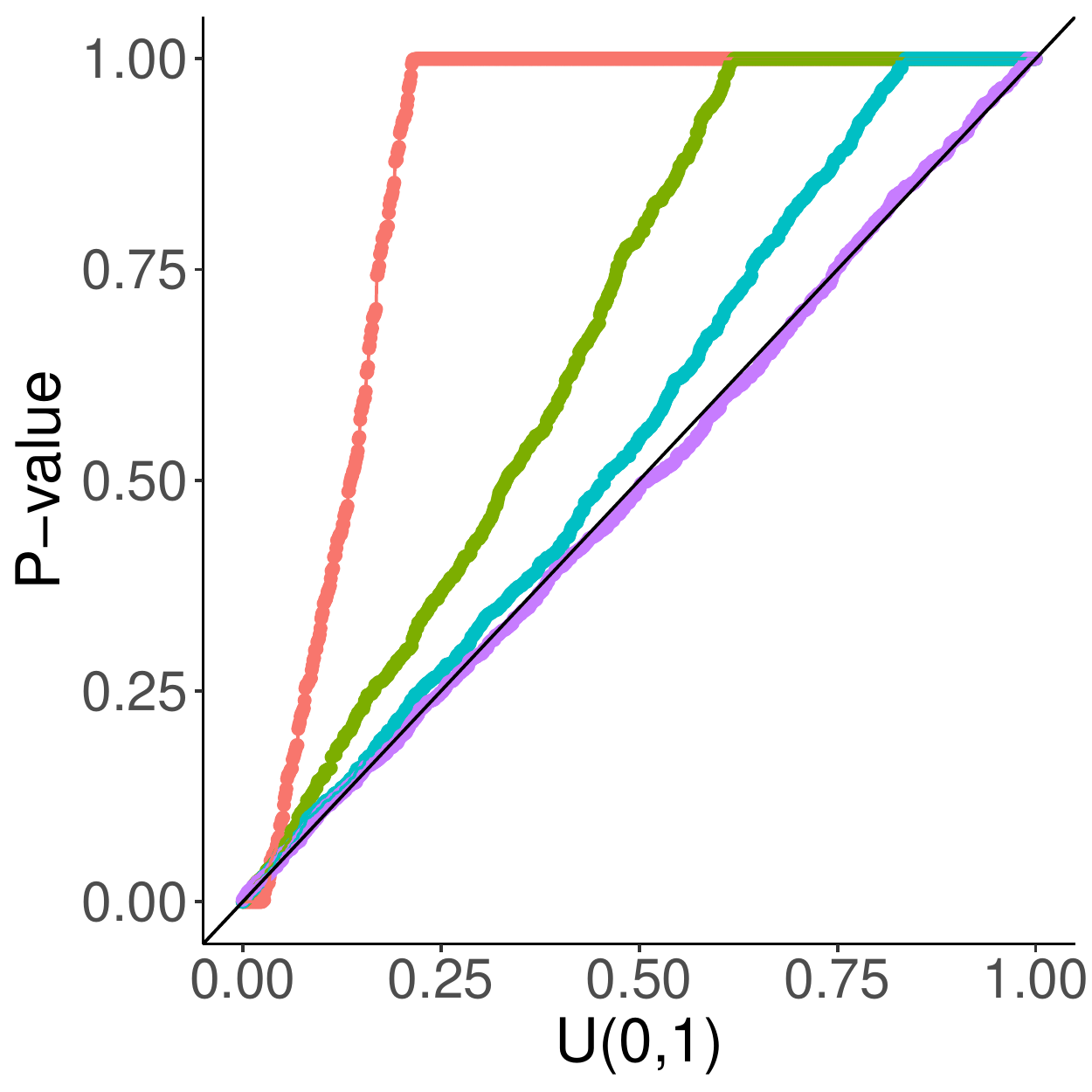}
        \put(-50,-7){\footnotesize{(b)}}
        \put(-56,107){\footnotesize{$h = 20$}}
    \includegraphics[width=0.25\linewidth]{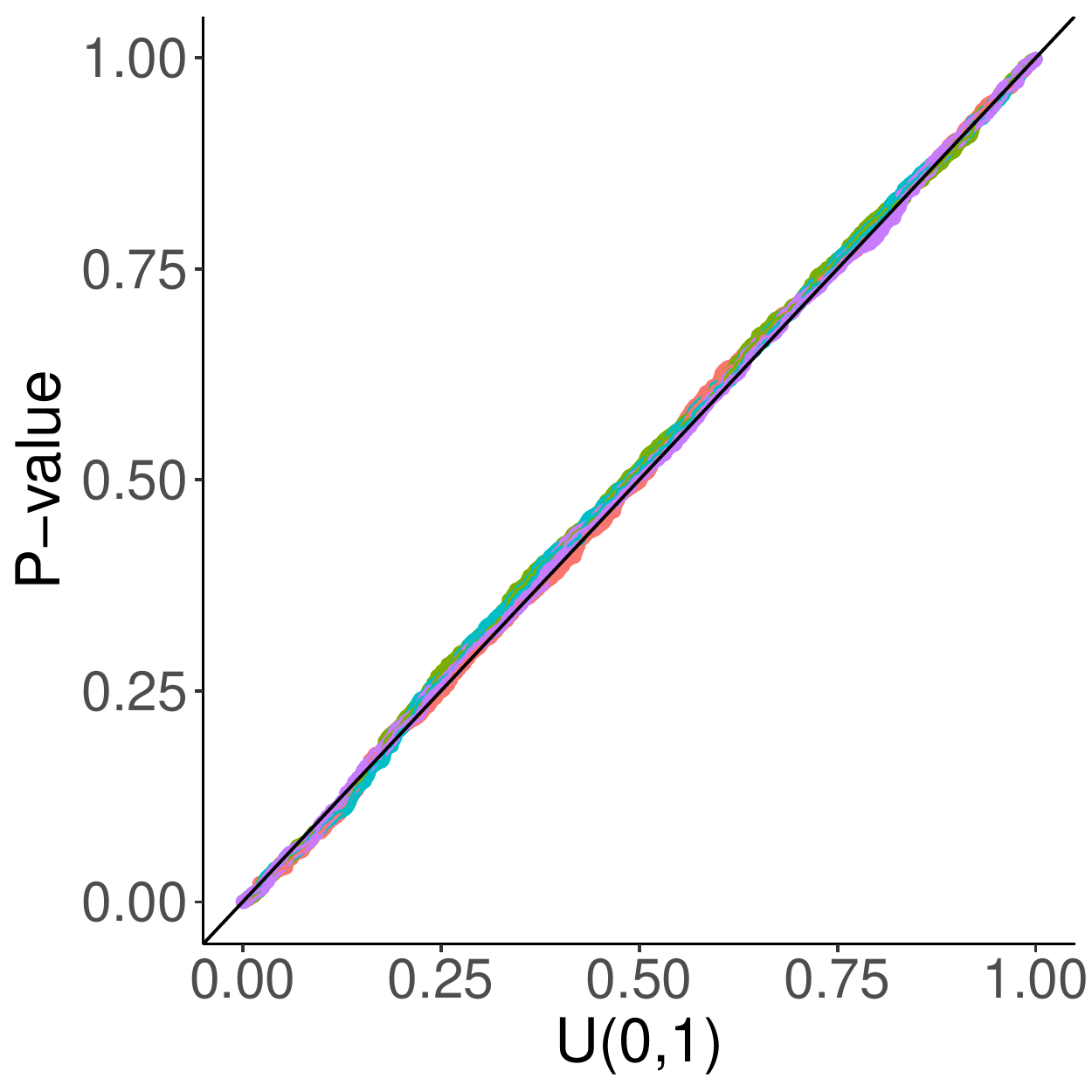}
        \put(-50,-7){\footnotesize{(c)}}
        \put(-56,107){\footnotesize{$h = 10$}}
    \includegraphics[width=0.25\linewidth]{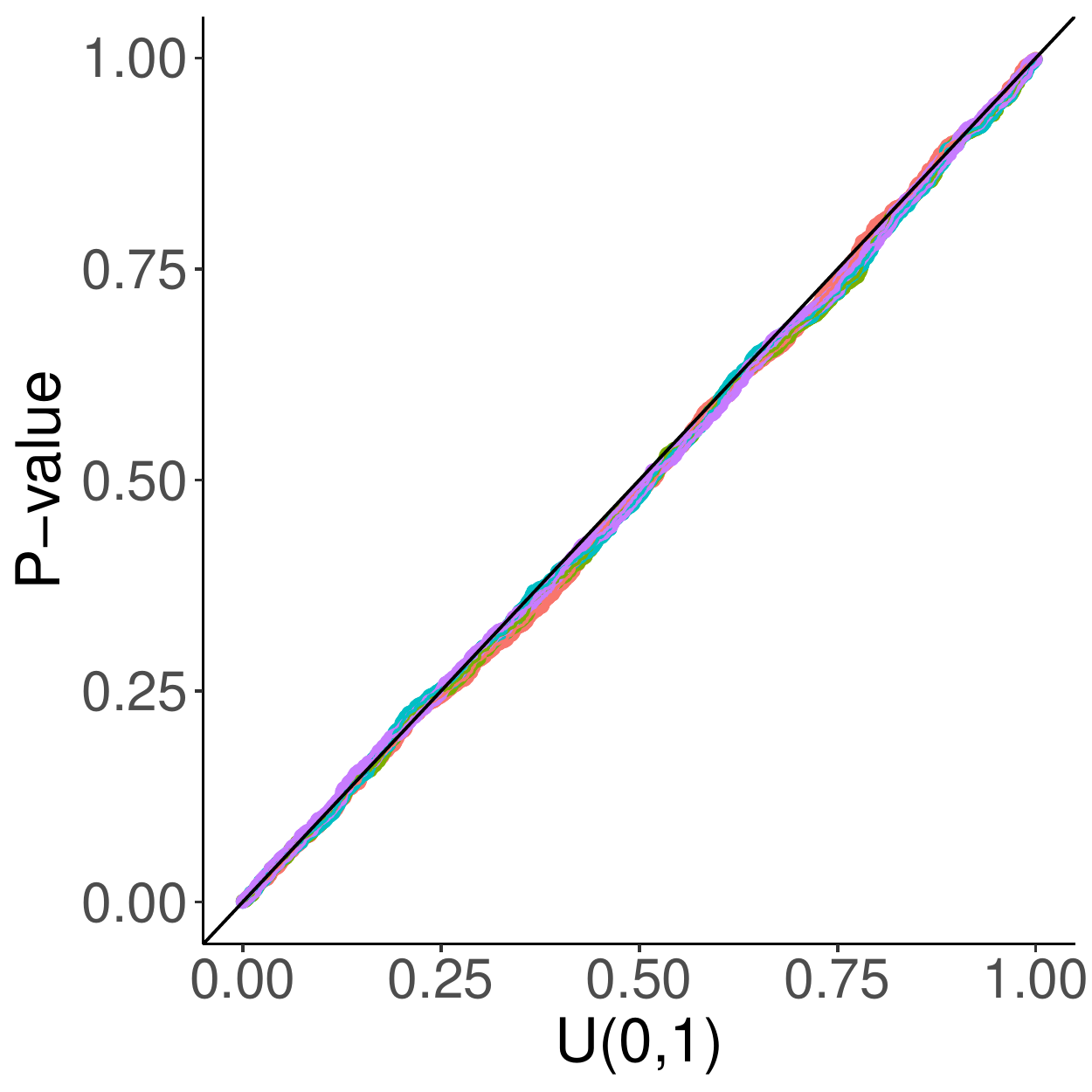}
        \put(-50,-7){\footnotesize{(d)}}
        \put(-56,107){\footnotesize{$h = 20$}}
    \caption{\small{QQ plots of $p$-value estimates, simulated under $H_0$ with $T = 1000$, for different values of $h$ and $N$. On each plot the ordered $p$-values obtained using different values of $N$ ($N = 1, 5, 10, 50$) are plotted against theoretical quantiles from $U(0,1)$. In (a) and (b) the $p$-values are calculated as in Equation \ref{eq:pN}, where all $\boldsymbol{\psi}^{(j)}$'s are simulated randomly. In (c) and (d), we take $\boldsymbol{\psi}^{(1)} = \boldsymbol{U}^T \boldsymbol{X}_{obs}$. If $p$-values are valid, the points should lie approximately along the line $y = x$.}}
    \label{fig:qq-N}
\end{figure}

As $N \rightarrow \infty$ this Monte Carlo estimate will converge to the ideal post-selection $p$-value (\ref{eq:ideal_p}). However for finite $N$ it will not necessarily be a valid $p$-value, in that there is no guarantee that under the null, and conditional on choosing to test the null, that the $p$-value will be uniformly distributed on $[0,1]$.

To see this, we simulated this Monte Carlo $p$-value for different values of $N$: see Figure \ref{fig:qq-N}(a) and (b). We see that, in particular, there is a non-trivial probability that some $\hat{p}_{\psi^{(j)}} = 1$, as for some values of $\boldsymbol{\psi}$ we have $\mathcal{S}_{\boldsymbol{\psi}} \subset \{ \phi : |\phi| \geq |\boldsymbol{\nu}_{\hat{\tau}}^T \boldsymbol{X}_{obs}| \}$.

Remarkably, we can overcome these issues by just setting one of the $\boldsymbol{\psi}$ values to be the value for the observed data. Remember that $\boldsymbol{\psi} = \boldsymbol{U}^T \boldsymbol{X}$. Let $\boldsymbol{\psi}^{(1)} = \boldsymbol{U}^T \boldsymbol{X}_{obs}$. 
Simulate $\boldsymbol{\psi}^{(2)},\ldots,\boldsymbol{\psi}^{(N)}$ independently from the null distribution for $\boldsymbol{\psi}$, and calculate the $p$-value as $\hat{p}_N$ in (\ref{eq:pN}). 

The following theorem shows that the resulting post-selection $p$-value will be distributed uniformly on $[0,1]$ under the null, for any value of $N$.

\begin{theorem}
  \label{th:p-values}
    Let
    \begin{equation*}
        \hat{p}_N = \frac{1}{\sum_{j=1}^N w_j} \sum_{j=1}^N w_j p_{\psi^{(j)}},
    \end{equation*}
    where $w_j = \Pr(\phi \in \mathcal{S}_{\psi^{(j)}})$.
    Given that there is one $j^* \in \{1, \ldots, N\}$ such that $\psi^{(j^*)}$ corresponds to the observed data, and that other $\psi^{(j)}$ are drawn independently from their distribution under the null, then under $H_0$, $\hat{p}_N \sim U(0,1)$.
\end{theorem}

Figure \ref{fig:qq-N} (c) and (d) show empirical validation of this result. An important consequence is that $\hat{p}_N$ is a valid $p$-value for any value of $N$, even if computational constraints limit $N$ to be small. Below, in Section \ref{sec:sims}, we show that even small to moderate values of $N$ can lead to a substantial increase in power.




\subsection{Extension to other null hypotheses} \label{sec:other-H0}

So far, we have calculated $p$-values for the change in mean model, based on the assumption that there are no other changepoints within a fixed window $h$ of $\hat{\tau}$. However, we can also apply our method to a range of other scenarios, such as different null hypotheses and different types of changepoint model. This may lead to different choices for $\phi$ and $\mathcal{F}(\boldsymbol{X})$, but as long as we can define $\boldsymbol{\psi}$ and have a method for calculating $\mathcal{S}_{\boldsymbol{\psi}}$, we can still use our method. We outline some examples below.

We discuss first the choice of null hypothesis. 
Rather than assume that the mean is constant within a fixed window of size $h$ on either side of $\hat{\tau}$, we can consider the more general case where $H_0$ is of the form
\begin{equation*}
    \mu_{\hat{\tau} - h_1 + 1} = \ldots = \mu_{\hat{\tau}} = \mu_{\hat{\tau} + 1} = \ldots = \mu_{\hat{\tau} + h_2}
\end{equation*}
for some integers $h_1, h_2$, where these can be either fixed in advance or selected based on the locations of neighbouring changepoints. We define $\boldsymbol{\nu}_{\hat{\tau}}$ as
\begin{equation*}
    (\boldsymbol{\nu}_{\hat{\tau}})_t = \begin{cases}
        \frac{1}{h_1} & \text{if } \hat{\tau} - h_1 < t \leq \hat{\tau} \\
        - \frac{1}{h_2} & \text{if } \hat{\tau} < t \leq \hat{\tau} + h_2 \\
        0 & \text{if } t \leq \hat{\tau} - h_1 \text{ or } t > \hat{\tau} + h_2.
    \end{cases}
\end{equation*}
The null hypothesis we have previously considered is the special case that $h_1 = h_2 = h$.

A modification which avoids the problem having of other detected changepoints within the region of interest is to choose a pre-determined $h$ but truncate the region if another changepoint is detected within $h$ of $\hat{\tau}$. Denoting the changepoint of interest as $\hat{\tau}_j$ (where $j \in \{ 1, \ldots, K \}$), we hence take $h_1 = \min\{h, \hat{\tau}_j - \hat{\tau}_{j - 1}\}$ and $h_2 = \min \{h, \hat{\tau}_{j + 1} - \hat{\tau}_j \}$. Alternatively, we can allow the region of interest to be determined by the changepoints on either side of $\hat{\tau}_j$: either by taking it to be all points between the two neighbouring changepoints $(\hat{\tau}_{j-1}, \hat{\tau}_{j+1})$, thus $h_1 = \hat{\tau}_j - \hat{\tau}_{j-1}$ and $h_2 = \hat{\tau}_{j+1} - \hat{\tau}_j$ -- this is the scenario used in \cite{hyun-post-selection}, and is also considered in \cite{jewell-testing} --
or by taking it to be the set of points which are closer to $\hat{\tau}_j$ than to neighbouring changepoints, so $h_1 = \lfloor \frac{\hat{\tau}_j - \hat{\tau}_{j-1}}{2} \rfloor$ and $h_2 = \lfloor \frac{\hat{\tau}_j+1 - \hat{\tau}_{j}}{2} \rfloor$. In general, the choice of null hypothesis will depend on what is appropriate for the application.

In any of these cases, we can apply our method in the same way as for the special case so far considered. The one possible difference is the choice of $\mathcal{F}$: if $H_0$ depends on the locations of other estimated changepoints besides $\hat{\tau}$, then we must condition on the locations of all estimated changepoints rather than just whether $\hat{\tau}$ is included in the model. So, in this case we replace the condition $\hat{\tau} \in \mathcal{M}(\boldsymbol{X})$ in Equation \ref{eq:ideal_p} with $\mathcal{M}(\boldsymbol{X}'(\phi, \boldsymbol{\psi})) = \mathcal{M}(\boldsymbol{X}_{obs})$, to get the $p$-value
\begin{equation*}
    p = \Pr_{\phi} \left( |\phi| \geq |\boldsymbol{\nu}_{\hat{\tau}}^T \boldsymbol{X}_{obs}| ~~ | ~~ \mathcal{M}(\boldsymbol{X}'(\phi, \boldsymbol{\psi})) = \mathcal{M}(\boldsymbol{X}_{obs}) \right).
\end{equation*}
This can be seen as a special case, since, given $\hat{\tau} \in \mathcal{M}(\boldsymbol{X}_{obs})$, $\{\phi ~ | ~ \mathcal{M}(\boldsymbol{X}'(\phi, \boldsymbol{\psi})) = \mathcal{M}(\boldsymbol{X}_{obs}) \} \subseteq \{\phi ~ | ~ \hat{\tau} \in \mathcal{M}(\boldsymbol{X}'(\phi, \boldsymbol{\psi})) \}$. \cite{jewell-testing} provides a method to compute this set for fixed $\boldsymbol{\psi}$, so we can apply the same method as in Section \ref{sec:p-value-ests}.

It is also possible to apply our methods to different sorts of models. For example, we consider the model of \cite{chen2023quantifying}. They use a model of the form
\begin{equation*}
    X_t = c_t + \epsilon_t, ~~ t = 1, \ldots, T,
\end{equation*}
where $\epsilon_t \sim N(0, \sigma^2)$ and $c_t = \gamma c_{t-1} + z_t$, with $z_t = 0$ except at changepoints and $\gamma$ assumed known.

In the paper they develop a selective inference procedure similar to the methods in \cite{jewell-testing}, where they fix a window of size $h$ around an estimated change $\hat{\tau}$, and take as the null hypothesis that there are no changes within this window: i.e.
\begin{equation*}
    \left( c_{\hat{\tau}-h+1}, c_{\hat{\tau}-h+2}, \ldots, c_{\hat{\tau}}, \ldots, c_{\hat{\tau}+h} \right) = \left( \gamma^{-h+1}, \gamma^{-h+2}, \ldots, \gamma^0, \ldots, \gamma^h \right) c_{\hat{\tau}}.
\end{equation*}
This leads to a test statistic $\boldsymbol{\nu}^T \boldsymbol{X}$, where $\boldsymbol{\nu}$ is defined as
\begin{equation*}
    \boldsymbol{\nu} = \begin{cases}
        - \frac{\gamma (\gamma^2 - 1)}{\gamma^2 - \gamma^{-2h+2}} \gamma^{t - \hat{\tau}} & \text{if } \hat{\tau} - h < t \leq \hat{\tau} \\
        \frac{\gamma^2 - 1}{\gamma^{2h} - 1} \gamma^{t - \hat{\tau} - 1} & \text{if } \hat{\tau} < t \leq \hat{\tau} + h \\
        0 & \text{if } t \leq \hat{\tau} - h \text{ or } t > \hat{\tau} + h.
    \end{cases}
\end{equation*}
Letting $\phi = \boldsymbol{\nu}^T \boldsymbol{X}$ as before and conditioning on $\boldsymbol{\Pi}_{\hat{\tau}} \boldsymbol{X} = \boldsymbol{\Pi}_{\hat{\tau}} \boldsymbol{X}_{obs}$, they show how to calculate the set $\mathcal{S} = \{ \phi : \hat{\tau} \in \mathcal{M}(\boldsymbol{X}'(\phi)) \}$, and so to calculate the $p$-value
\begin{equation*}
    \Pr \left( \phi \geq \boldsymbol{\nu}^T \boldsymbol{X}_{obs} | \phi \in \mathcal{S} \right),
\end{equation*}
where $\boldsymbol{X}'(\phi)$ is defined as in Equation \ref{eq:x-phi}.


As before, in our approach we condition on the values of $\boldsymbol{X}$ outside of the window $(\hat{\tau} - h + 1, \hat{\tau} + h)$; this is equivalent to fixing the value of $\boldsymbol{B}^T \boldsymbol{X}$ where $\boldsymbol{B}$ is the $T \times \left( T - 2h \right)$ matrix obtained by removing  the columns corresponding to $\{ \hat{\tau} - h + 1, \ldots, \hat{\tau} + h \}$ from the $T \times T$ identity matrix. We also need to account for the unknown parameter $c_{\hat{\tau}}$ within this window. Letting $\boldsymbol{a}$ be defined as
\begin{equation*}
    a_t = 
    \begin{cases}
        \gamma^{t - \hat{\tau}} & \text{if } \hat{\tau} - h + 1 \leq t \leq \hat{\tau} + h \\
        0 & \text{otherwise},
    \end{cases}
\end{equation*}
we condition on $\hat{c}_{\hat{\tau}} = \frac{1}{||\boldsymbol{a}||_2} \boldsymbol{a}^T \boldsymbol{X}$, which is a sufficient statistic for $c_{\hat{\tau}}$.
(It can be shown that $\boldsymbol{a}$ and $\boldsymbol{\nu}$ as defined here are orthogonal.)

Having defined $\boldsymbol{\nu}, \boldsymbol{a}$ and $\boldsymbol{B}$, we can then write, as in Section \ref{sec:ideal},
\begin{equation*}
    \boldsymbol{X} = \boldsymbol{ZX} + \frac{1}{||\boldsymbol{\nu}||_2^2} \boldsymbol{\nu \nu}^T \boldsymbol{X} + \left( \frac{1}{||\boldsymbol{a}||_2^2} \boldsymbol{a a}^T + \boldsymbol{B B}^T \right) \boldsymbol{X}_{obs},
\end{equation*}
where $\boldsymbol{Z} = \boldsymbol{I} - \left( \boldsymbol{BB}^T + \frac{1}{||\boldsymbol{a}||_2^2} \boldsymbol{a a}^T + \frac{1}{||\boldsymbol{\nu}_{\hat{\tau}}||_2^2} \boldsymbol{\nu}_{\hat{\tau}} \boldsymbol{\nu}_{\hat{\tau}}^T \right)$, and hence
\begin{equation*}
    \boldsymbol{X}'(\phi, \boldsymbol{\psi}) = \boldsymbol{U \psi} + \frac{1}{||\boldsymbol{\nu}||_2^2} \boldsymbol{\nu} \phi + \left( \frac{1}{||\boldsymbol{a}||_2^2} \boldsymbol{a a}^T + \boldsymbol{B B}^T \right) \boldsymbol{X}_{obs},
\end{equation*}
with the $p$-value
\begin{equation*}
    p = \Pr_{\phi} \left( \phi \geq \boldsymbol{\nu}_{\hat{\tau}}^T \boldsymbol{X}_{obs} | \mathcal{M}(\boldsymbol{X}'(\phi, \boldsymbol{\psi})) = \mathcal{M}(\boldsymbol{X}_{obs}) \right).
\end{equation*}
Since \cite{chen2023quantifying} provides a method for computing $\mathcal{S}$ given $\boldsymbol{\psi}$, it is then straightforward to simulate $\boldsymbol{\psi}$ and calculate the estimated $p$-value as in Equation \ref{eq:pN}.


Algorithm \ref{alg:1} gives the algorithm to calculate $p$-values, as in Equation \ref{eq:pN}, for a general case. To implement the algorithm, we first must select a changepoint algorithm and a null hypothesis (e.g. that there are no changepoints within a window $h$ of the estimated changepoint), and have a method for calculating $\mathcal{S}_{\psi}$ given both of these. The algorithm is then straightforward to implement.

\begin{algorithm}[h]
    \caption{\small{General algorithm for calculating $p$-values. 
    }}
    \label{alg:1}
    \begin{algorithmic}
        \State Implement changepoint algorithm to obtain $\mathcal{M}(\boldsymbol{X})$.
        \State $j = 1$.
        \State $\boldsymbol{\psi}^{(1)} = \boldsymbol{\psi}_{obs}$
        \State Calculate $\mathcal{S}_{\psi^{(1)}}$.
        \State Calculate $p^{(1)} = \Pr(
        |\phi| \geq |\boldsymbol{\nu}_{\hat{\tau}}^T \boldsymbol{X}_{obs}| ~|~ \phi \in \mathcal{S}_{\psi^{(1)}})$ and $w^{(1)} = \Pr(\phi \in \mathcal{S}_{\psi^{(1)}})$.
        \While {$j < N$}
            \State $j = j+1$.
            \State Sample $\boldsymbol{\psi}^{(j)} \sim N_{2h-2}(\boldsymbol{0}, \sigma^2 \boldsymbol{I})$.
            \State Calculate $\mathcal{S}_{\psi^{(j)}}$.
            \State Calculate $p^{(j)}$ and $w^{(j)}$.
        \EndWhile
        \State Calculate $p$-value estimate:
            \begin{equation*}
                \hat{p}_N = \frac {\frac{1}{N} \sum_{j=1}^N w^{(j)} p^{(j)}} {\frac{1}{N} \sum_{j=1}^N w^{(j)}}.
            \end{equation*}
    \end{algorithmic}
\end{algorithm}

\section{Simulation study} \label{sec:sims}

\subsection{Power simulations}

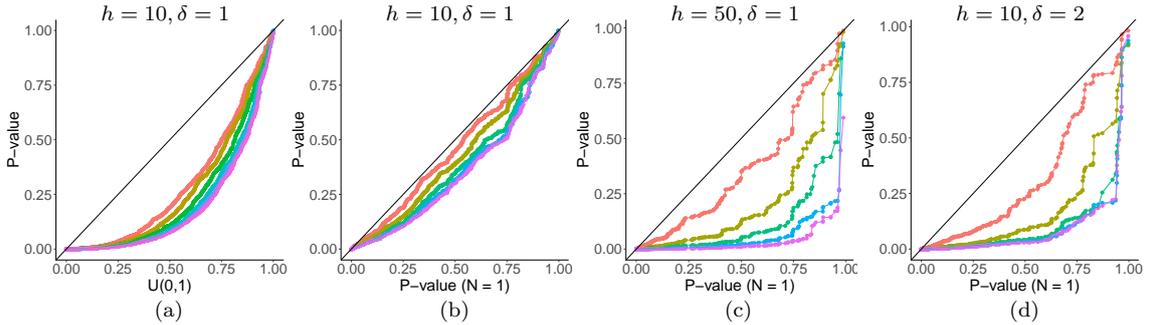
\begin{figure}
    \centering
    \includegraphics[width=0.25\linewidth]{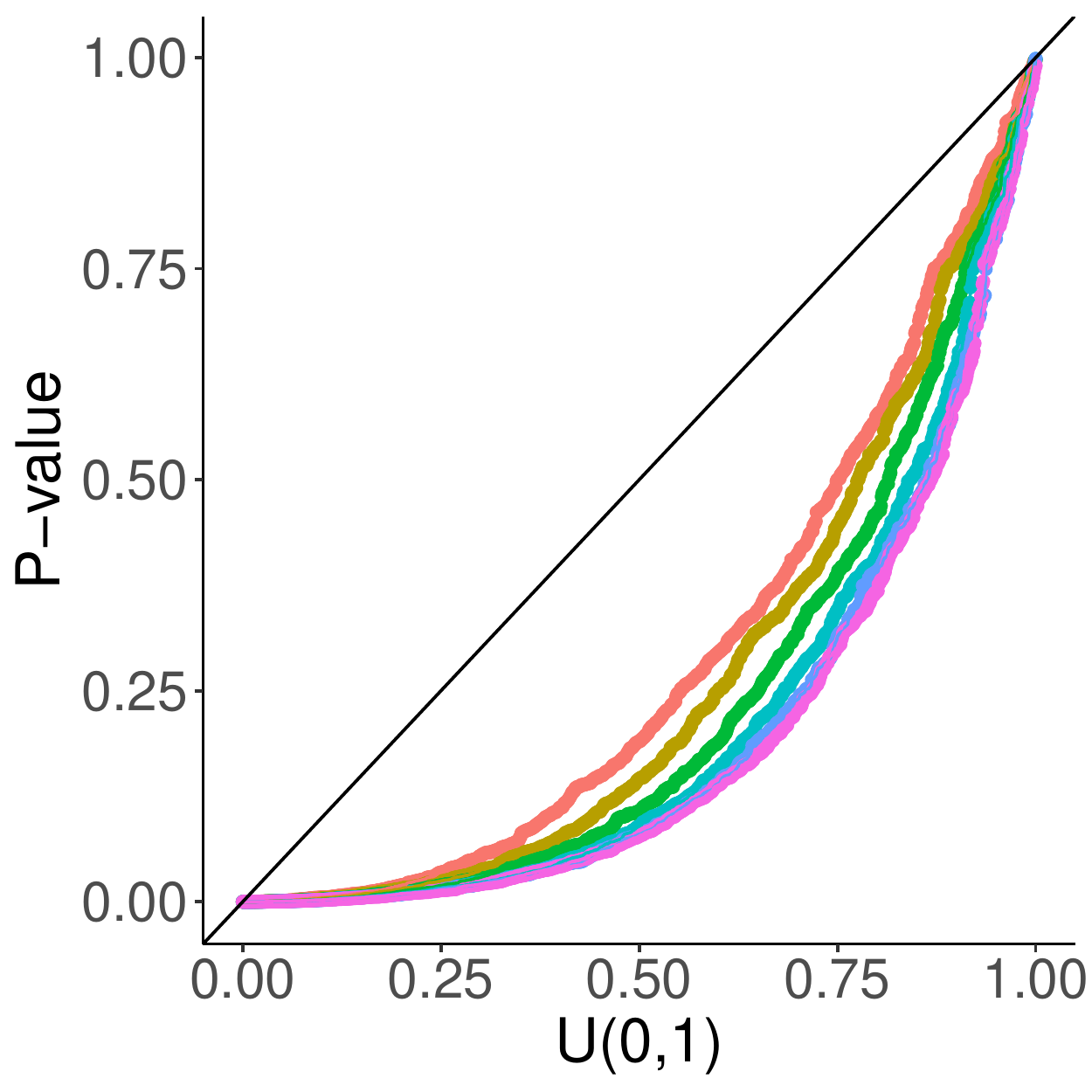}
        \put(-50,-7){\footnotesize{(a)}}
        \put(-70,105){\footnotesize{$h = 10, \delta=1$}}
    \includegraphics[width=0.25\linewidth]{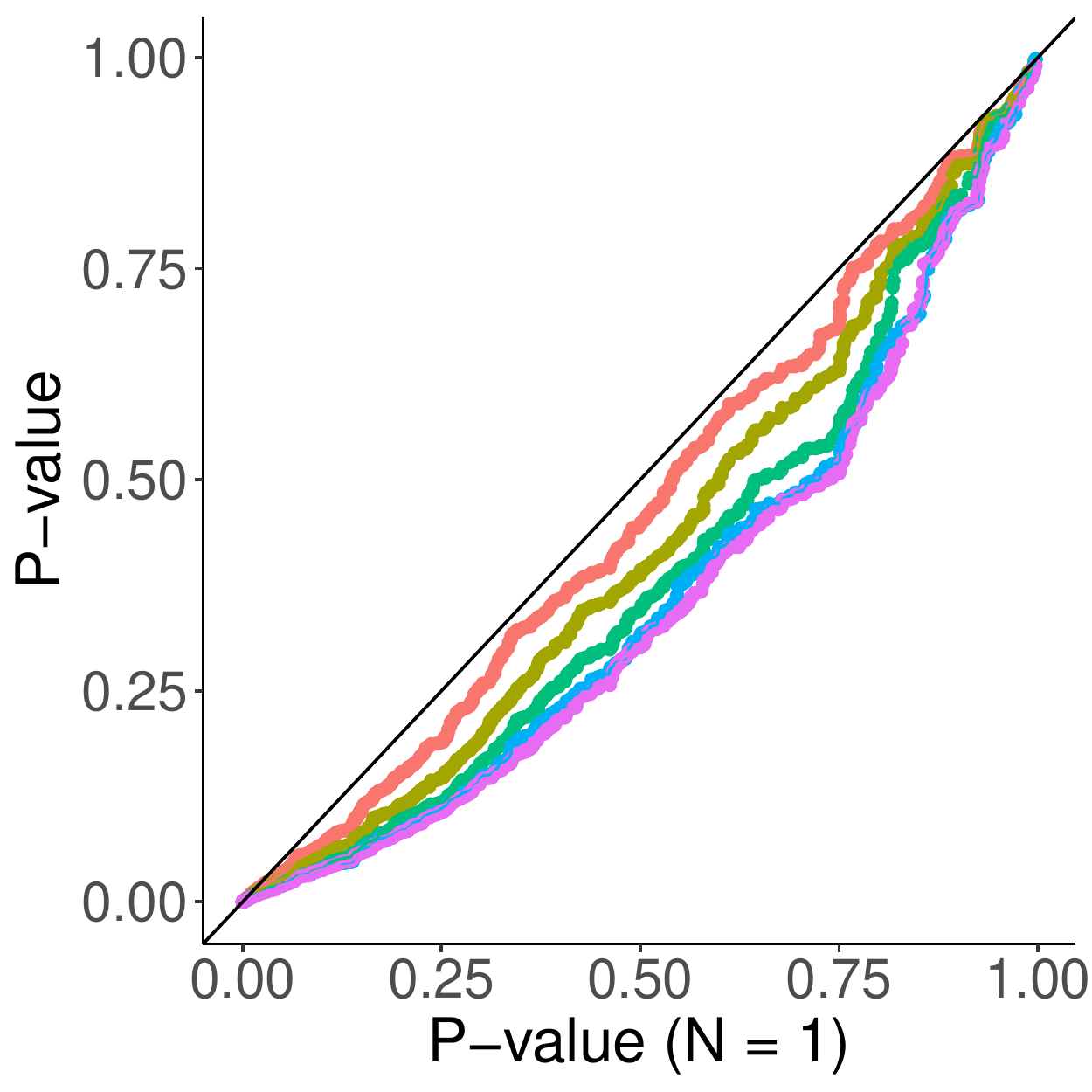}
        \put(-50,-7){\footnotesize{(b)}}
        \put(-70,105){\footnotesize{$h = 10, \delta=1$}}
    \includegraphics[width=0.25\linewidth]{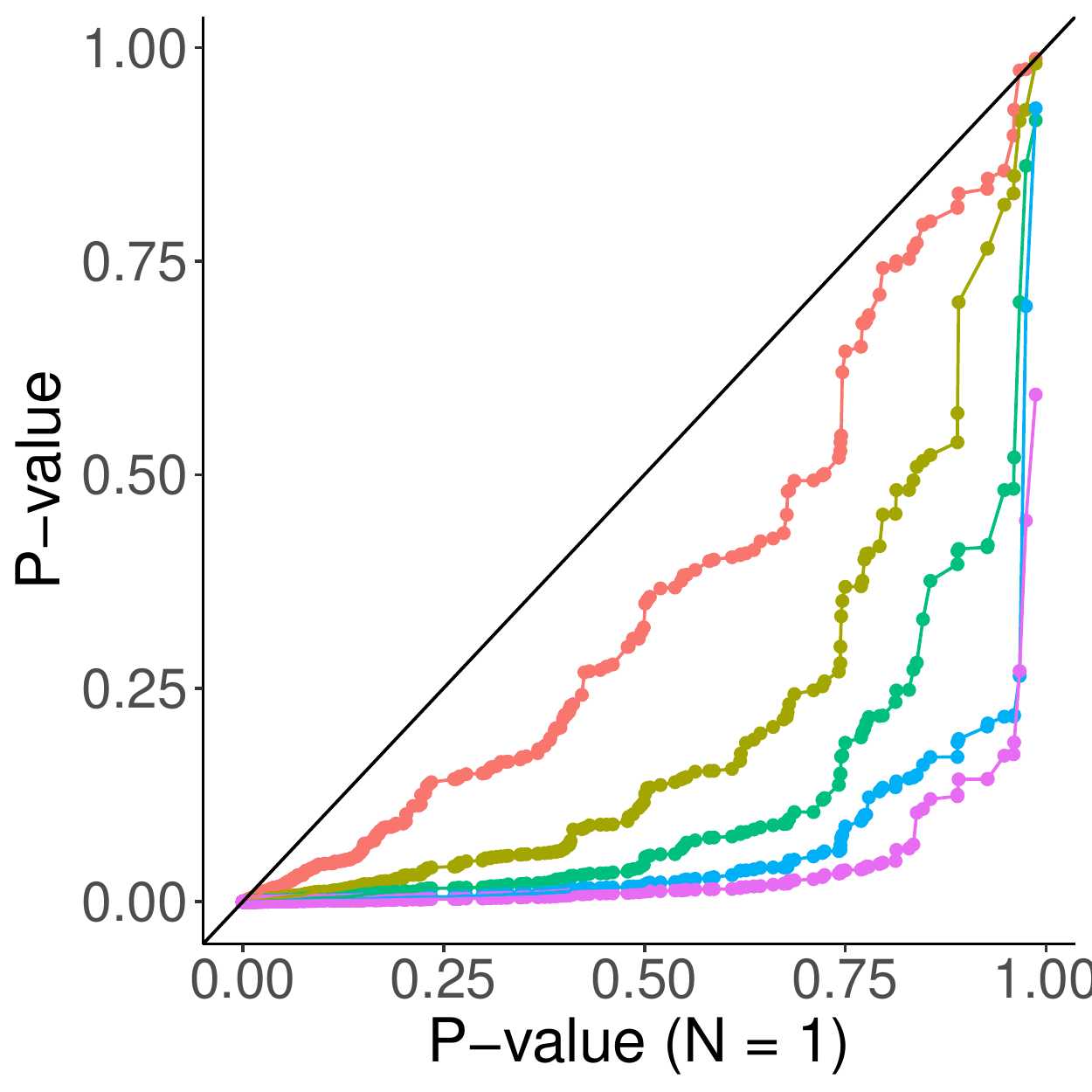}
        \put(-50,-7){\footnotesize{(c)}}
        \put(-70,105){\footnotesize{$h = 50, \delta=1$}}
    \includegraphics[width=0.25\linewidth]{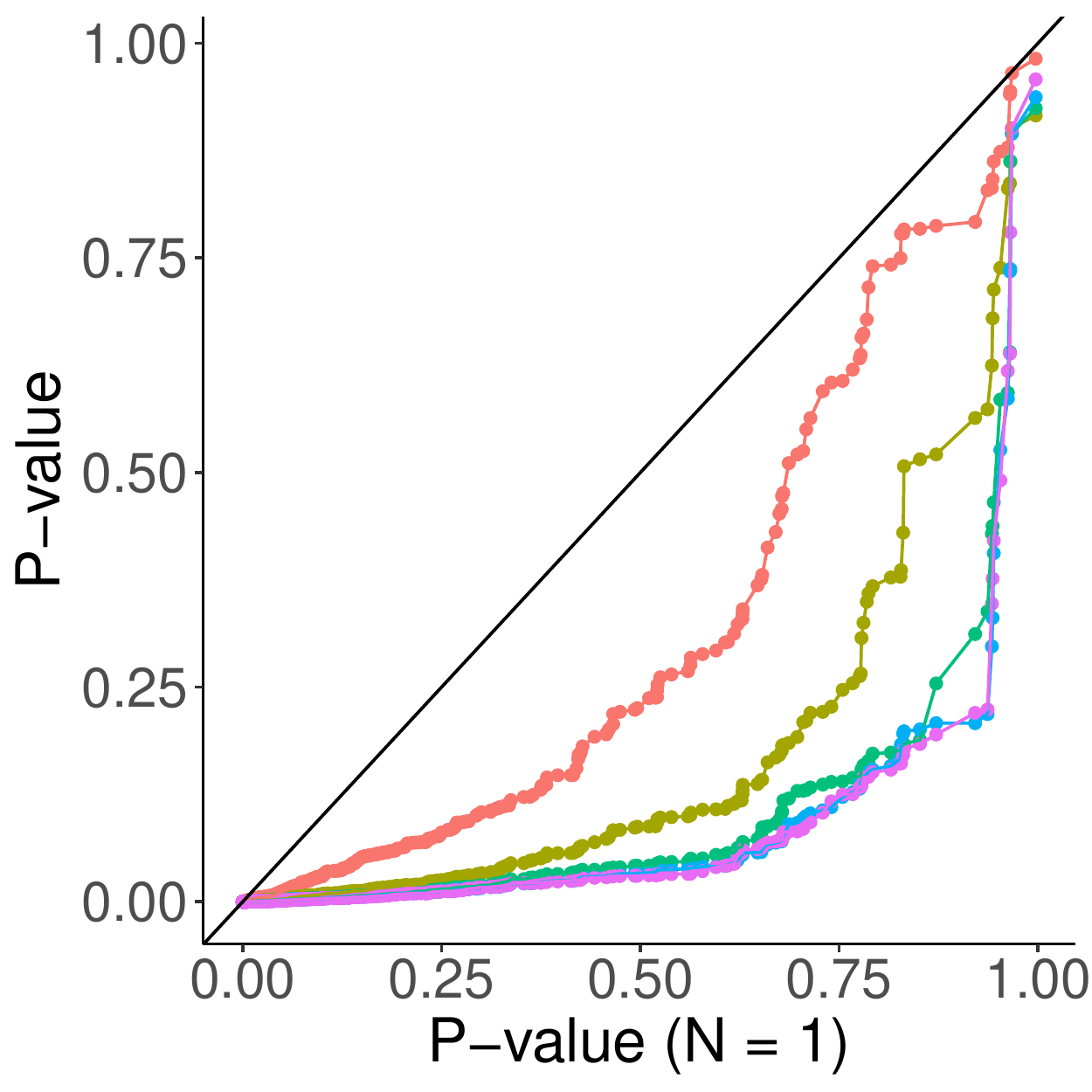}
        \put(-50,-7){\footnotesize{(d)}}
        \put(-70,105){\footnotesize{$h = 10, \delta=2$}}
    \caption{\small{QQ plots of $p$-values for changepoints obtained using binary segmentation: $h$ is the window size and $\delta$ the size of the change in the model from which we simulate. In (a), $p$-values from our method are plotted against theoretical quantiles from $U(0,1)$ for $N = 1, 2, 5, 10, 20, 50$. (b), (c) and (d) show QQ plots of $p$-values calculated using our method (with $N = 2, 5, 10, 20, 50$) against $p$-values from the method of \cite{jewell-testing} (equivalent to $N = 1$).}}
    \label{fig:qq-bs-h1}
\end{figure}

In Section \ref{sec:p-value-ests}, 
we showed that our $p$-value estimates were valid $p$-values under $H_0$. In this section we show that, under $H_1$, our method has greater power to detect changepoints than that of \cite{jewell-testing} for binary segmentation, 
and we investigate how the power changes with window size $h$, size of change $\delta$, and number of samples $N$. All simulations are conducted in R; the code is available at \url{https://github.com/rachelcarrington/changepointsR}.

\begin{figure}
    \centering
    \includegraphics[width=0.32\linewidth]{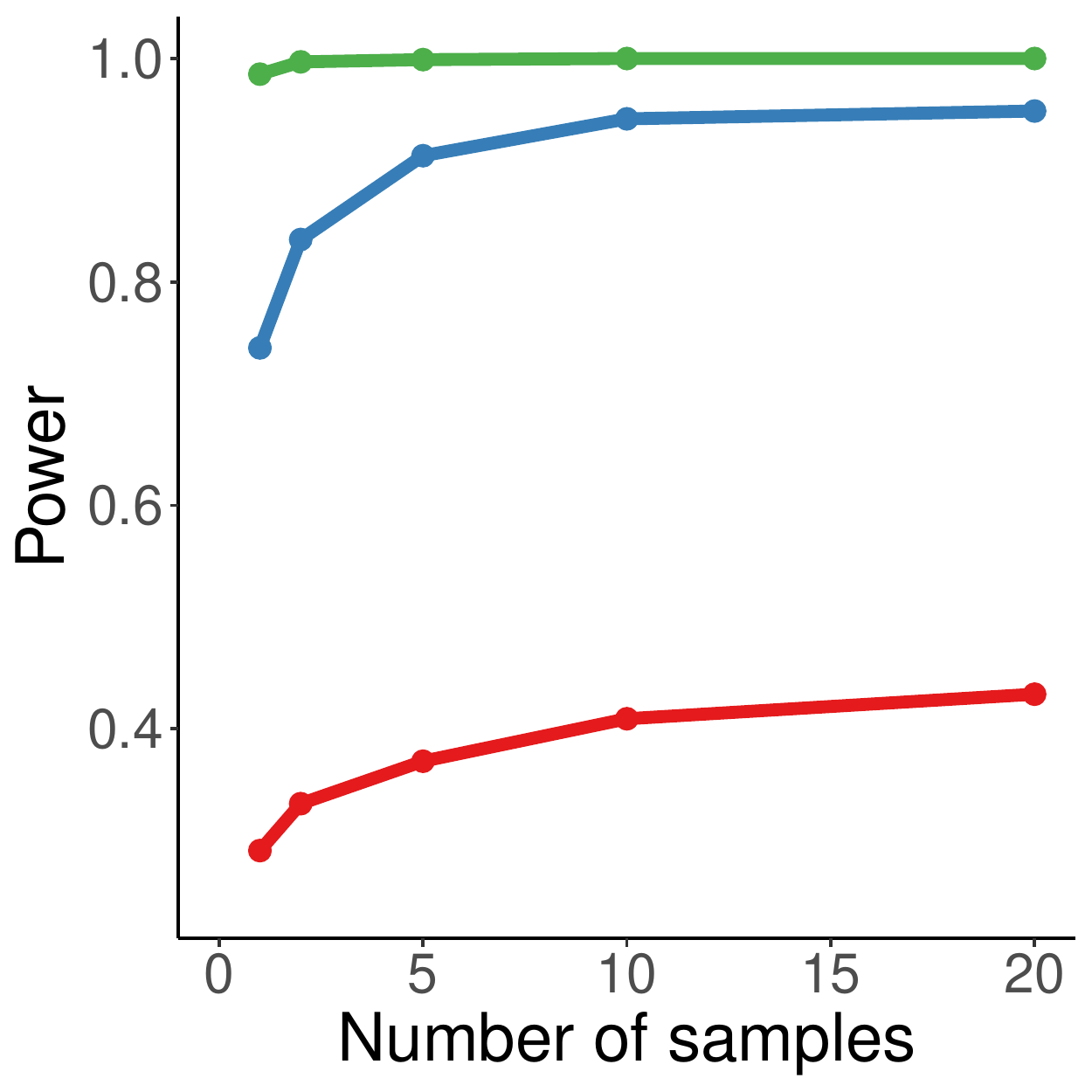}
        \put(-70,140){\footnotesize{$h = 10$}}
    \includegraphics[width=0.32\linewidth]{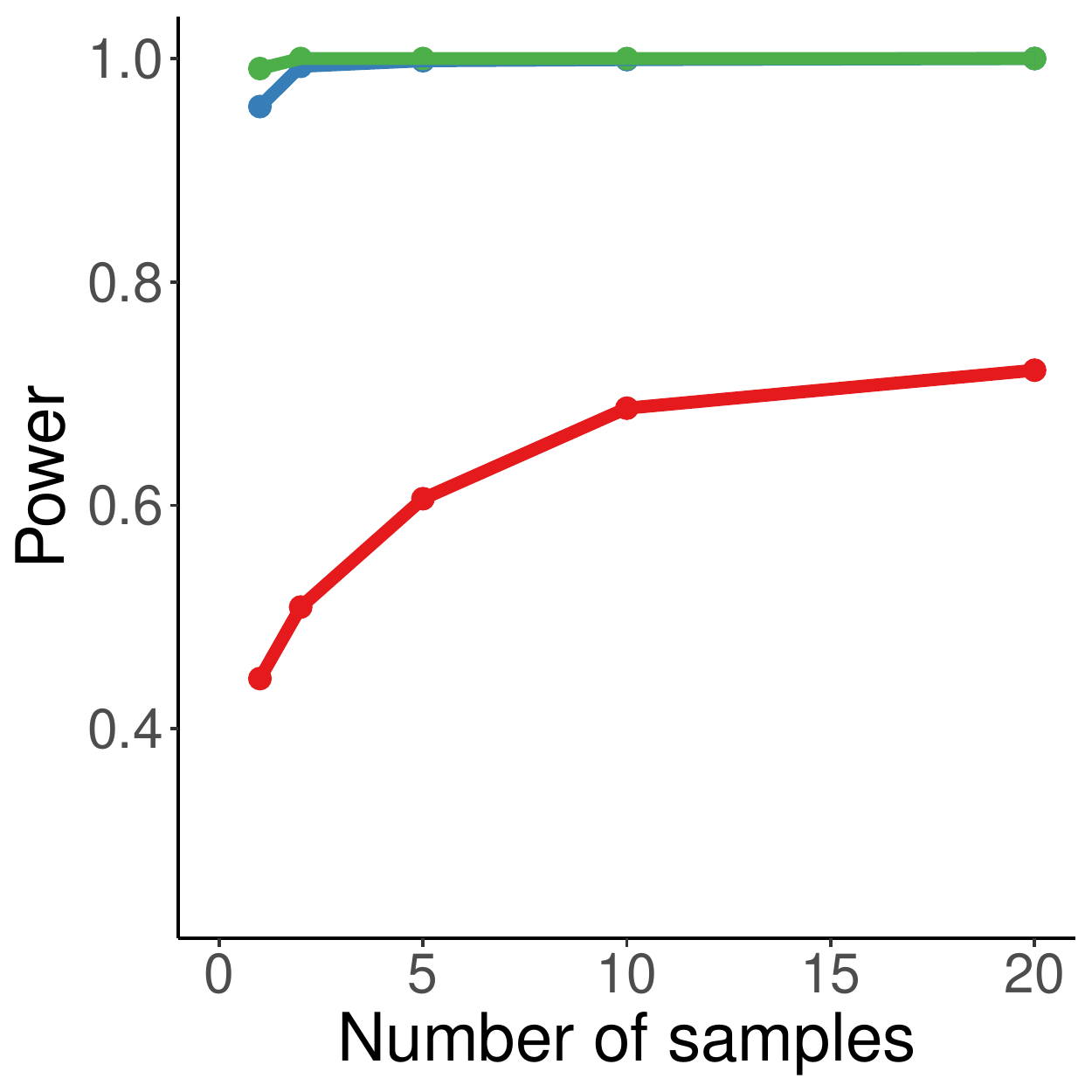}
        \put(-70,140){\footnotesize{$h = 20$}}
    \includegraphics[width=0.32\linewidth]{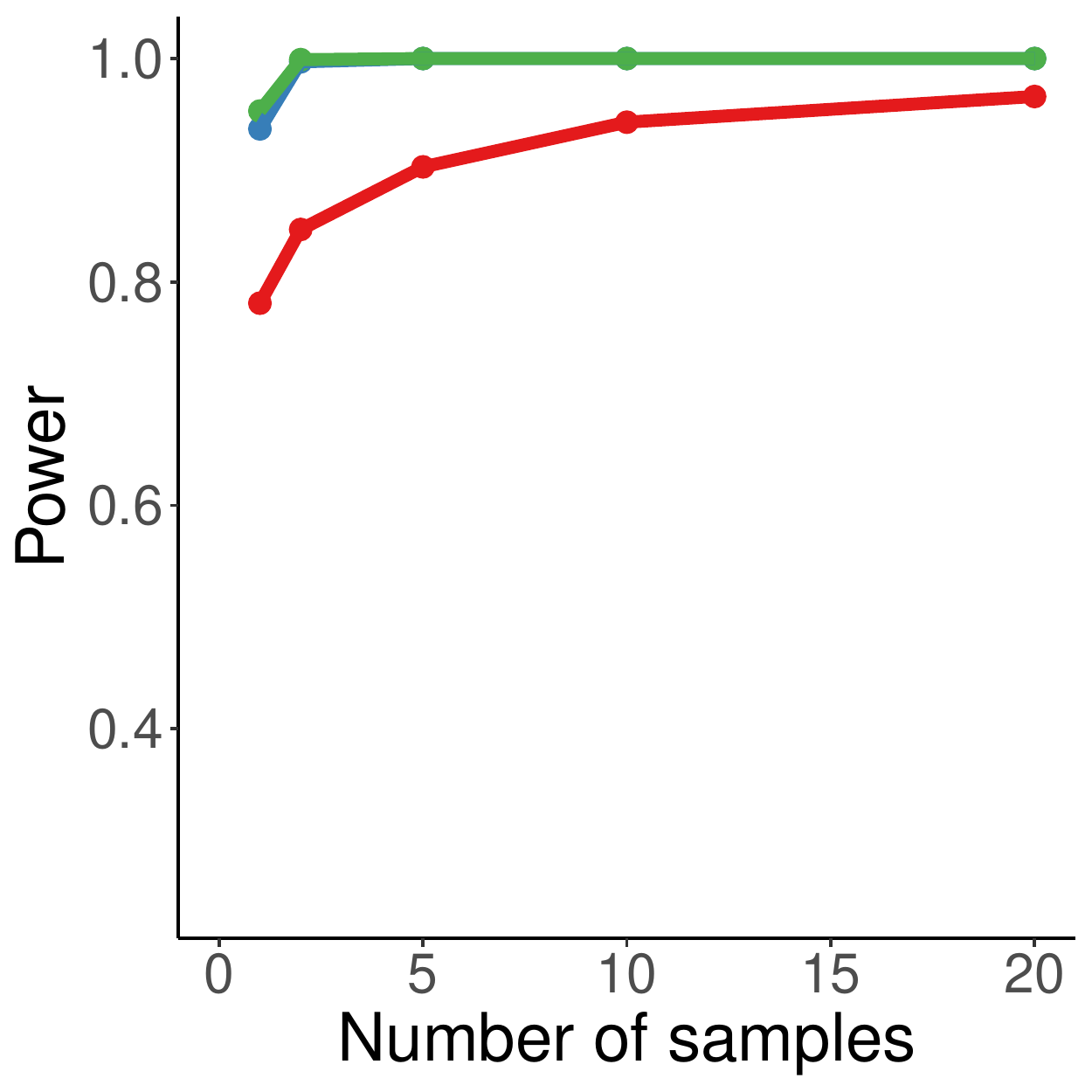}
        \put(-70,140){\footnotesize{$h = 50$}}
    \caption{\small{Rejection rates of $H_0$ for binary segmentation, plotted against $N$. On each plot the three lines show the proportion of samples where the $p$-value was below $0.05$, leading $H_0$ to be rejected. Each line corresponds to a different size of change $\delta$: green corresponds to $\delta = 3$, blue to $\delta = 2$, and red to $\delta = 1$.}}
    \label{fig:power-bs}
\end{figure}

\begin{figure}
    \centering
    \includegraphics[width=0.32\linewidth]{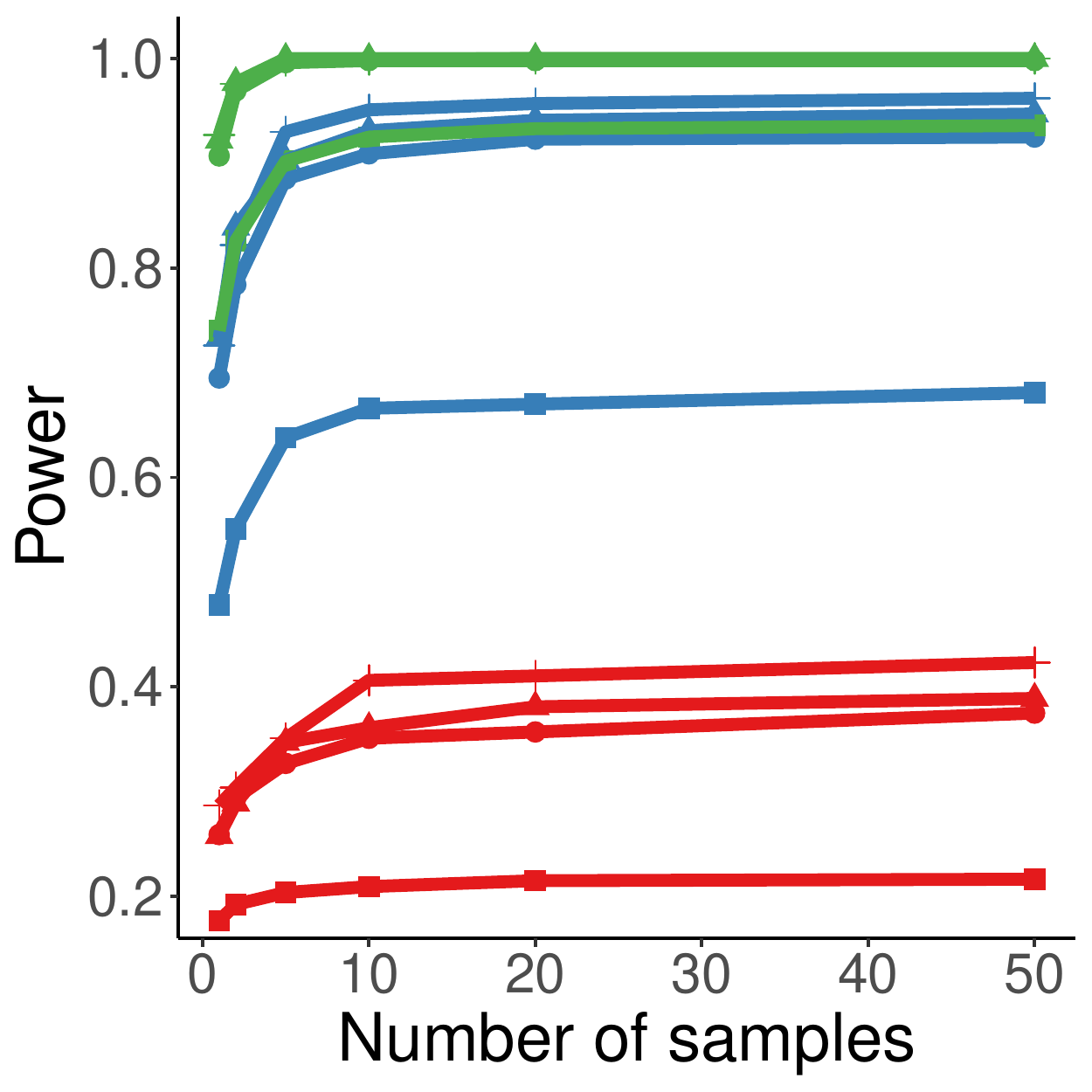}
        \put(-70,138){\footnotesize{$h = 10$}}
    \includegraphics[width=0.32\linewidth]{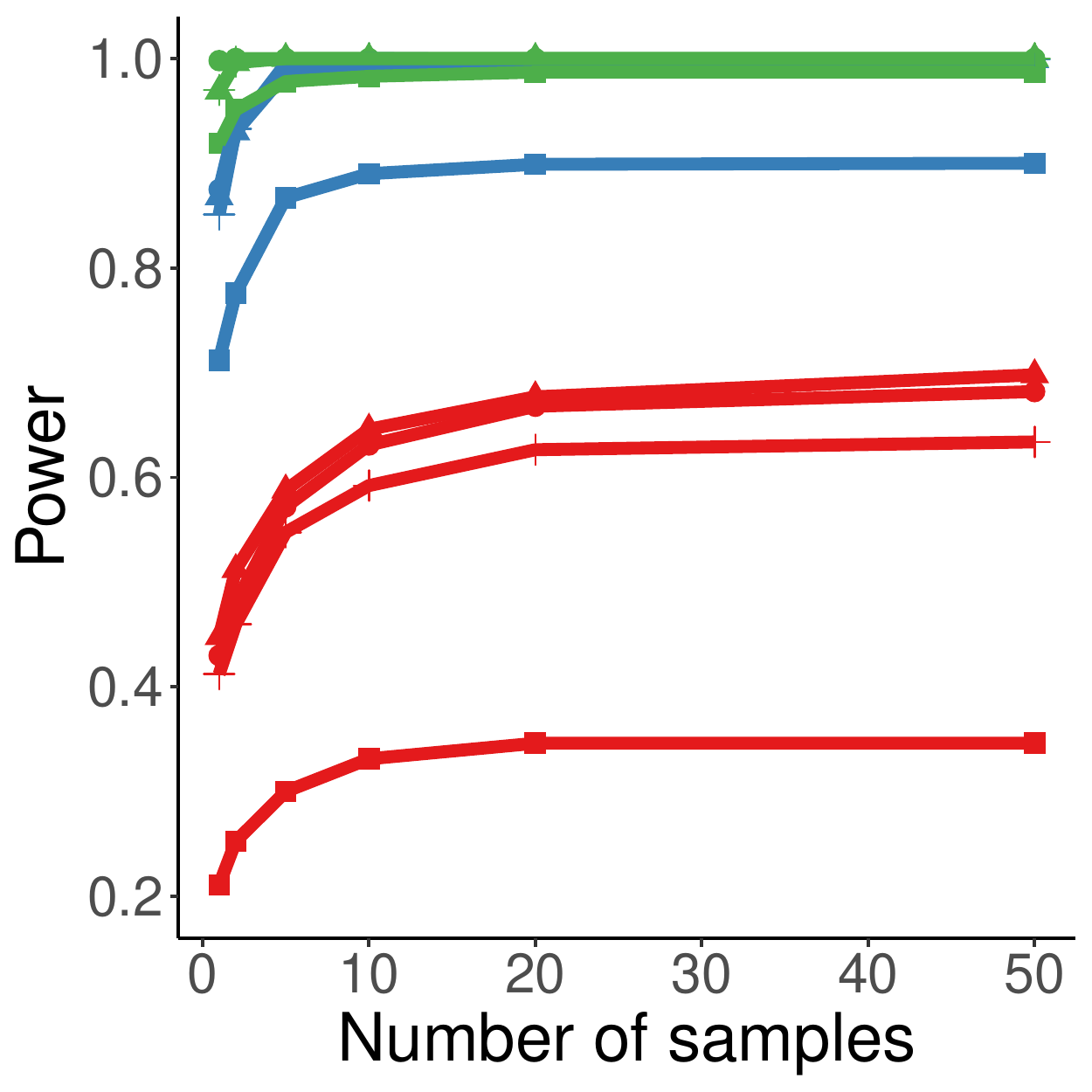}
        \put(-70,138){\footnotesize{$h = 20$}}
    \includegraphics[width=0.32\linewidth]{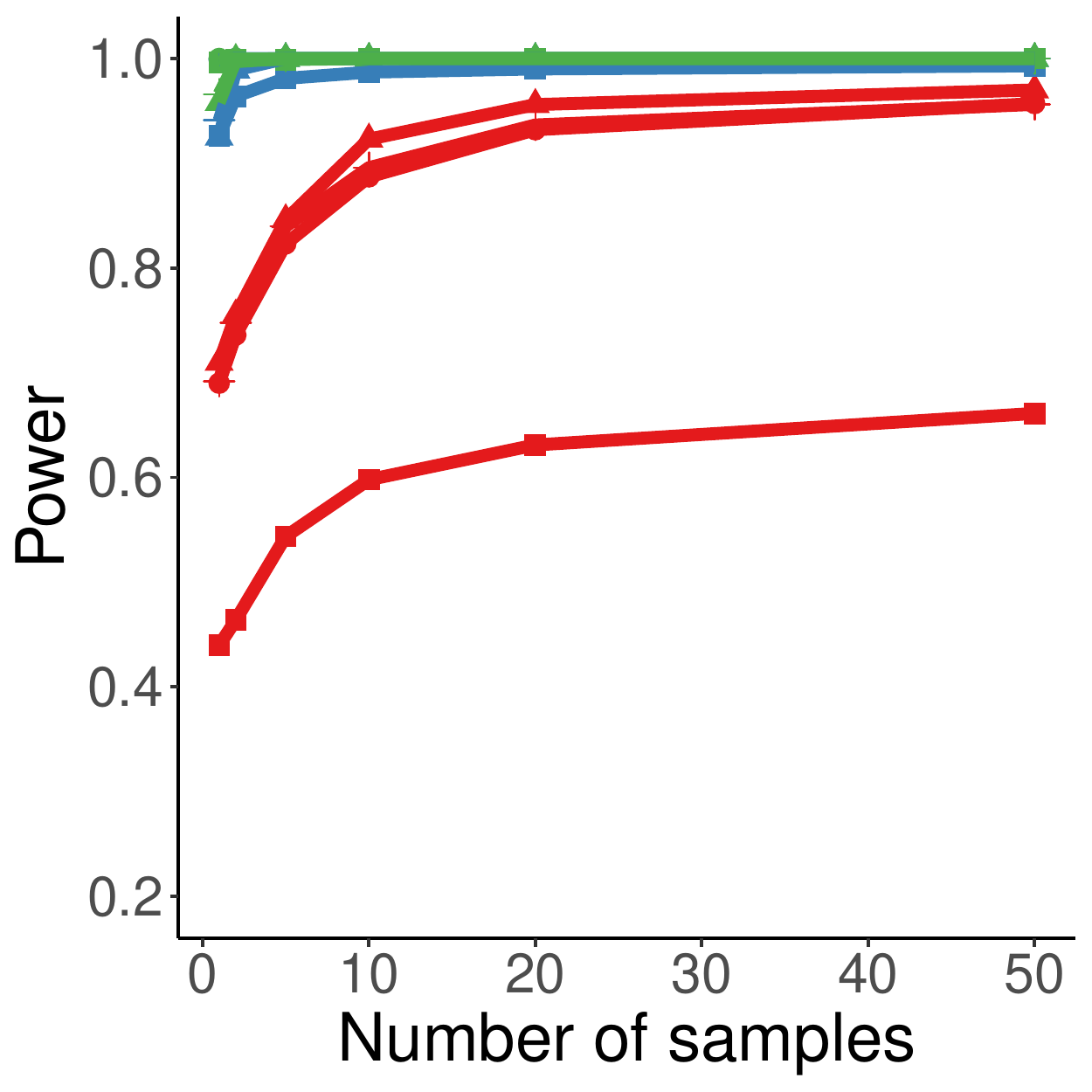}
        \put(-70,138){\footnotesize{$h = 30$}} \\
    \caption{\small{Rejection rates of $H_0$ for different values of $h$, $\delta$, and $N$, when we simulate from a model with 4 changepoints, and apply binary segmentation with 4 changepoints. For each $h$, the process is run three times with changes of size $\delta = 1, 2, 3$, which are shown on the plots as red, blue, and green lines respectively. Each line corresponds to a changepoint.
    }}
    \label{fig:power-bs-k4}
\end{figure}

In each case, we set the number of data points $T = 1000$, and take $\sigma^2 = 1$. For Figures \ref{fig:qq-bs-h1} and \ref{fig:power-bs}, we simulate from a model with a single change at $t = 500$, where the change is of size $\delta$: we consider $\delta = 1, 2, 3$. In each case, we estimate changepoints using binary segmentation, and calculate the $p$-value for the first detected changepoint. Simulations where the changepoint algorithm returns no changepoints are discarded. In Figure \ref{fig:power-bs-k4} we simulate from a model with 4 changes, at $t = 100, 400, 500, 700$, and test for changes at the first 4 estimated changepoints.

Figure \ref{fig:qq-bs-h1} shows QQ plots of $p$-values for binary segmentation, where we simulate data with a single change of size $\delta$. Figure \ref{fig:qq-bs-h1} (a) shows that our test has power when we simulate from $H_1$, and the power increases with the number of samples $N$. In Figure \ref{fig:qq-bs-h1} (b)-(d), $p$-values from the method of \cite{jewell-testing} (equivalent to $N=1$) are plotted against $p$-values from our method with different values of $N$. The $p$-values generated from our method are generally smaller, indicating increased power, particularly so as $h$ and $\delta$ increase. 
Figure \ref{fig:power-bs} shows plots of the power for different values of $h$ and $\delta$: we see that initially increasing the number of Monte Carlo samples $N$ leads to substantial increases in power, but this levels off as $N$ continues to increase. 
Figure \ref{fig:power-bs-k4} shows the same thing when simulating from a model with 4 changepoints.

Further plots, which show similar results for wild binary segmentation and $L_0$ segmentation, are given in Appendix \ref{sec:l0-wbs}.

\subsection{Multiple testing} \label{sec:multiple-testing}

We often detect multiple changepoints in a given data set, and as we conduct hypothesis tests at each estimated changepoint, we need to account for the fact that we carry out multiple tests. There are two common approaches to do so. One is to use a Bonferroni correction, or the Holm-Bonferroni method \citep{holm1979simple}. These provide control over the family wise error rate, i.e. the rate of one or more true null hypotheses, and do not make any assumptions on the dependence of the $p$-values.



Alternatively we may wish to control the false discovery rate (FDR), the proportion of rejected null hypotheses that are true. This can be done using the Benjamini-Hochberg procedure \citep{benjamini1995controlling}, but this procedure assumes independent $p$-values. If we are testing the presence for a change in non-overlapping regions of the data, then the unconditional test-statistics are independent. However, the conditioning on the selection can induce dependence. We investigated empirically the level of dependence, see for example Appendix \ref{App:p-values}, where we observed that there is non-zero but small positive correlation in $p$-values (with e.g. correlation less than $0.05$ across the scenarios). This suggests the use of Benjamini-Hochberg may be appropriate in practice, and below we consider both this and the Holm-Bonferroni method.
For comparison, we also provide the false discovery rate 
we obtain using a MOSUM approach: for consistency we set the window sizes in each case to be the same, i.e. $G = h$.


Table \ref{tab:us-vs-mosum-h0} shows the number of false positives we get when we simulate from $H_0$ with $T = 1000$, and adjust the $p$-values using the Holm-Bonferroni and Benjamini-Hochberg methods, for the method of \citep{jewell-testing} and our method; we also calculate this for MOSUM. In each case the mean number of false positives is below $0.05$, suggesting that our tests are slightly conservative, although it is lower for MOSUM than for our method. Table \ref{tab:us-vs-mosum-h1} shows the mean number of true positives and the error rate (family-wise error rate for the Holm-Bonferroni method and false discovery rate for Benjamini-Hochberg) obtained when simulating under $H_1$, with $K = 1, 4, 9$ equally spaced changepoints. A true positive occurs if a changepoint with $p$-value below 0.05 is detected within the region $(\tau - h, \tau + h)$ for some $\tau$ in the set of true changepoints (we set $h = 10$). Our method outperforms both that of \cite{jewell-testing} and MOSUM at finding true changepoints. All three methods have very few false discoveries; our method has a lower false discovery rate than MOSUM when the true number of changepoints is small, although it is higher (but still small) when the true number of changepoints is higher. For binary segmentation, the error rate will depend in part on the changepoint threshold, which here we set to 3. This is a relatively high threshold, which means that most detected changepoints correspond to true changes, and hence the overall error rates are substantially below the 5\% threshold. Appendix \ref{sec:tables} shows results we obtain using different choices of threshold, as well as for different values of $T$; these show similar numbers of true positives for a range of thresholds, whilst all having an error rate below 5\%.

\begin{table}
    \centering
    \caption{\label{tab:us-vs-mosum-h0} Simulating from $H_0$ with $T = 1000$, we applied binary segmentation and MOSUM and calculated how many times $H_0$ was falsely rejected. The table reports the mean number of estimated changepoints with adjusted $p$-value below 0.05 across 1000 iterations, and the proportion of runs in which there was at least one false discovery, for both the Holm-Bonferroni and Benjamini-Hochberg procedures.}
    \begin{tabular}{lllllll}
      \hline
        & \multicolumn{3}{c}{Mean false positives} & \multicolumn{3}{c}{Proportion with $\geq 1$ false positive} \\
      \hline
           & H-B & B-H & & H-B & B-H & \\
      \hline
        BS, $N = 1$ & 0.03 & 0.03 & & 0.03 & 0.03 & \\
        BS, $N = 10$ & 0.03 & 0.04 & & 0.03 & 0.03 & \\
        MOSUM & & & 0.01 & & & 0.01 \\
      \hline
    \end{tabular}
\end{table}

\begin{table}
\caption{\label{tab:us-vs-mosum-h1} Mean number of true positives and error rate when simulating from a model with $T = 1000$ and $K$ equally spaced changepoints, with $\mu$ alternating between $1$ and $-1$, using binary segmentation and MOSUM. The error rate given is the family-wise error rate for the Holm-Bonferroni method, and the false discovery rate for the Benjamini-Hochberg method and for MOSUM. This is calculated as $FDR = \frac{\text{false positives}}{\text{false positives } + \text{ true positives}}$.}
\centering
    \begin{tabular}{llcccccc}
      \hline
        & & \multicolumn{3}{c}{Mean true positives} & \multicolumn{3}{c}{Error rate (\%)} \\
      \hline
            & & H-B & B-H & & H-B & B-H & \\
      \hline
                & BS, $N = 1$ & 0.79 & 0.79 & & 0.60 & 0.53 & \\
        $K = 1$ & BS, $N = 10$ & 0.94 & 0.94 & & 0.50 & 0.29 & \\
                & MOSUM & & & 0.58 & & & 1.20 \\
      \hline
                & BS, $N = 1$ & 2.78 & 2.92 & & 0.90 & 0.42 & \\
        $K = 4$ & BS, $N = 10$ & 3.42 & 3.51 & & 1.50 & 0.46 & \\
                & MOSUM & & & 2.32 & & & 0.22 \\
      \hline
                & BS, $N = 1$ & 5.56 & 6.38 & & 1.50 & 0.42 & \\
        $K = 9$ & BS, $N = 10$ & 7.23 & 7.80 & & 2.10 & 0.51 & \\
                & MOSUM & & & 5.30 & & & 0.09 \\
      \hline
    \end{tabular}
\end{table}

\begin{figure}
    \centering
    \includegraphics[width=0.32\linewidth]{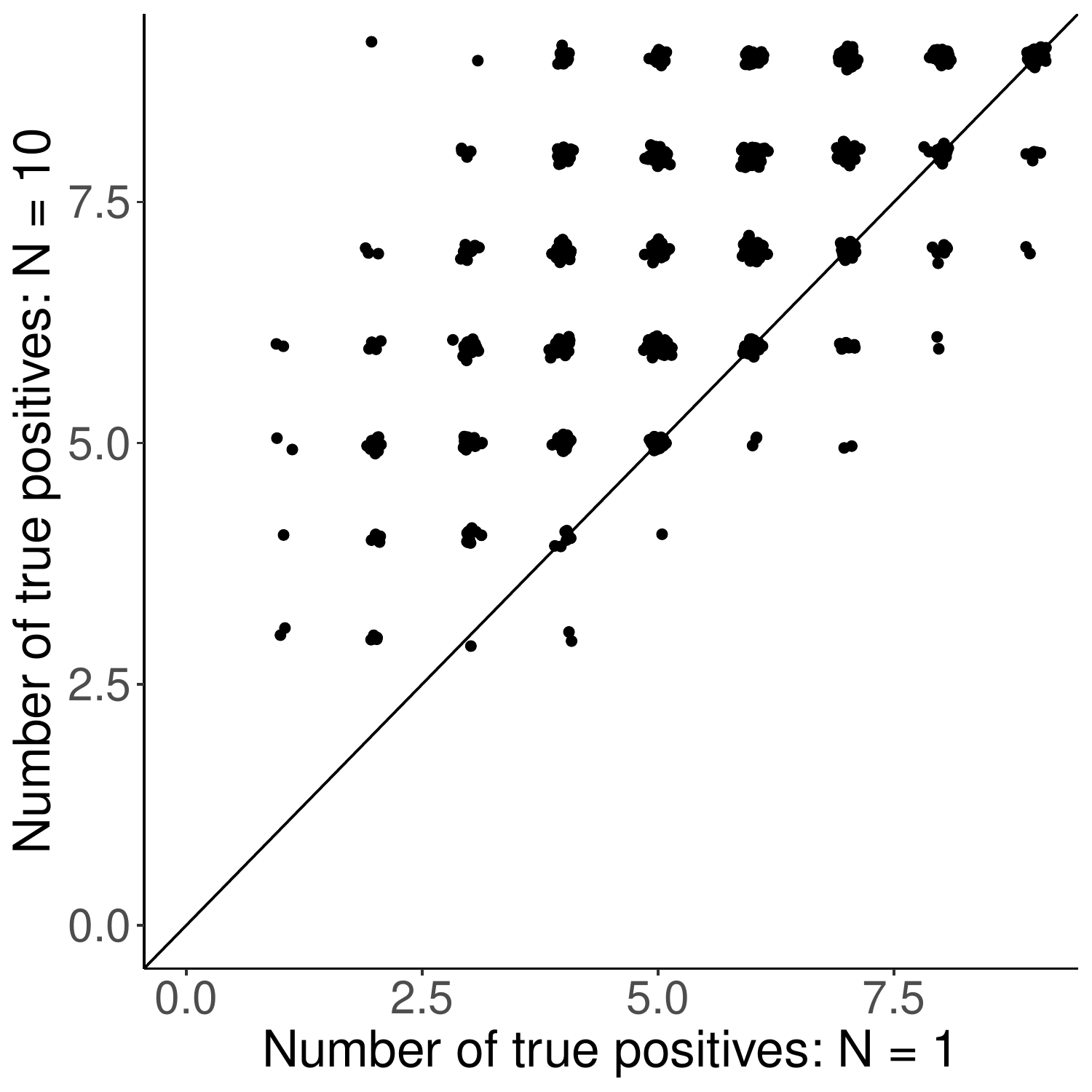}
    \includegraphics[width=0.32\linewidth]{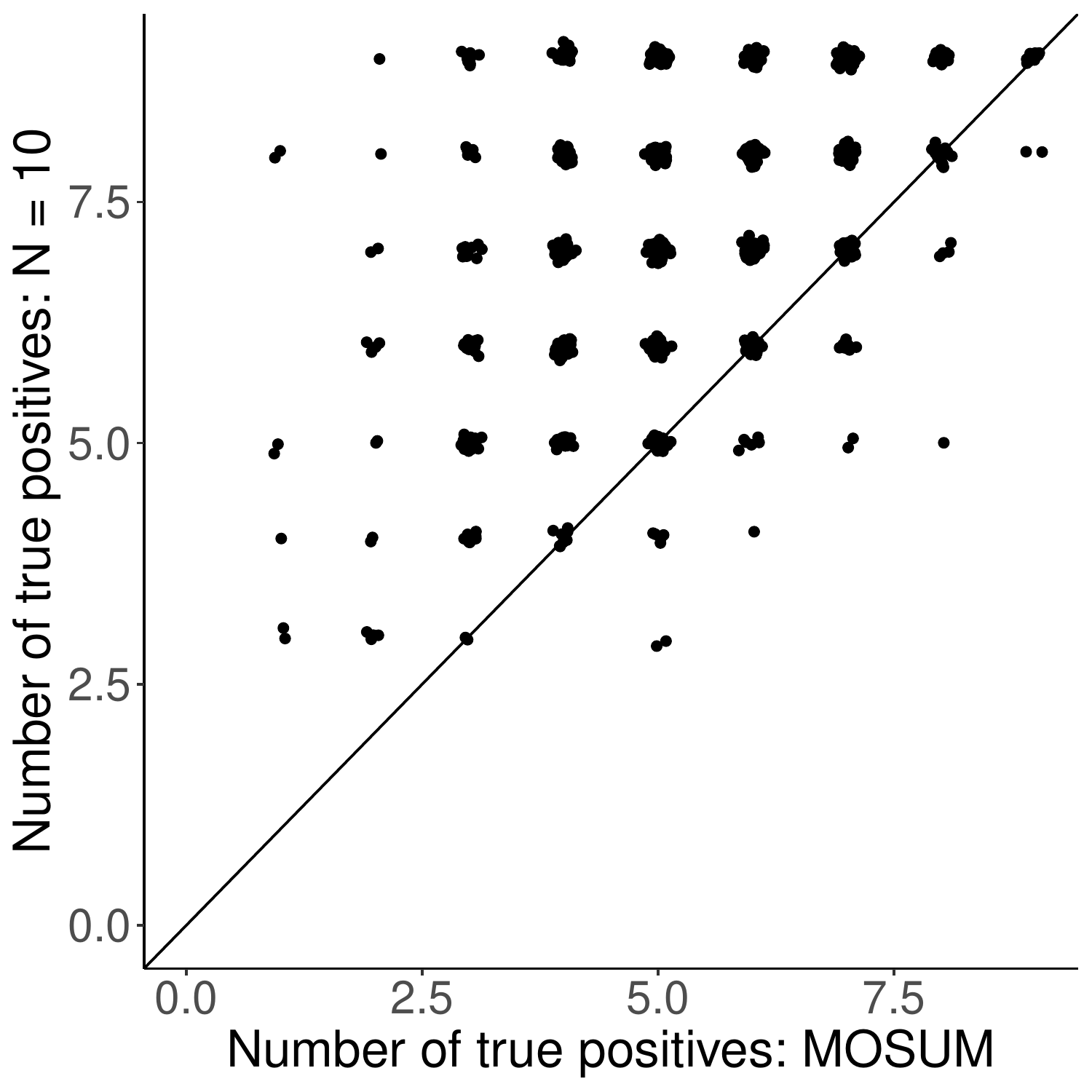}
    \caption{\small{Plots showing the number of true positives found in 1000 simulations using our method with $N = 10$ against the number found using the method of \cite{jewell-testing} (left) and MOSUM (right). A small amount of noise has been added so that points corresponding to different simulated data sets can be distinguished.}}
    \label{fig:us-vs-jewell-mosum}
\end{figure}

Figure \ref{fig:us-vs-jewell-mosum} shows plots of the number of true positives found using our method wit $N = 10$, against the number of true positives found using $N = 1$ (equivalent to the method of \cite{jewell-testing}) and MOSUM, for each of 1000 simulations from a model with 9 true positives. A small amount of noise has been added to points to allow for visualisation. We see that in both cases, our method finds the same number or more true positives in most cases.

\subsection{Robustness}

We now investigate how well our method works if the assumptions that the noise in the data is i.i.d. and Gaussian with known variance, do not hold. To get some intuition it is helpful to consider which aspects of our test depend on the distributional assumptions. Our test is based on two parts: the first is the distribution of $\phi$ and $\boldsymbol{\psi}$ conditional on the data outside the region of our test and the mean of the data within the region; the second are, for each value of $\boldsymbol{\psi}$, the set of $\phi$ values that would lead to the same information used to choose the test. It is only the first part that depends on the distributional assumptions. If these do not hold then our test statistic has a different distribution but truncated to the same range of values for each $\boldsymbol{\psi}$.

\begin{figure}
    \centering
    \includegraphics[width=0.32\linewidth]{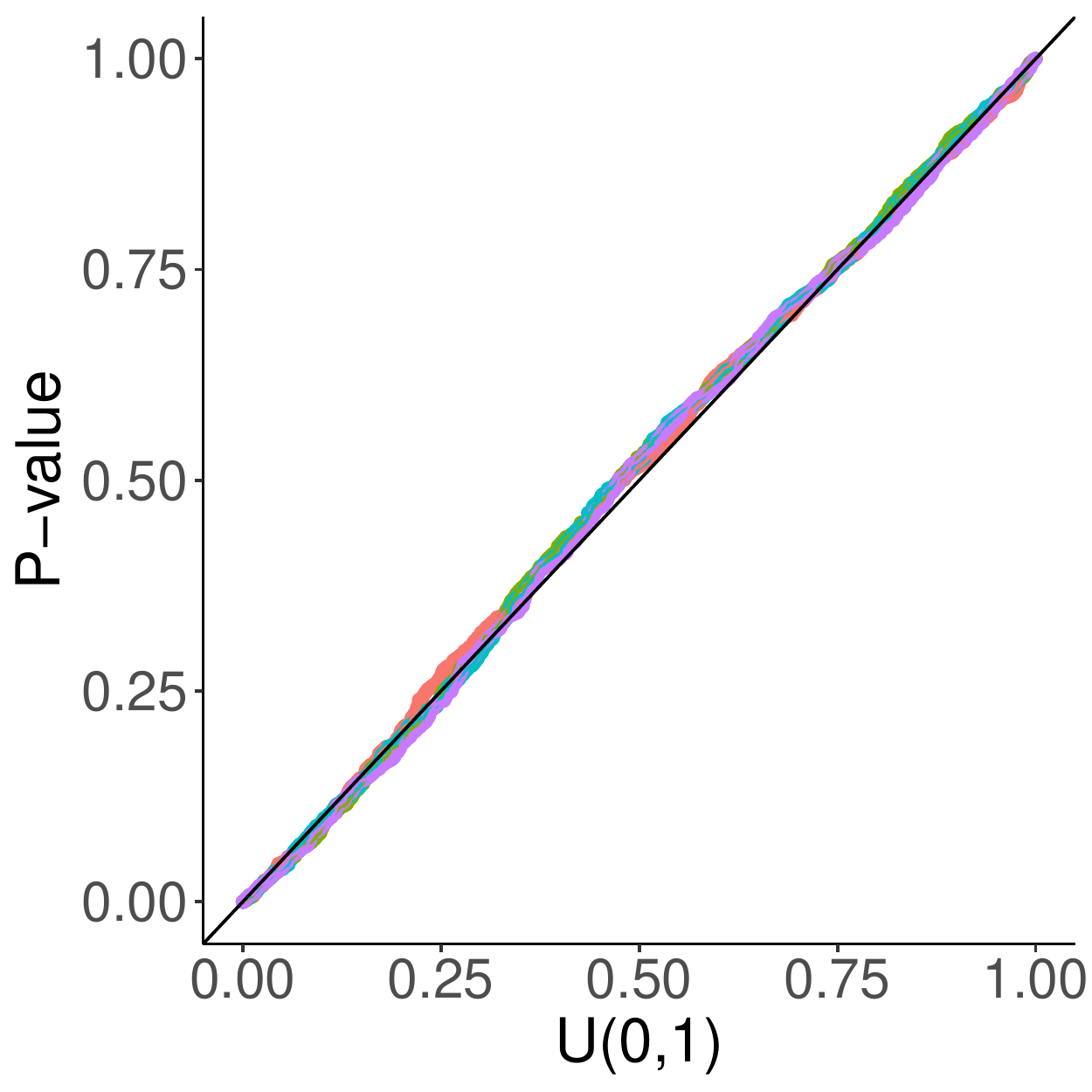}
        \put(-61,135){\footnotesize{$H_0$}}
    \includegraphics[width=0.32\linewidth]{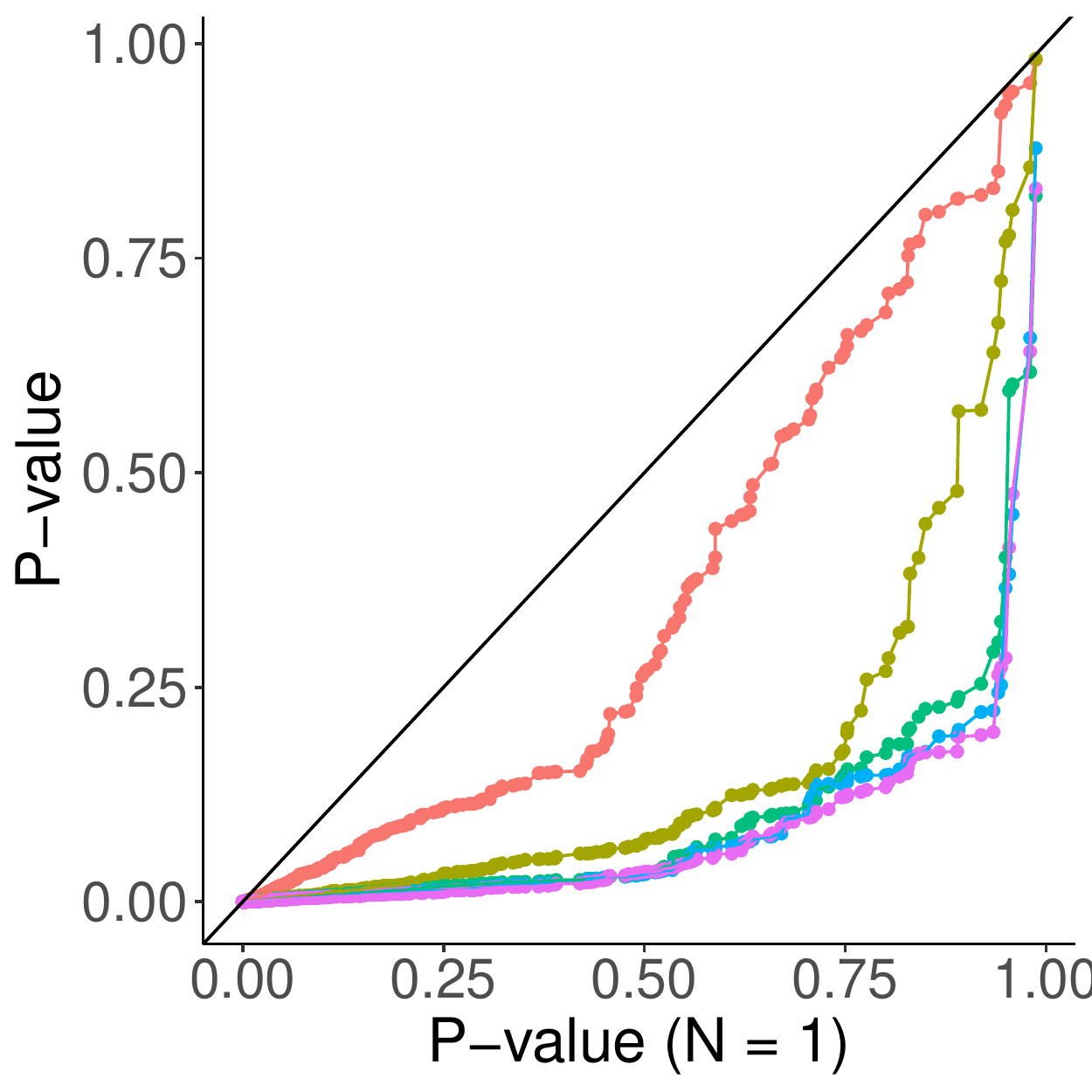}
        \put(-75,135){\footnotesize{$H_1, \delta=2$}}
    \caption{\small{QQ plots of $p$-values when $\sigma^2$ is estimated using median absolute deviation, rather than being assumed known. In each case $T = 1000$ and $h = 10$ and we estimate 1 changepoint using binary segmentation and calculate the associated $p$-value. In the left-hand plot we simulate from $H_0$ with no changes; in the right-hand plot we simulate from a model with a single change of size $\delta = 2$ at $\tau = 500$.}}
    \label{fig:qq-mad}
\end{figure}

Thus in situations where we have a limiting regime whereby the distribution of $\phi$ and $\boldsymbol{\psi}$ converge to their assumed distribution, our method will give asymptotically correct $p$-values. The most important example of this is where the independent Gaussian assumption for the noise is correct but the variance is unknown. In this case if we use any consistent (in data size $T$) estimator for the variance, our approach will be asymptotically valid as $T \rightarrow \infty$ \cite[see][]{jewell-testing}.

We investigated the finite sample performance, with Figure \ref{fig:qq-mad} showing QQ plots of the $p$-values obtained when we simulate from a model with no changepoints (left) and with one changepoint (right), apply binary segmentation with one changepoint, and estimate the variance of $\boldsymbol{X}$ using median absolute deviation. We see that the distribution of $p$-values under $H_0$ is very close to $U(0, 1)$, and under $H_1$ the distribution is close to that when we assume known variance (compare with Figure \ref{fig:qq-bs-h1}d).

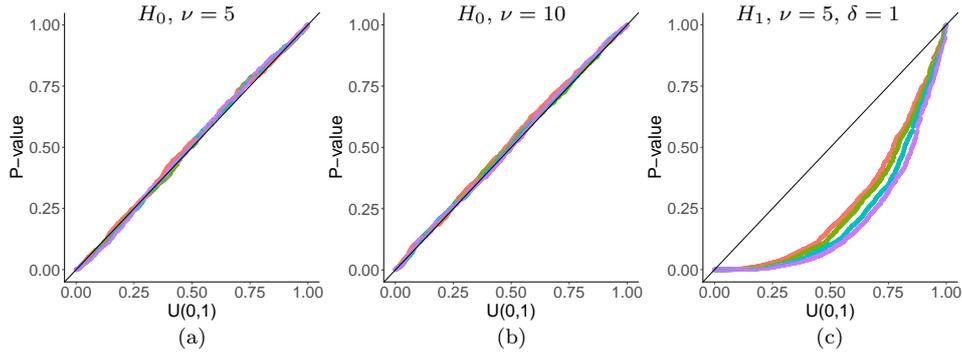
\begin{figure}
    \centering
    \includegraphics[width=0.28\linewidth]{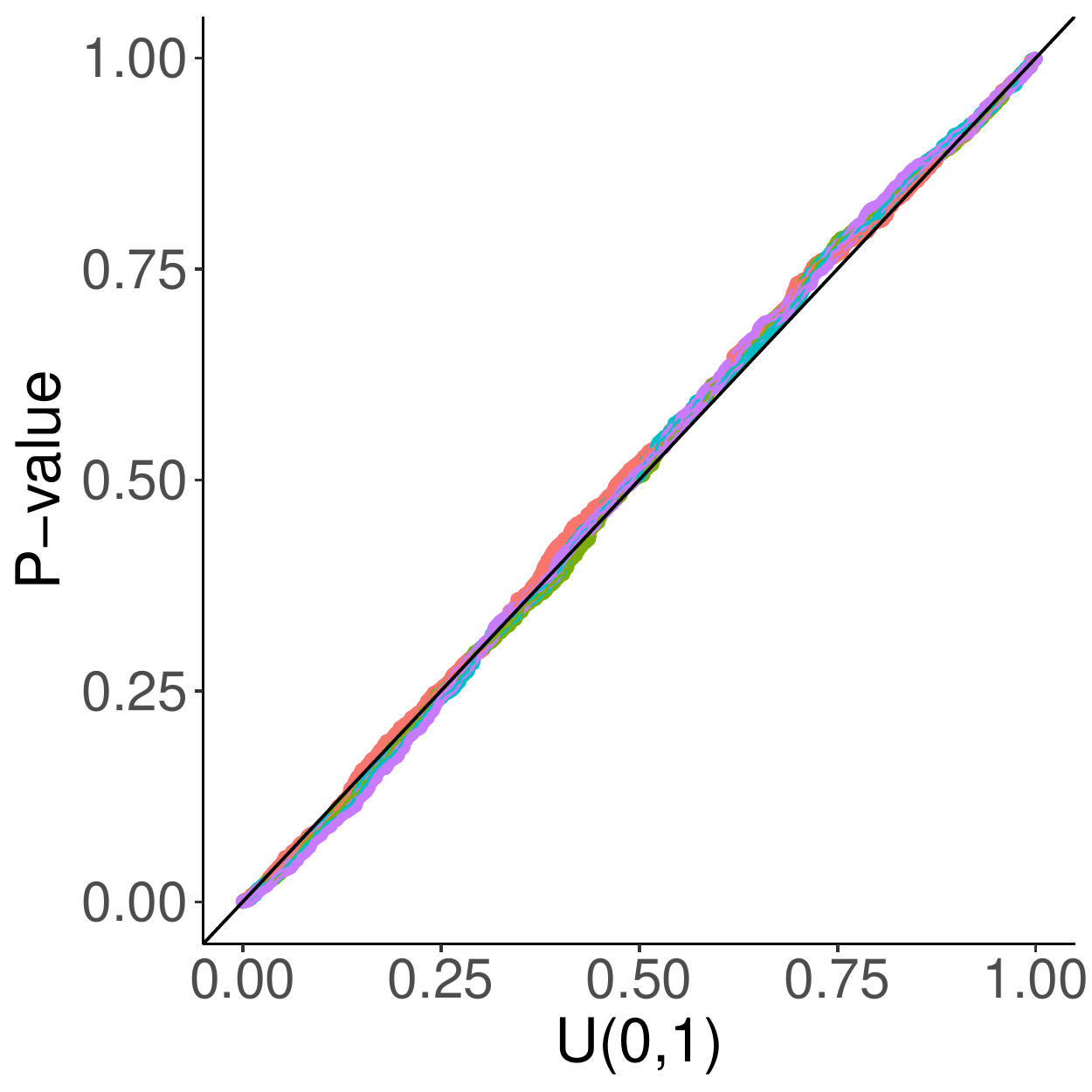}
        \put(-55,-7){\footnotesize{(a)}}
        \put(-70,115){\footnotesize{$H_0$, $\nu = 5$}}
    \includegraphics[width=0.28\linewidth]{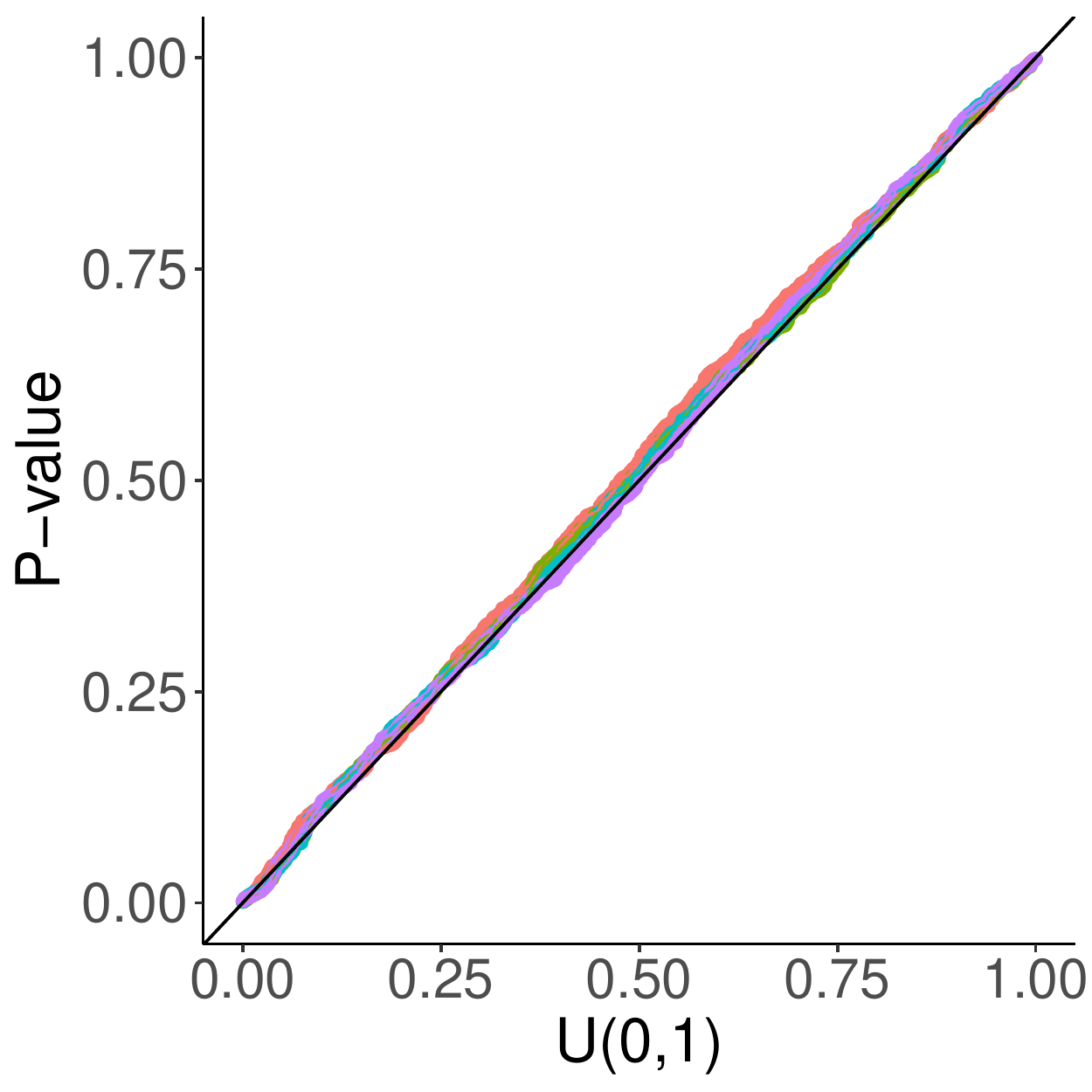}
        \put(-55,-7){\footnotesize{(b)}}
        \put(-70,115){\footnotesize{$H_0$, $\nu = 10$}}
    \includegraphics[width=0.28\linewidth]{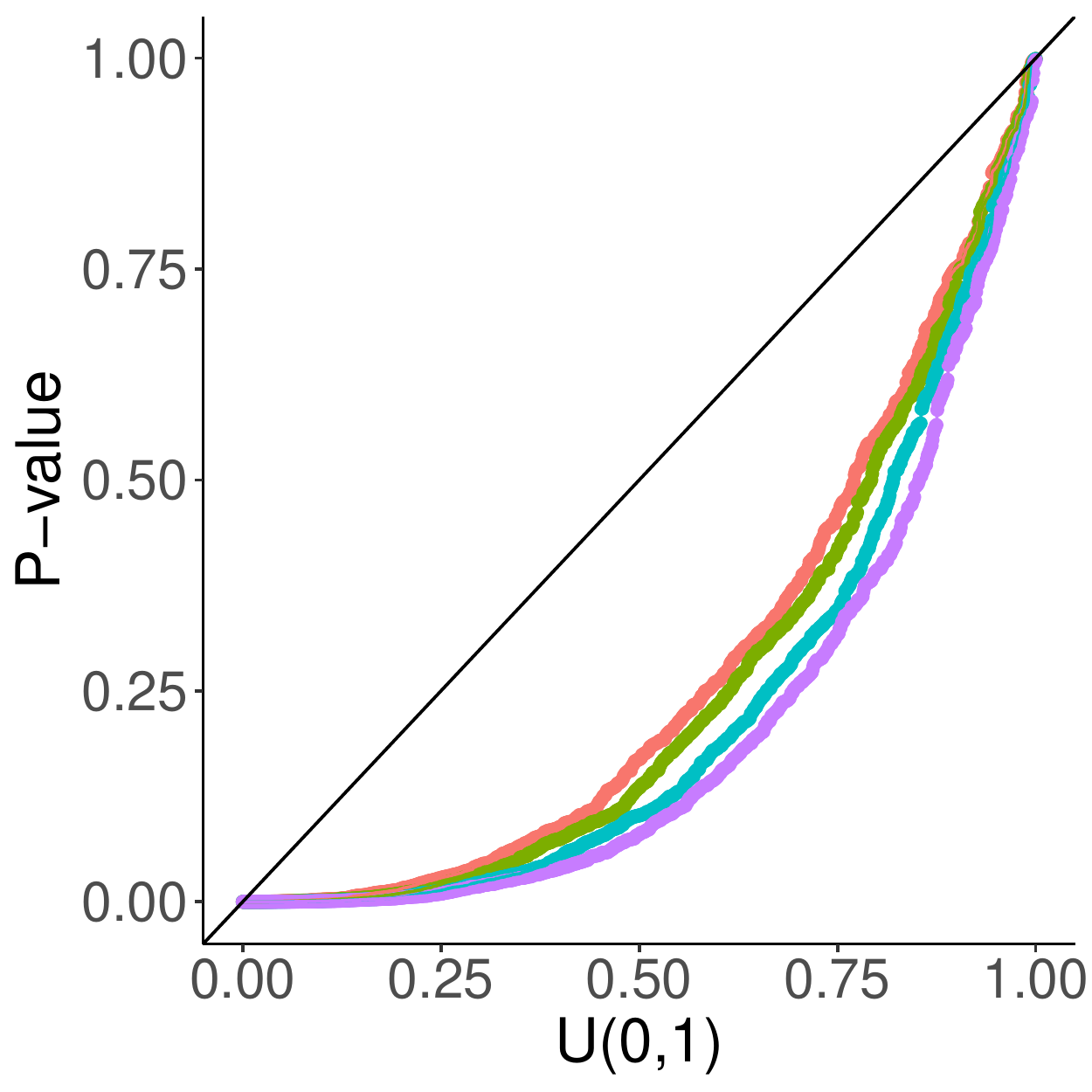}
        \put(-55,-7){\footnotesize{(c)}}
        \put(-85,115){\footnotesize{$H_1$, $\nu = 5$, $\delta = 1$}}
    \caption{\small{QQ plots of $p$-values obtained when simulating from a $t_\nu$ distribution with $\nu$ degrees of freedom.}}
    \label{fig:t-sims}
\end{figure}

The discussion above suggests that our approach may also be appropriate in situations where a central limit theorem holds for our test statistic, so that a Gaussian approximation for the distribution of $\phi$ is reasonable. To investigate this, we simulate data $\boldsymbol{X}$ with heavy-tailed noise, simulating from a student's $t$ distribution. Figure \ref{fig:t-sims} shows QQ plots we obtain when data is simulated from a $t_\nu$ distribution with $\nu = 5, 10$ degrees of freedom. In panels (a) and (b) we simulate from $H_0$ and find that the $p$-values approximately follow the expected distribution $U(0, 1)$. In panel (c) we simulate data with a single change of size $\delta = 1$; in this case the $p$-values are smaller than under $H_0$, so the test has power to detect deviation from $H_0$. Figure \ref{fig:laplace-sims} in Appendix \ref{App:Laplace} shows similar results when we simulate $\boldsymbol{X}$ with Laplace noise.

\section{Application to genomic data} \label{sec:real}

\begin{figure}
    \centering
    \includegraphics[width=0.98\linewidth]{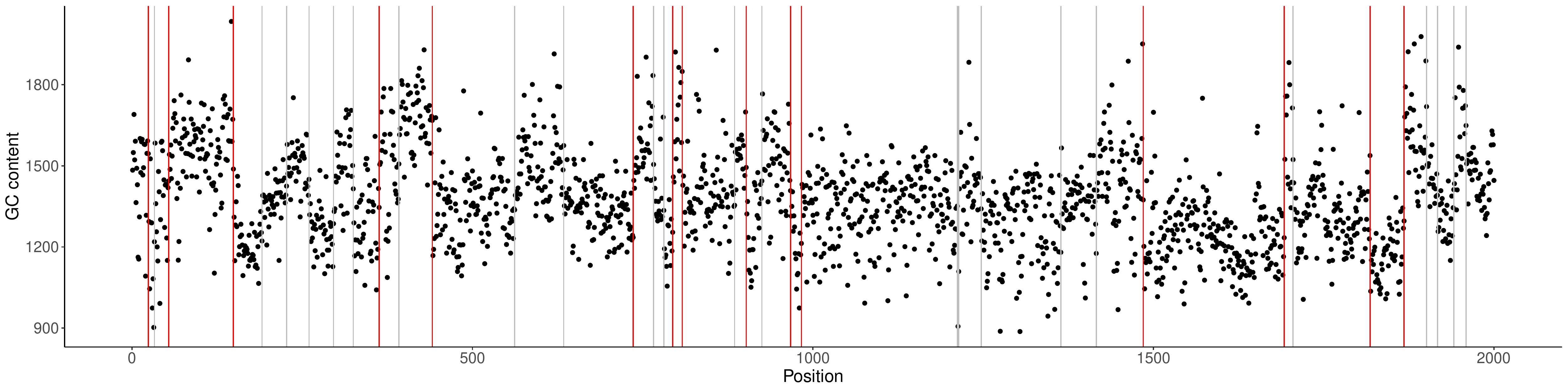}
    \includegraphics[width=0.98\linewidth]{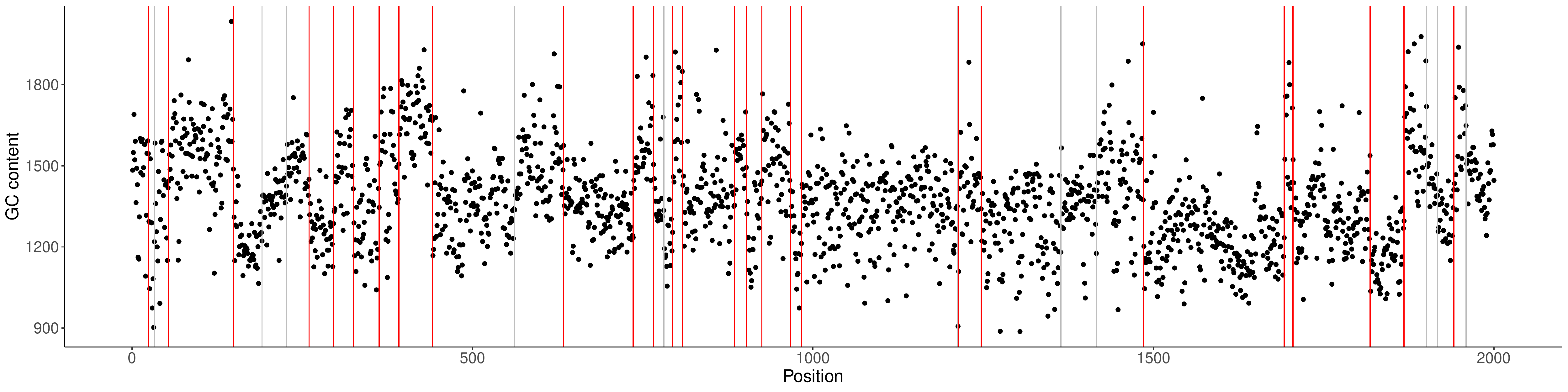}
    \caption{\small{Estimated changepoints in GC content data. Binary segmentation was used to estimate 38 changepoints, and we set $h = 10$. Each vertical line corresponds to an estimated changepoint; changepoints found to be significant at significance level $\alpha = 0.05$ are shown in red, with others shown in grey. In the top panel, we used $N = 1$ (equivalent to the method of \cite{jewell-testing}) to calculate $p$-values; in the bottom panel, we used $N = 10$. 
    }}
    \label{fig:hc1}
\end{figure}

We now apply our method to genomic data consisting of GC content in 3kb windows along the human chromosome, using both binary segmentation and $L_0$ segmentation. 
The data is available in the \texttt{R} package \texttt{changepoint}. As in \cite{jewell-testing}, only the first $2000$ data points are used, and we set the detection thresholds 
so that the number of changepoints detected is 38.
For each changepoint, we calculate a $p$-value using both the method of \cite{jewell-testing} (equivalent to our method with $N = 1$) and our method (with $N = 10$ and $N = 20$), using a window size of $h = 10$. The variance of the data is estimated using median absolute deviation. Figure \ref{fig:hc1} shows plots of the data and estimated changepoints for binary segmentation, where red lines correspond to estimated changepoints with $p$-values smaller than $0.05$ (after controlling for multiple testing by using the Holm-Bonferroni correction), and grey lines to changepoints with $p$-values above $0.05$. The corresponding figure for $L_0$ segmentation is included in Appendix \ref{sec:l0-wbs}. In both cases, a greater proportion of changepoints are deemed to be significant using our method, indicating that it has greater power to detect changes.

Table \ref{tab:hc1} gives the number of significant changepoints found in each case. To account for the fact that we are conducting multiple tests, we report results after controlling for multiple testing using the Holm-Bonferroni and Benjamini-Hochberg procedures. For both binary segmentation and $L_0$ segmentation, we get a greater number of significant changepoints when $N = 10$ than when $N = 1$, although increasing $N$ beyond this does not yield much further improvement. Hence, in practice, only a moderate number of Monte Carlo samples is sufficient to get an increase in power. 

Interestingly for this application the MOSUM approach finds more significant changepoints, with 45 changes with a $p$-value less than 0.05. This may be due to binary segmentation or $L_0$ penalisation approaches detecting too few changes. 

\begin{table}
    \caption{\label{tab:hc1} Number of changepoints in GC content data with $p$-values below $\alpha = 0.05$, where we control for multiple testing using the Holm-Bonferroni and Benjamini-Hochberg methods. We implement our method with binary segmentation and $L_0$ segmentation, where $N = 1$ is equivalent to the method of \cite{jewell-testing}.
    }
    \centering
    \begin{tabular}{lrrr}
      \hline
            & \multicolumn{3}{c}{Number of $p$-values $< 0.05$} \\
      \hline
        Changepoint algorithm & & Holm-Bonferroni & Benjamini-Hochberg \\
      \hline
                                      & $N = 1$  & 15 & 21 \\
        Binary segmentation, $K = 38$ & $N = 10$ & 27 & 30 \\
                                      & $N = 20$ & 27 & 30 \\
      \hline
                                      & $N = 1$  & 20 & 26 \\
        Binary segmentation, $K = 45$ & $N = 10$ & 32 & 36 \\
                                      & $N = 20$ & 31 & 36 \\
      \hline
                                     & $N = 1$  & 16 & 25 \\
        $L_0$ segmentation, $K = 38$ & $N = 10$ & 25 & 28 \\
                                     & $N = 20$ & 28 & 29 \\
      \hline
                                     & $N = 1$  & 20 & 27 \\
        $L_0$ segmentation, $K = 46$ & $N = 10$ & 26 & 32 \\
                                     & $N = 20$ & 27 & 32 \\
      \hline
    \end{tabular}
\end{table}


\section{Discussion} \label{sec:diss}

We have introduced a method for increasing the power of changepoint inference procedures, by reducing the amount of information we condition on. We have shown that this method is effective in increasing power compared to existing methods, both in simulated and real-world data sets.

Whilst our approach has been developed for changepoint problems, the general idea can be applied to other scenarios such as clustering \cite[]{gao2022selective,chen2023selective} or regression tress \cite[]{neufeld2022tree}. For example, current methods for post-selection inference after clustering are based on a test statistic that compares the mean of the cluster, and fixes the projection of the data that is orthogonal to this. However we could reduce this to conditioning just on the sample mean of one of the clusters, and the data in the clusters that are not being combined. Our method would then re-simulate the perturbations of each data point about its clustered mean, apply the existing post-selection inference method to each data set, and calculate the weighted average as we do in this paper. We believe that this approach would have similar properties, of being a valid $p$-value regardless of the Monte Carlo sample size, and of having larger power as the Monte Carlo sample size increases. Our approach for constructing valid $p$-values when using Monte Carlo to estimate post-selection $p$-values \cite[e.g.][]{saha2022inferring} may also be applicable more widely.

\section*{Acknowledgement}

This work is supported by EPSRC grant number EP/V053590/1.

\bibliography{references.bib}
\bibliographystyle{apalike}

\newpage

\appendix

\section{Proofs}

\subsection{Proof of Theorem \ref{thm:ideal}} \label{sec:prooft1}
\begin{proof}
Fix $\alpha$, and consider maximising $\Pr( P^*\leq \alpha)$. This corresponds to choosing a rejection region $R_\alpha$ for $(\phi,\boldsymbol{\psi})$ that maximises
\[
\int_{R_\alpha} \tilde{k}(|\phi|) f(\phi)\prod_{i=1}^{2h-2}g(\psi_i) \mbox{d}\phi \mbox{d}\boldsymbol{\psi},
\]
subject to
\[
\int_{R_\alpha} f(\phi)\prod_{i=1}^{2h-2}g(\psi_i) \mbox{d}\phi \mbox{d}\boldsymbol{\psi} \leq \alpha \int_\mathcal{S} f(\phi)\prod_{i=1}^{2h-2}g(\psi_i) \mbox{d}\phi \mbox{d}\boldsymbol{\psi}.
\]
As $\tilde{k}$ is an increasing function of $\phi$, it is straightforward that this is achieved for the region
\[
R_\alpha=\{ (\phi,\boldsymbol{\psi}): |\phi|\geq c_\alpha \},
\]
with $c_\alpha$ defined by
\[
\int_{R_\alpha} f(\phi)\prod_{i=1}^{2h-2}g(\psi_i) \mbox{d}\phi \mbox{d}\boldsymbol{\psi} = \alpha \int_\mathcal{S} f(\phi)\prod_{i=1}^{2h-2}g(\psi_i) \mbox{d}\phi \mbox{d}\boldsymbol{\psi}.
\]
This is precisely the form of $P_I$, as $P_I
\leq \alpha$ corresponds to $|\phi|\geq c_\alpha$. The result follows directly.

We now show that the condition of the theorem holds if under the alternative the we have a common mean for $X_{\hat{\tau}-h+1},\ldots,X_{\hat{\tau}}$, and a different common mean for $X_{\hat{\tau}+1},\ldots,X_{\hat{\tau}+h}$, and that the density for the difference in means is symmetric about 0. Under this model, denote the difference in means at $\hat{\tau}$ by $Y$, and denotes its density function by $h(y)$, which by assumption is symmetric about $0$. First note that for this model $\phi$ is independent of $\boldsymbol{\psi}_{1:2h-2}$ as the covariance of $(\phi,\boldsymbol{\psi}_{1:2h-2})$ is still diagonal, and it is straightforward to hence show that the conditional distribution of $\boldsymbol{\psi}_{1:2h-2}$  given $\phi$ is unchanged. So we need just check the condition on the marginal distribution for $\phi$.

Under $H_1$,
\begin{equation*}
    \phi | Y \sim N(Y, ||\boldsymbol{\nu}_{\hat{\tau}}||_2^2 \sigma^2).
\end{equation*}
and hence
\begin{equation*}
    f_{H_1}(\phi) = \int_Y f_{H_1}(\phi | y) h(y) dy,
\end{equation*}
where $f_{H_1}$ denotes the density of $\phi$ under $H_1$. Since $h$ is symmetric in $Y$, we get
\begin{equation*}
    \begin{split}
        h(\phi) & = \int_Y f_{H_1}(\phi|y) h(y) dy \\
            & = \int_{Y > 0} f_{H_1}(\phi|y) h(y) dy + \int_{Y < 0} f_{H_1}(\phi|y) h(y)dy \\
            & = \int_{Y > 0} \left( f_{H_1}(\phi|y) + f_{H_1}(\phi|-y) \right) h(y) dy.
    \end{split}
\end{equation*}
Note that $f_{H_1} (\phi | y) = f_{H_1} (\phi - y|0) = f(\phi - y)$, so
\begin{equation*}
    f_{H_1}(\phi) = \int_{Y > 0} \left( f(\phi - y) + f(\phi + y)\right) h(y) dy,
\end{equation*}
and so
\begin{equation*}
    k(\phi) = \frac{\int_{Y > 0} \left( f(\phi - y) + f(\phi + y)\right) h(y) dy}{f(\phi)} = \int_{Y > 0} \frac{f(\phi - y) + f(\phi + y)}{f(\phi)} h(y) dy.
\end{equation*}
Let 
\begin{equation*}
    \begin{split}
        \tilde{k}(\phi) & = \frac{f(\phi - y) + f(\phi + y)}{f(\phi)} \\
            & = \frac{e^{-(\phi - y)^2/2 ||\boldsymbol{\nu}_{\hat{\tau}}||_2^2 \sigma^2} + e^{-(\phi + y)^2/2 ||\boldsymbol{\nu}_{\hat{\tau}}||_2^2 \sigma^2}}{e^{-\phi^2/2 ||\boldsymbol{\nu}_{\hat{\tau}}||_2^2 \sigma^2}} \\
            & = e^{-y^2/2 ||\boldsymbol{\nu}_{\hat{\tau}}||_2^2 \sigma^2} \left(e^{\phi y/2 ||\boldsymbol{\nu}_{\hat{\tau}}||_2^2 \sigma^2} - e^{-\phi y/2 ||\boldsymbol{\nu}_{\hat{\tau}}||_2^2 \sigma^2}\right).
    \end{split}
\end{equation*}
This is monotone increasing in $|\phi|$ for every $y > 0$. Therefore, this must also be true of $k(\phi)$.
\end{proof}

\subsection{Proof of Theorem \ref{th:p-values}}

\begin{proof}
    The $p$-value, $\hat{p}_N$, is invariant to shuffling the labels of the $\boldsymbol{\psi}^{(j)}$s. Let $\boldsymbol{\psi}^{(1:N)}$ denote the set of $\boldsymbol{\psi}^{(j)}$ values after shuffling, and $I$ the label of $\boldsymbol{\psi}^{(j)}$ that corresponds to the observed data.

    The proof follows by calculating $\Pr(\hat{p}_N > \alpha | \boldsymbol{\psi}^{(1:N)})$. This requires calculating the distribution of $\phi$ given $\boldsymbol{\psi}^{(1:N)}$.

    As before, let $f$ and $g$ represent the pdfs under the null of $\phi$ and each component of $\boldsymbol{\psi}$, respectively. 
    Then
    \begin{equation*}
        f(\phi, \boldsymbol{\psi}^{(1:N)}, I | \mathcal{S}) \propto f(\phi) \left( \prod_{i=1}^N g(\boldsymbol{\psi}^{(i)}) \right) \mathbb{I}_{ \{ \phi \in \mathcal{S}_{\boldsymbol{\psi}^{(I)}} \} }.
    \end{equation*}
    If we condition on $\boldsymbol{\psi}^{(1:N)}$, then we get
    \begin{equation*}
        f(\phi, I | \boldsymbol{\psi}^{(1:N)}, \mathcal{S}) \propto f(\phi) \mathbb{I}_{ \{ \phi \in \mathcal{S}_{\boldsymbol{\psi}^{(I)}} \} }.
    \end{equation*}
    To normalize this, evaluate
    \begin{equation*}
        W = \sum_{i=1}^N \int_{\phi} f(\phi) \mathbb{I}_{ \{ \phi \in \mathcal{S}_{\boldsymbol{\psi}^{(i)}} \} } d\phi = \sum_{i=1}^N \Pr(\phi \in \mathcal{S}_{\boldsymbol{\psi}^{(i)}}) = \sum_{i=1}^N w_i,
    \end{equation*}
    where $w_i$ is as defined above.
    Thus,
    \begin{equation*}
        f(\phi, I | \boldsymbol{\psi}^{(1:N)}, \mathcal{S}) = \frac{1}{W} f(\phi) \mathbb{I}_{ \{ \phi \in \mathcal{S}_{\boldsymbol{\psi}^{(I)}} \} }.
    \end{equation*}
    We can now marginalise out $I$ to get
    \begin{equation*}
        f(\phi | \boldsymbol{\psi}^{(1:N)}, \mathcal{S}) = \sum_{I=1}^N f(\phi, I | \boldsymbol{\psi}^{(1:N)}, \mathcal{S}).
    \end{equation*}
    So,
    \begin{equation*}
        f(\phi | \boldsymbol{\psi}^{(1:N)}, \mathcal{S}) = \frac{1}{W} \sum_{I=1}^N f(\phi) \mathbb{I}_{ \{ \phi \in \mathcal{S}_{\boldsymbol{\psi}^{(I)}} \} } = \frac{1}{W} \sum_{I=1}^N w_I \frac{ f(\phi) \mathbb{I}_{ \{ \phi \in \mathcal{S}_{\boldsymbol{\psi}^{(I)}} \} } }{w_I} = \frac{1}{W} \sum_{I=1}^N w_I f(\phi | \phi \in \mathcal{S}_{\boldsymbol{\psi}^{(I)}}).
    \end{equation*}
    So
    \begin{equation*}
        \Pr(|\phi| > \alpha | \boldsymbol{\psi}^{(1:N)}, \mathcal{S}) = \frac{1}{W} \sum_{I=1}^N w_i \Pr(|\phi| > \alpha | \boldsymbol{\psi}^{(I)}, \mathcal{S}).
    \end{equation*}
    This is the form of $\hat{p}_N$, but with $\alpha$ replaced by $\phi_{obs}$. So by the probability inverse transform, $\hat{p}_N$ will have a uniform distribution on $[0,1]$.    
\end{proof}

\section{Additional Simulations}

\subsection{Results for $L_0$-penalised and Wild Binary Segmentation} \label{sec:l0-wbs}

In Section \ref{sec:sims}, we investigated the performance of our method using binary segmentation as the changepoint algorithm. Here, we show similar results for $L_0$-penalised segmentation and wild binary segmentation.

\begin{figure}
    \centering
    \includegraphics[width=0.25\linewidth]{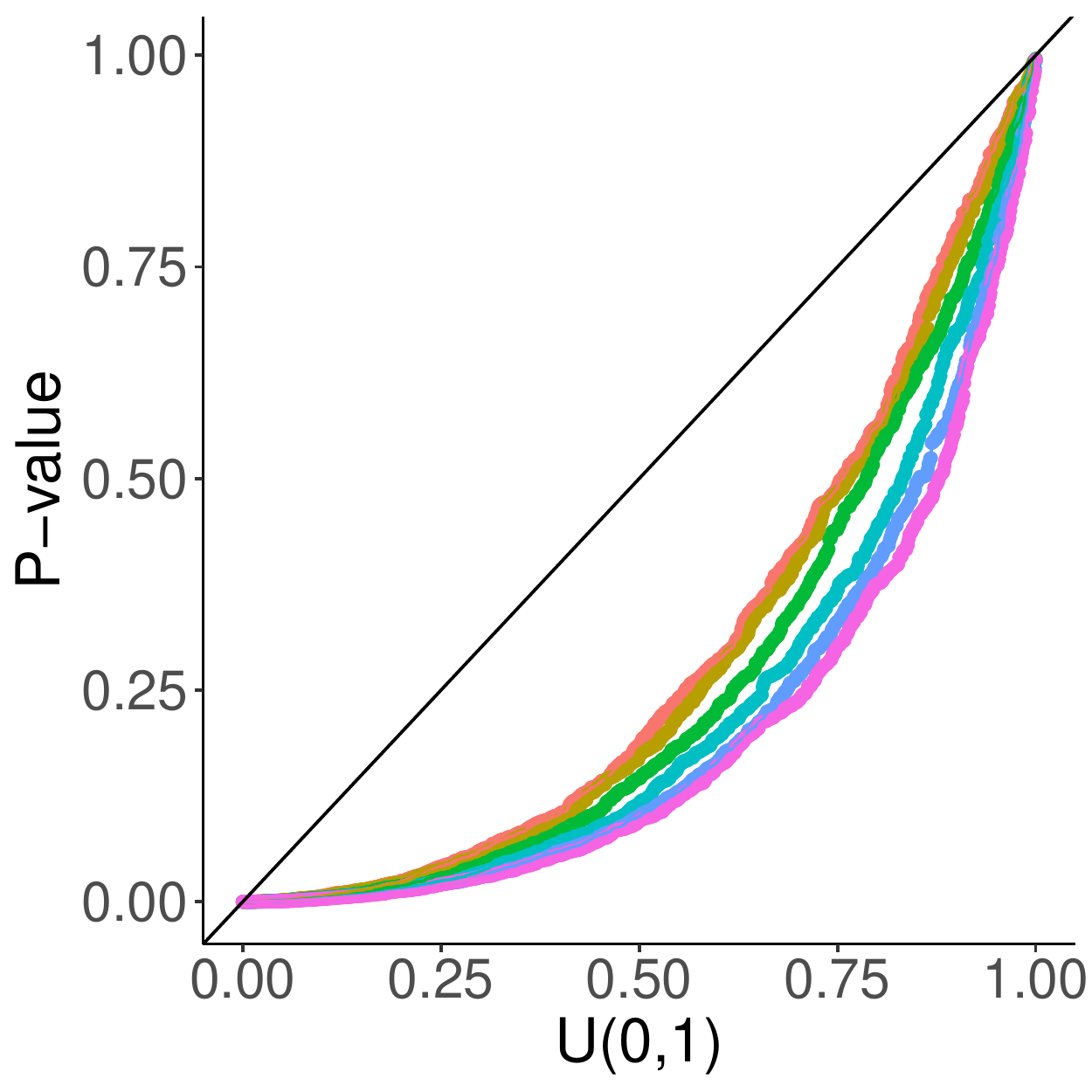}
        \put(-50,-7){\footnotesize{(a)}}
        \put(-70,105){\footnotesize{$h = 10, \delta=1$}}
    \includegraphics[width=0.25\linewidth]{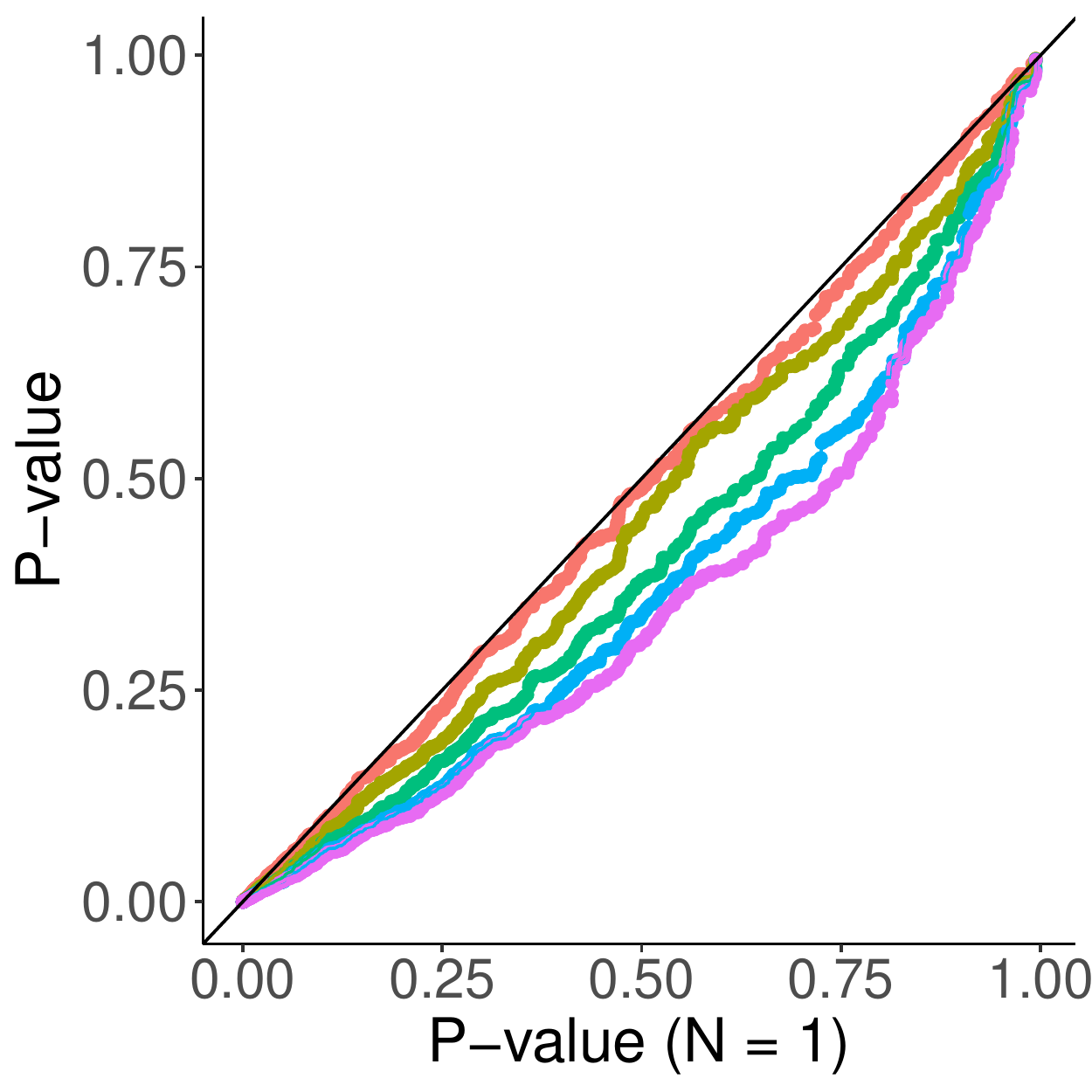}
        \put(-50,-7){\footnotesize{(b)}}
        \put(-70,105){\footnotesize{$h = 10, \delta=1$}}
    \includegraphics[width=0.25\linewidth]{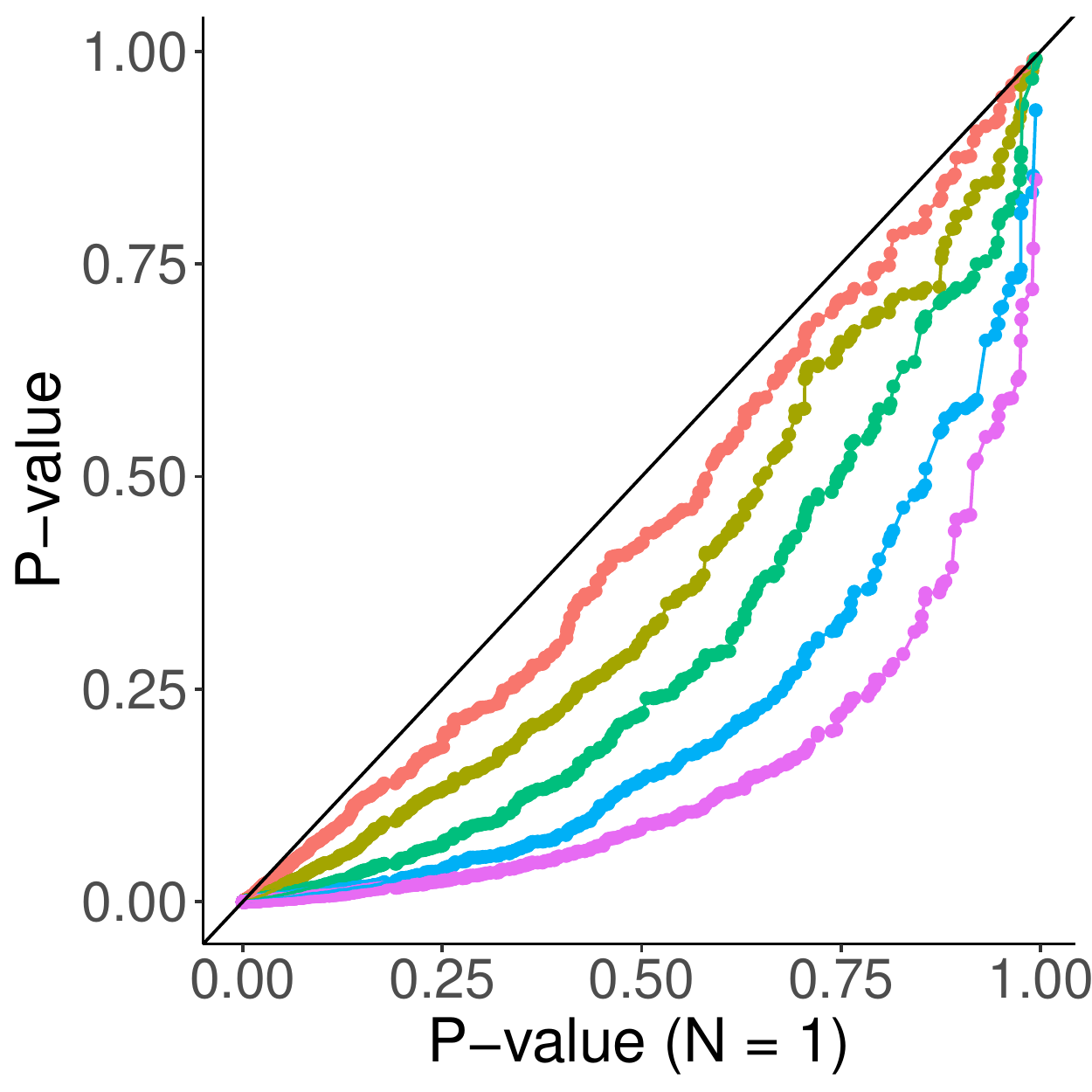}
        \put(-50,-7){\footnotesize{(c)}}
        \put(-70,105){\footnotesize{$h = 30, \delta=1$}}
    \includegraphics[width=0.25\linewidth]{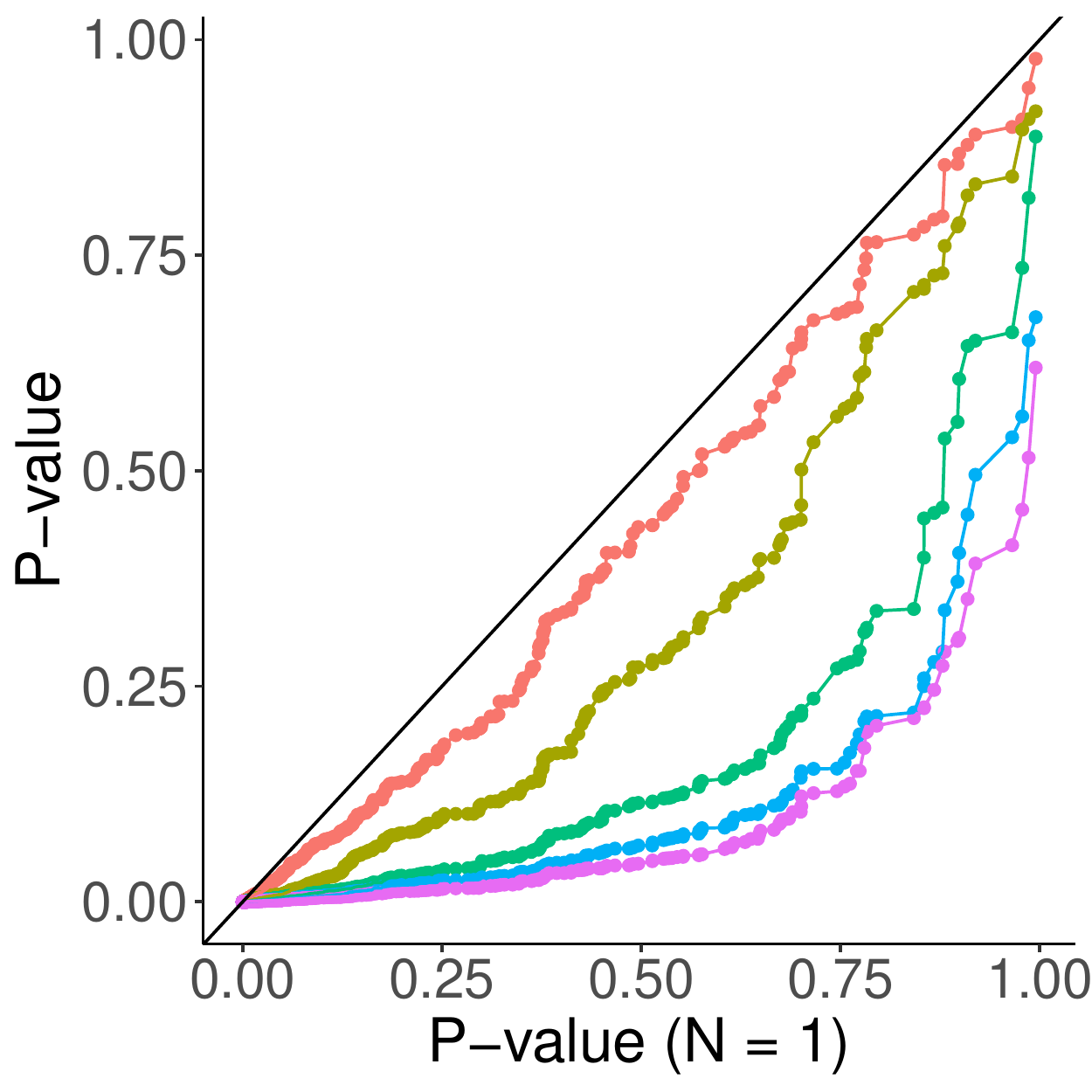}
        \put(-50,-7){\footnotesize{(d)}}
        \put(-70,105){\footnotesize{$h = 10, \delta=2$}}
    \caption{\small{QQ plot of $p$-values for $L_0$ segmentation; $h$ is the window size and $\delta$ the size of the change in the model from which we simulate. In (a), $p$-values from our method are plotted against theoretical quantiles from $U(0,1)$ for $N = 1, 2, 5, 10, 20, 50$. (b), (c) and (d) show QQ plots of $p$-values calculated using our method (with $N = 2, 5, 10, 20, 50$) against $p$-values from the method of \cite{jewell-testing} (equivalent to $N = 1$).}}
    \label{fig:qq-l0-h1}
\end{figure}

\begin{figure}
    \centering
    \includegraphics[width=0.32\linewidth]{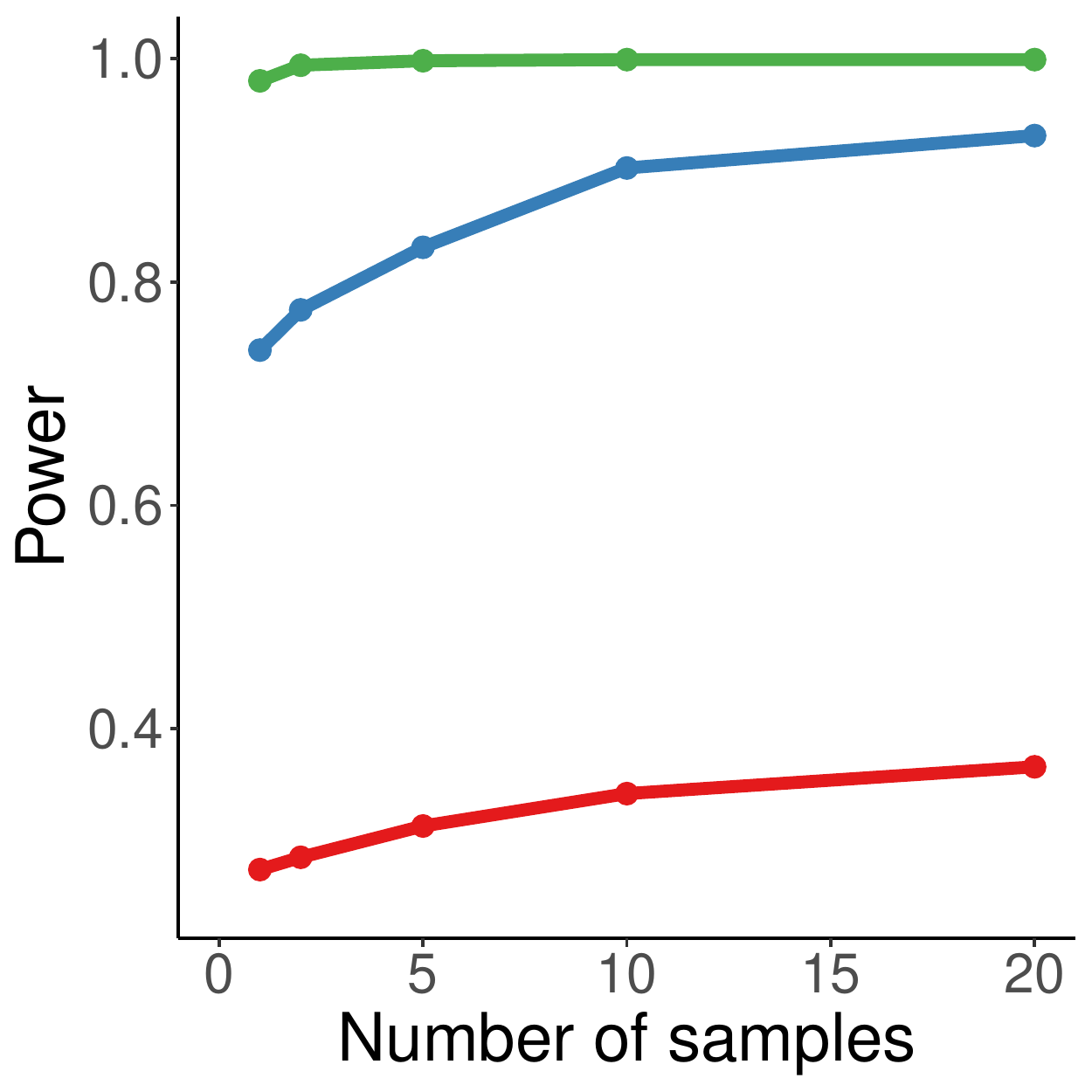}
        \put(-70,140){\footnotesize{$h = 10$}}
    \includegraphics[width=0.32\linewidth]{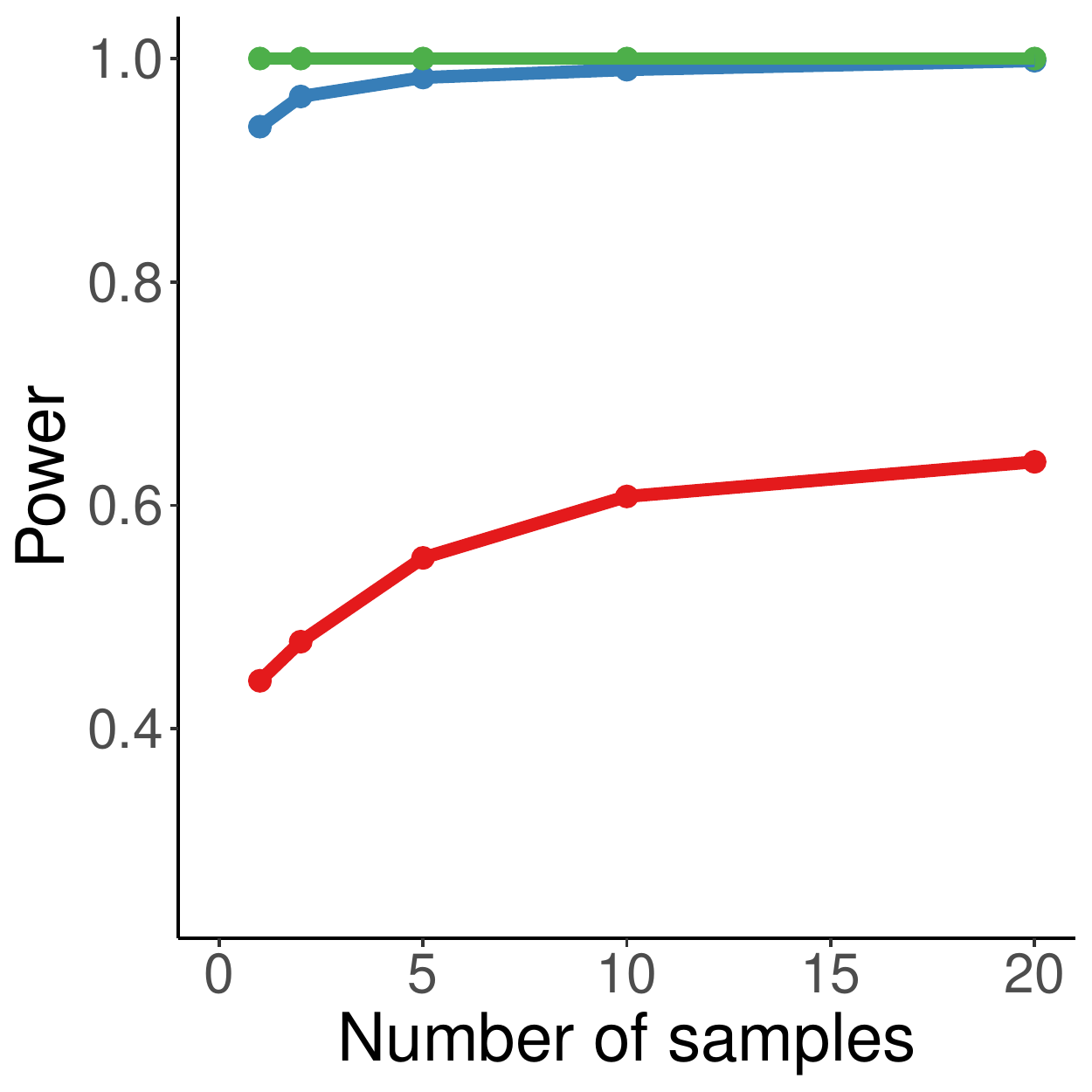}
        \put(-70,140){\footnotesize{$h = 20$}}
    \includegraphics[width=0.32\linewidth]{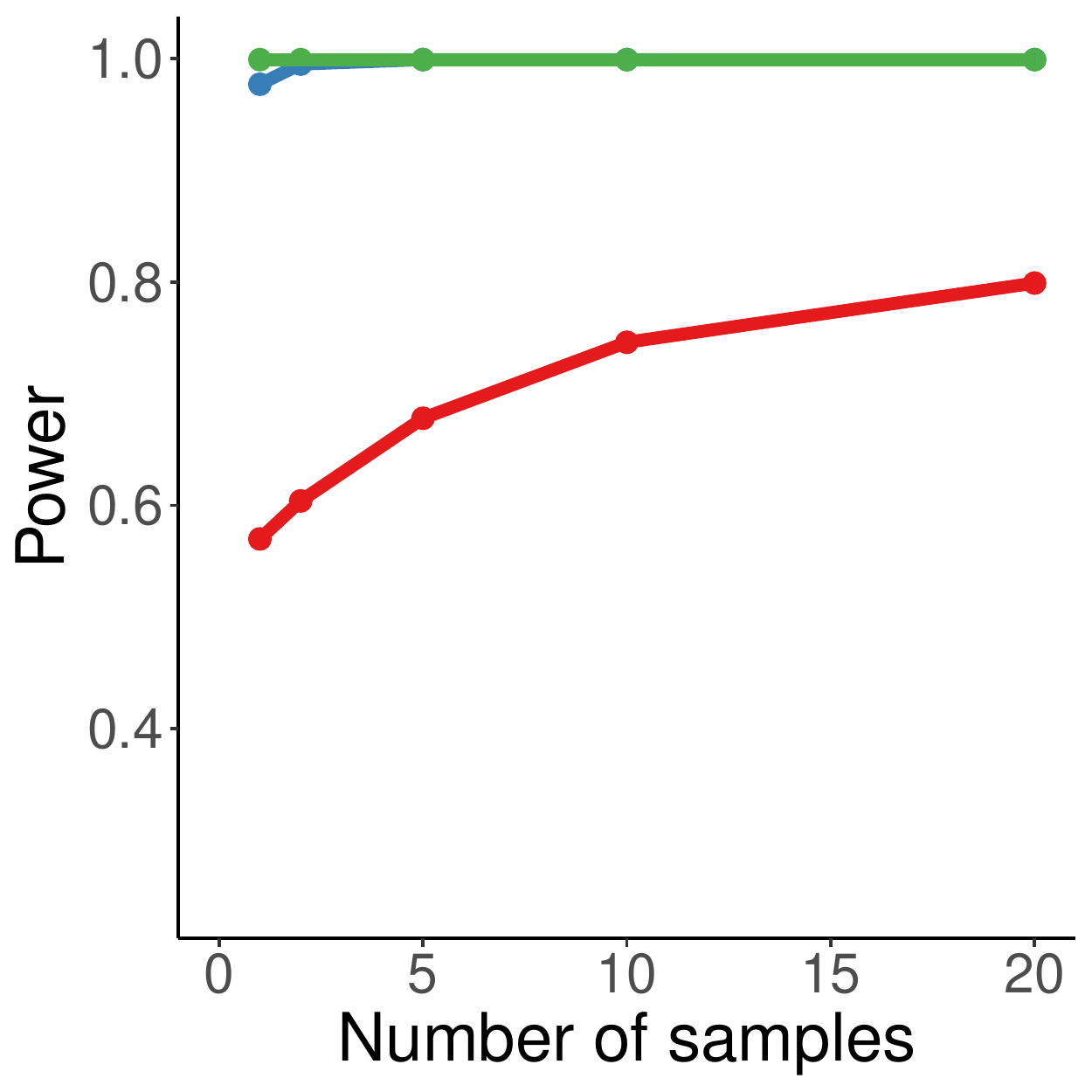}
        \put(-70,140){\footnotesize{$h = 30$}}
    \caption{\small{Rejection rates of $H_0$ for $L_0$ segmentation. On each plot the three lines show the proportion of samples (of 1000 total) where the $p$-value was below $0.05$, resulting in $H_0$ being rejected, for different values of $N$. Each line corresponds to a different size of change $\delta$: green corresponds to $\delta = 3$, blue to $\delta = 2$, and red to $\delta = 1$.}}
    \label{fig:power-l0}
\end{figure}

\begin{figure}
    \centering
    \includegraphics[width=0.32\linewidth]{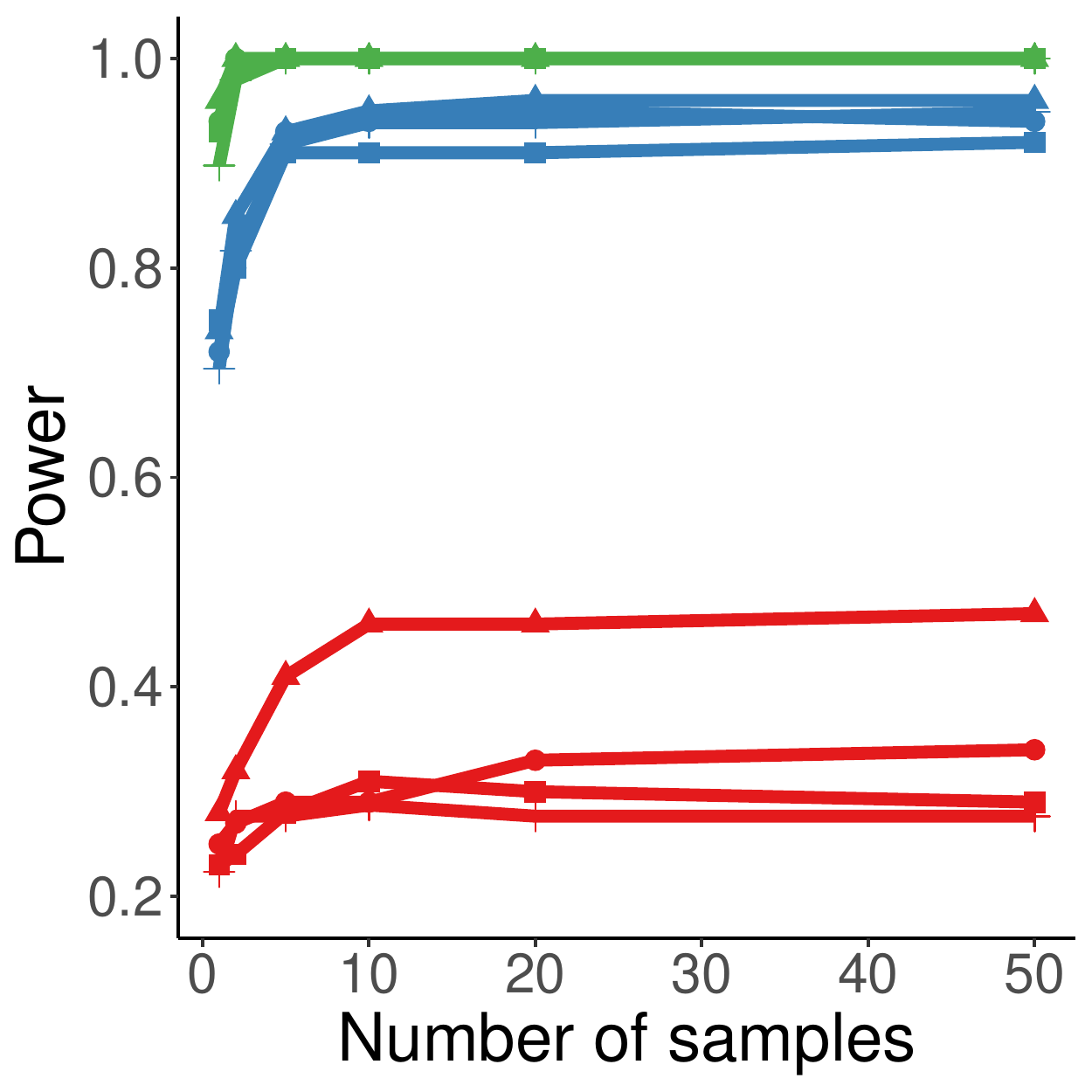}
        \put(-70,138){\footnotesize{$h = 10$}}
    \includegraphics[width=0.32\linewidth]{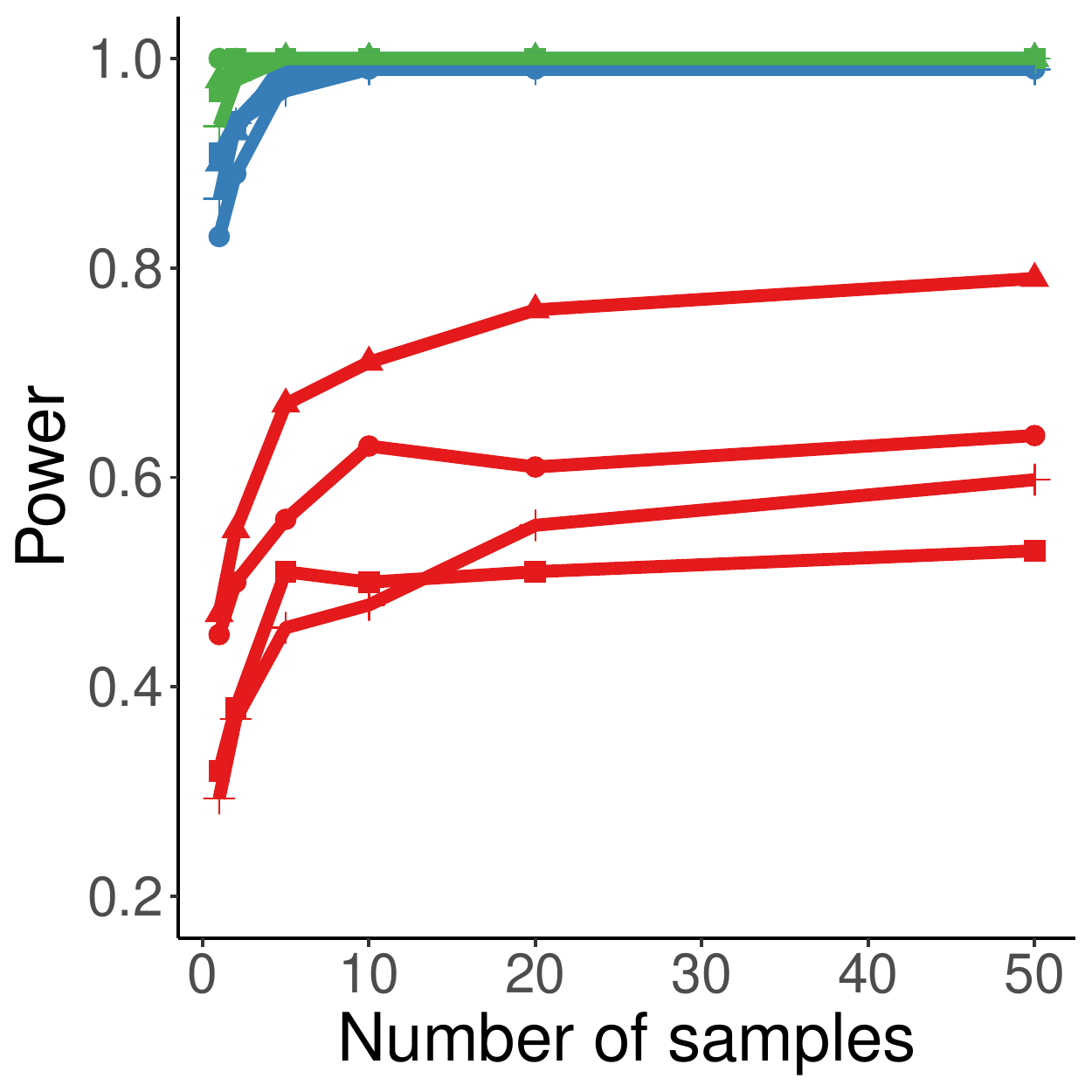}
        \put(-70,138){\footnotesize{$h = 20$}}
    \includegraphics[width=0.32\linewidth]{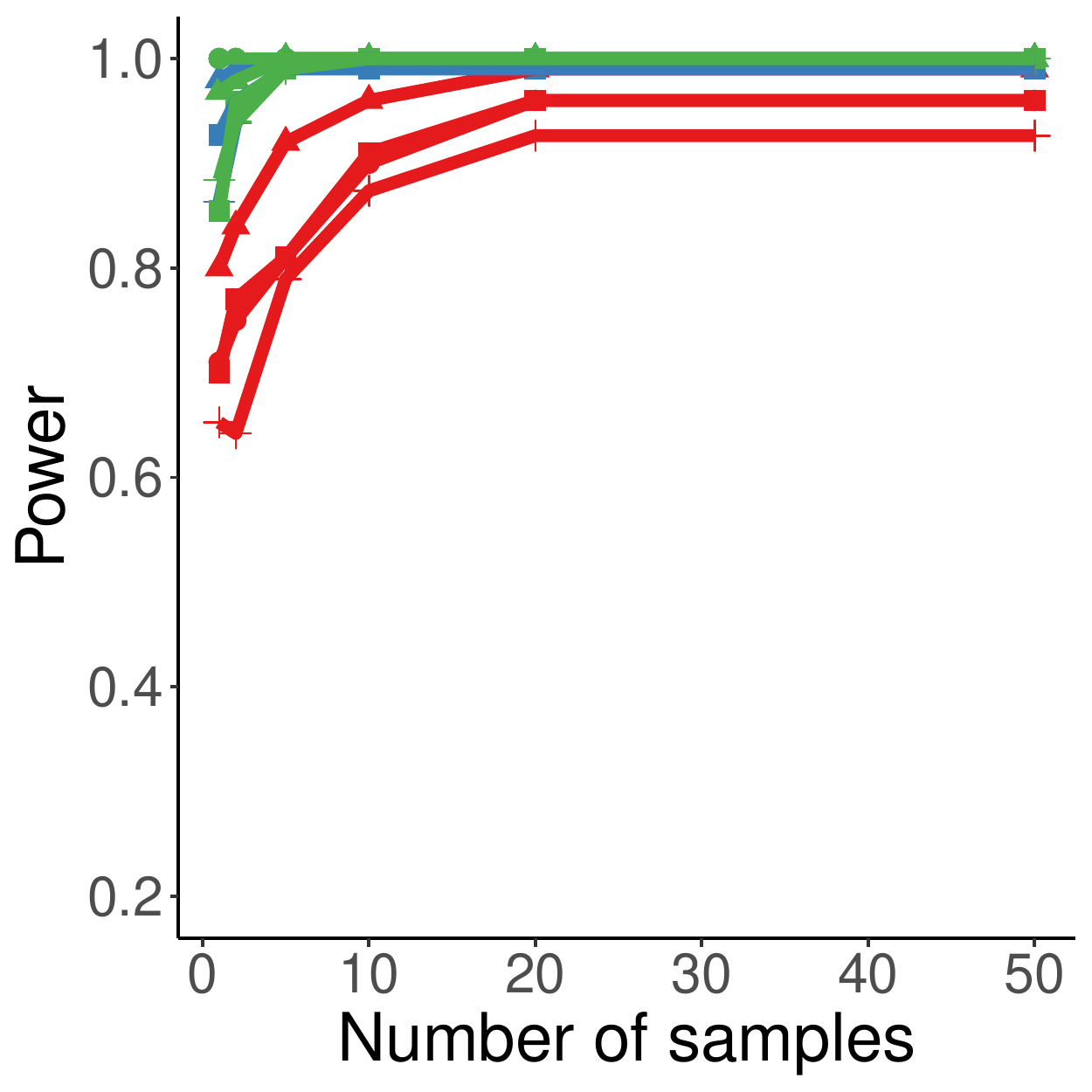}
        \put(-70,138){\footnotesize{$h = 30$}} \\
    \caption{\small{Rejection rates of $H_0$ for different values of $h$, $\delta$, and $N$, when we simulate from a model with 4 changepoints, and apply wild binary segmentation with 4 changepoints. For each $h$, the process is run three times with changes of size $\delta = 1, 2, 3$, which are shown on the plots as red, blue, and green lines respectively. Each line corresponds to a changepoint.
    }}
    \label{fig:power-wbs-k4}
\end{figure}

Figure \ref{fig:qq-l0-h1} shows equivalent plots to Figure \ref{fig:qq-bs-h1}, where we estimate changepoints using $L_0$ segmentation rather than binary segmentation. In each case we simulate 1000 data sets, each with a single changepoint, and calculate the $p$-values. Figure \ref{fig:power-l0} shows plots of the power when we simulate under a model with a single changepoint, and apply $L_0$ segmentation. As for binary segmentation, the power increases with $h$ and $\delta$. Figure \ref{fig:power-wbs-k4} shows similar plots for wild binary segmentation when there are 4 changepoints.

\begin{figure}
    \centering
    \includegraphics[width=0.98\linewidth]{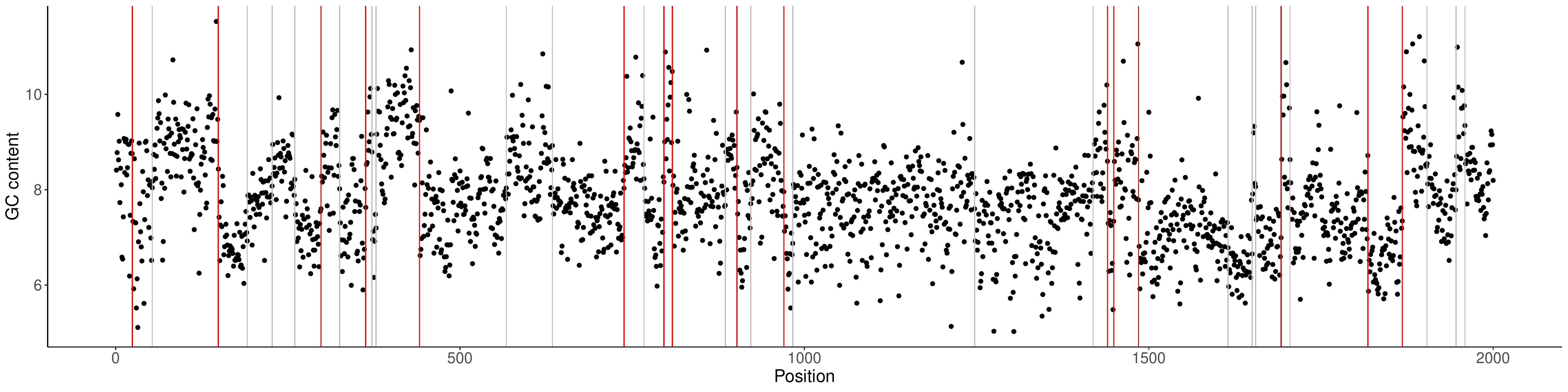}
    \includegraphics[width=0.98\linewidth]{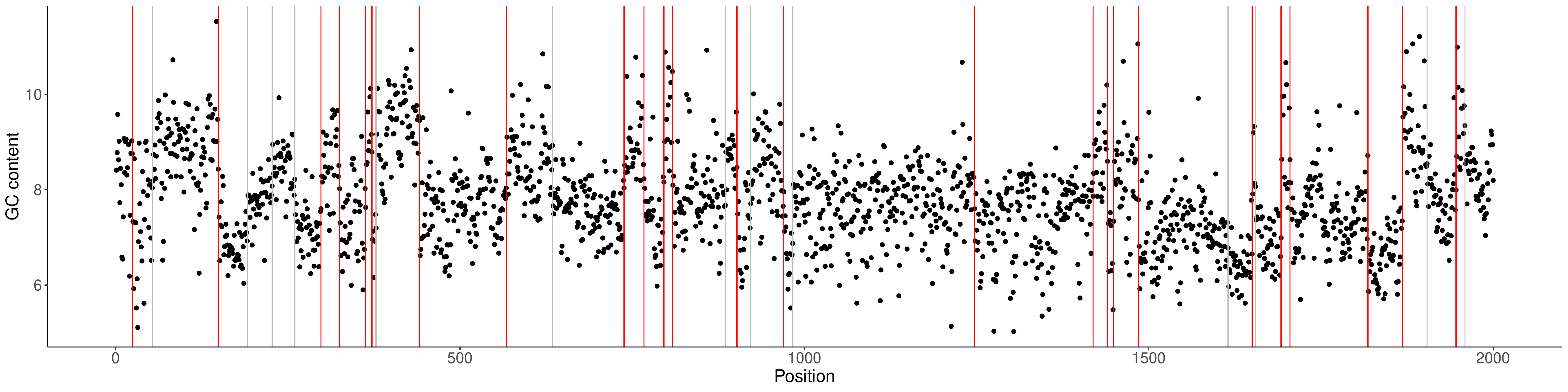} \\
    \includegraphics[width=0.98\linewidth]{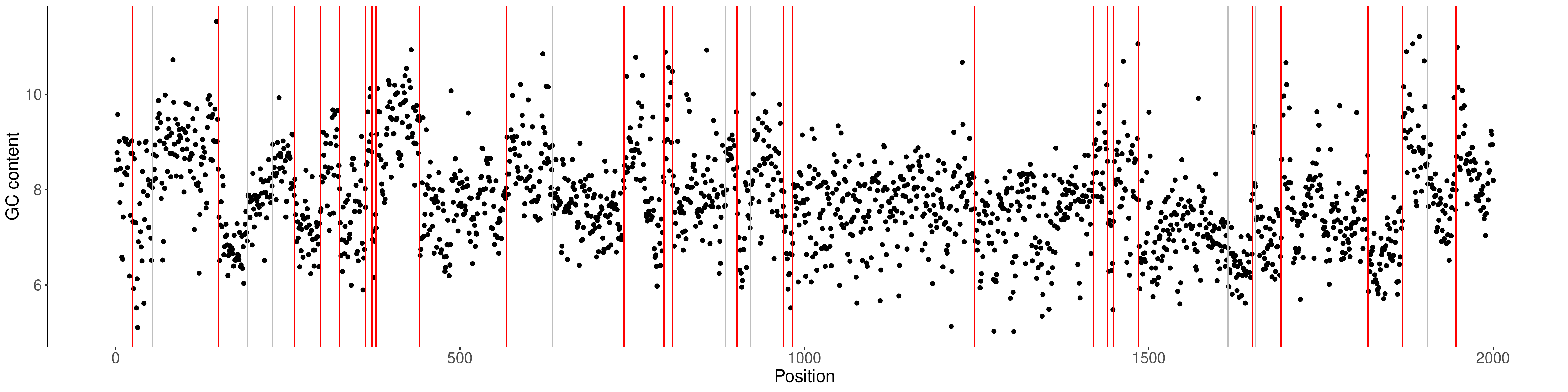}
    \caption{\small{Estimated changepoints in GC content data. $L_0$ segmentation was used to estimate 38 changepoints, and we set $h = 10$. Each vertical line corresponds to an estimated changepoint; changepoints found to be significant at significance level $\alpha = 0.05$ are shown in red, with others shown in grey. In the top panel, we used $N = 1$ (equivalent to the method of \cite{jewell-testing}) to calculate $p$-values. The middle and bottom panels show results obtained using $N = 10$ and $N = 20$ respectively.}}
    \label{fig:hc1-l0-h10}
\end{figure}

Figure \ref{fig:hc1-l0-h10} shows the results of applying $L_0$ segmentation to the GC content data set we used in Section \ref{sec:real}. The $p$-values correspond to those in Table \ref{tab:hc1}.

\subsection{Correlation of $p$-values} \label{App:p-values}

To investigate empirically whether $p$-values at distinct locations (i.e. non-overlapping regions of interest) are uncorrelated, we simulated from a model with $T = 400$ and three changepoints at $\tau = 100, 200, 300$. Using binary segmentation, changepoints were detected at $\hat{\tau} = 92, 200, 297$. We set $h = 10$, so that the regions of interest around each $\hat{\tau}$ were non-overlapping. For each $\hat{\tau}$, we sampled $\phi$ from its (conditional) distribution 1000 times. For each $\phi^{(j)}$, we calculated $\boldsymbol{X}'(\phi^{(j)})$, applied binary segmentation, and calculated the $p$-values. We then calculated pairwise correlations between $p$-values, which are displayed in Table \ref{tab:p-value-cors}. We find that the correlations in all cases are very close to $0$, so empirically the $p$-values at different changepoints appear to be uncorrelated, and it seems reasonable to treat them as independent.

\begin{table}
    \centering
    \caption{\label{tab:p-value-cors} Investigating independence of $p$-values: in each case we sample different $\phi$ values at the changepoint of interest (keeping the rest of the data the same), re-apply the changepoint model, and calculate $p$-values at each changepoint. We then calculate the pairwise correlations between $p$-values.}
    \begin{tabular}{lccc}
      \hline
        Changepoint of interest & $\rho(\hat{\tau}_1, \hat{\tau}_2)$ & $\rho(\hat{\tau}_1, \hat{\tau}_3)$ & $\rho(\hat{\tau}_2, \hat{\tau}_3)$ \\
      \hline
        $\hat{\tau}_1$ & 0.042 & 0.014 & 0.005 \\
        $\hat{\tau}_2$ & 0.017 & 0.001 & 0.007 \\
        $\hat{\tau}_3$ & 0.065 & 0.021 & 0.029 \\
      \hline
    \end{tabular}
\end{table}

\subsection{Robustness to Laplace noise} \label{App:Laplace}

\begin{figure}
    \centering
    \includegraphics[width=0.32\linewidth]{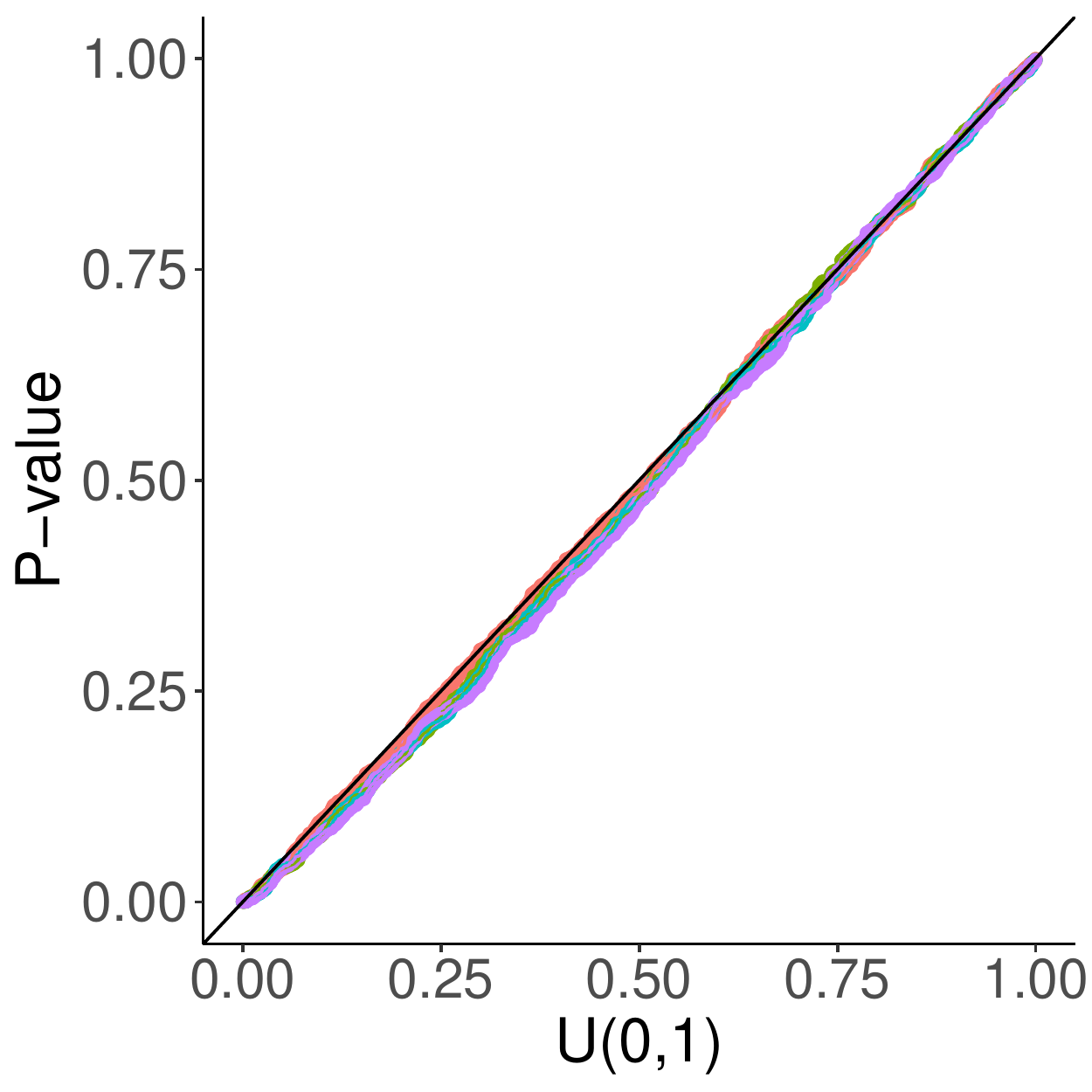}
        \put(-75, 135){$H_0$, $s = 0.5$}
    \includegraphics[width=0.32\linewidth]{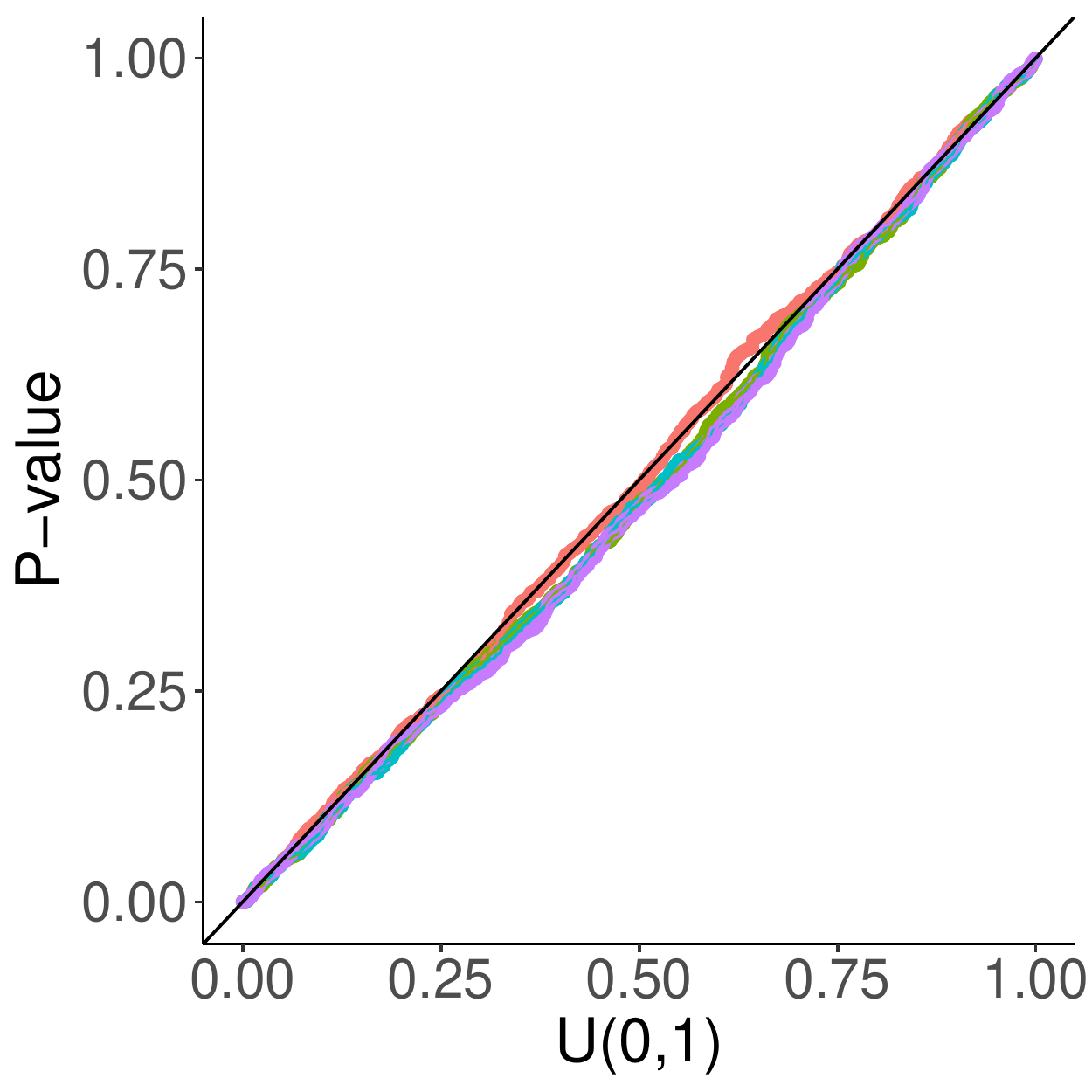}
        \put(-76, 135){$H_0$, $s = 1$}
    \includegraphics[width=0.32\linewidth]{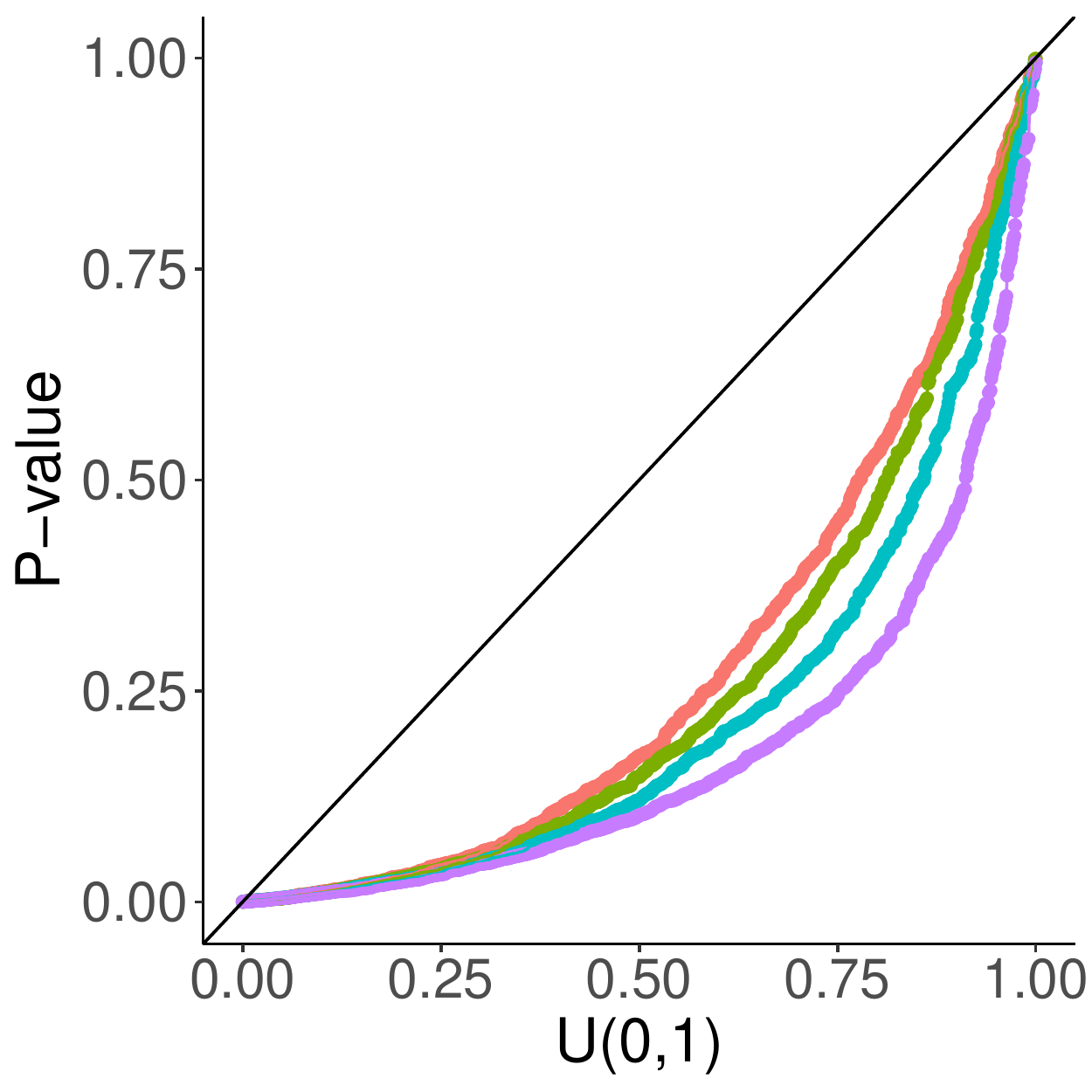}
        \put(-76, 135){$H_1$, $s = 0.5$}
    \caption{\small{QQ plots of $p$-values obtained when simulating from a Laplace distribution under $H_0$ and $H_1$; $s$ is the scale parameter.}}
    \label{fig:laplace-sims}
\end{figure}

Figure \ref{fig:laplace-sims} shows QQ-plots for post-selection $p$-values for data simulated with i.i.d. Laplace noise. Under $H_0$, the $p$-values are approximately uniform on $[0, 1]$.

\section{Additional Tables} \label{sec:tables}

We present here fuller versions of Tables \ref{tab:us-vs-mosum-h0} and \ref{tab:us-vs-mosum-h1}, giving results for $T = 250, 500, 1000$. Table \ref{tab:us-vs-mosum-h0-full} corresponds to Table \ref{tab:us-vs-mosum-h0} and Table \ref{tab:us-vs-mosum-h1-full} to Table \ref{tab:us-vs-mosum-h1}. The different value of $T$ give similar results for the numbers of true and false positives detected.

Table \ref{tab:bs-fdr} gives further results for binary segmentation for one of the scenarios in Table \ref{tab:us-vs-mosum-h1-full}, in the case where $T = 1000$ and $K = 4$. Here we compare different choices of threshold for the binary segmentation algorithm, to investigate the effect this has on the number of true positives discovered and on the error rate. The threshold used for previous results was 3. Here we see that using a lower threshold increases the number of false positives (as would be expected, since we are more likely to detect false changepoints), but the FWER and FDR are still below 5\%.

\begin{table}
    \centering
    \caption{\label{tab:us-vs-mosum-h0-full} Simulating from $H_0$, we applied binary segmentation and MOSUM and calculated how many times $H_0$ was falsely rejected. The table reports the mean number of estimated changepoints with adjusted $p$-value below 0.05 across 1000 iterations, and the proportion of runs in which there was at least one false discovery, for both the Holm-Bonferroni and Benjamini-Hochberg procedures.}
    \begin{tabular}{llllllll}
      \hline
        & & \multicolumn{3}{c}{Mean false positives} & \multicolumn{3}{c}{Proportion with $\geq 1$ false positive} \\
      \hline
           & & H-B & B-H & & H-B & B-H & \\
      \hline
        & BS, $N = 1$ & 0.03 & 0.03 & & 0.03 & 0.03 & \\
        $T = 250$ & BS, $N = 10$ & 0.04 & 0.04 & & 0.03 & 0.03 & \\
        & MOSUM & & & 0.01 & & & 0.01 \\
      \hline
        & BS, $N = 1$ & 0.03 & 0.03 & & 0.03 & 0.03 & \\
        $T = 500$ & BS, $N = 10$ & 0.03 & 0.03 & & 0.03 & 0.03 & \\
        & MOSUM & & & 0.01 & & & 0.01 \\
      \hline
        & BS, $N = 1$ & 0.03 & 0.03 & & 0.03 & 0.03 & \\
        $T = 1000$ & BS, $N = 10$ & 0.03 & 0.04 & & 0.03 & 0.03 & \\
        & MOSUM & & & 0.01 & & & 0.01 \\
      \hline
    \end{tabular}
\end{table}

\begin{table}
\caption{\label{tab:us-vs-mosum-h1-full} Mean number of true positives and error rate when simulating $\boldsymbol{X}$ with $T$ data points and $K$ equally spaced changepoints, with $\mu$ alternating between $1$ and $-1$, using binary segmentation and MOSUM. The error rate given is the family-wise error rate for the Holm-Bonferroni method, and the false discovery rate for the Benjamini-Hochberg method and for MOSUM. This is calculated as $FDR = \frac{\text{false positives}}{\text{false positives } + \text{ true positives}}$.}
\centering
    \begin{tabular}{llrrrrrr}
      \hline
        & & \multicolumn{3}{c}{Mean true positives} & \multicolumn{3}{c}{Error rate (\%)} \\
      \hline
            & & H-B & B-H & & H-B & B-H & \\
      \hline
                & BS, $N = 1$ & 0.78 & 0.78 & & 0.30 & 0.34 & \\
        $T = 250, K = 1$ & BS, $N = 10$ & 0.94 & 0.50 & & 0.27 & 0.34 & \\
                & MOSUM & & & 0.74 & & & 1.08 \\
      \hline
                & BS, $N = 1$ & 2.67 & 2.79 & & 0.30 & 0.16 & \\
        $T = 250, K = 4$ & BS, $N = 10$ & 3.45 & 3.55 & & 0.70 & 0.24 & \\
                & MOSUM & & & 2.83 & & & 0.18 \\
      \hline
                & BS, $N = 1$ & 5.07 & 5.92 & & 0.70 & 0.17 & \\
        $T = 250, K = 9$ & BS, $N = 10$ & 7.24 & 7.74 & & 0.50 & 0.12 & \\
                & MOSUM & & & 6.43 & & & 0.03 \\
      \hline
                & BS, $N = 1$ & 0.79 & 0.79 & & 1.10 & 0.99 & \\
        $T = 500, K = 1$ & BS, $N = 10$ & 0.95 & 0.95 & & 1.60 & 0.93 & \\
                & MOSUM & & & 0.63 & & & 1.26 \\
      \hline
                & BS, $N = 1$ & 2.75 & 2.89 & & 0.40 & 0.29 & \\
        $T = 500, K = 4$ & BS, $N = 10$ & 3.46 & 3.53 & & 1.00 & 0.35 & \\
                & MOSUM & & & 2.59 & & & 0.27 \\
      \hline
                & BS, $N = 1$ & 5.54 & 6.36 & & 1.10 & 0.31 & \\
        $T = 500, K = 9$ & BS, $N = 10$ & 7.38 & 7.87 & & 1.30 & 0.33 & \\
                & MOSUM & & & 5.89 & & & 0.07 \\
      \hline
                & BS, $N = 1$ & 0.79 & 0.79 & & 0.60 & 0.53 & \\
        $T = 1000, K = 1$ & BS, $N = 10$ & 0.94 & 0.94 & & 0.50 & 0.29 & \\
                & MOSUM & & & 0.58 & & & 1.20 \\
      \hline
                & BS, $N = 1$ & 2.78 & 2.91 & & 0.90 & 0.42 & \\
        $T = 1000, K = 4$ & BS, $N = 10$ & 3.42 & 3.51 & & 1.50 & 0.46 & \\
                & MOSUM & & & 2.32 & & & 0.22 \\
      \hline
                & BS, $N = 1$ & 5.56 & 6.38 & & 1.50 & 0.42 & \\
        $T = 1000, K = 9$ & BS, $N = 10$ & 7.23 & 7.80 & & 2.10 & 0.51 & \\
                & MOSUM & & & 5.30 & & & 0.09 \\
      \hline
    \end{tabular}
\end{table}

\begin{table}
\caption{\label{tab:bs-fdr} Mean number of true positives and error rate when simulating from a model with $T = 1000$ and $K = 4$ equally spaced changepoints, with $\mu$ alternating between $1$ and $-1$, using binary segmentation. The error rate given is the family-wise error rate for the Holm-Bonferroni method, and the false discovery rate for the Benjamini-Hochberg method. This is calculated as $FDR = \frac{\text{false positives}}{\text{false positives }+\text{ true positives}}$.}
\centering
    \begin{tabular}{cccccc}
      \hline
        & & \multicolumn{2}{c}{Mean true positives} & \multicolumn{2}{c}{Error rate (\%)} \\
      \hline
        Threshold & $N$ & H-B & B-H & H-B & B-H \\
      \hline
        2 & $N = 1$ & 2.70 & 3.02 & 3.10 & 2.89 \\
        3 & $N = 1$ & 2.78 & 2.91 & 0.90 & 0.42 \\
        4 & $N = 1$ & 2.49 & 2.63 & 0.30 & 0.11 \\
      \hline
        2 & $N = 10$ & 3.33 & 3.60 & 3.80 & 2.83 \\
        3 & $N = 10$ & 3.42 & 3.51 & 1.50 & 0.46 \\
        4 & $N = 10$ & 3.51 & 3.57 & 0.30 & 0.08 \\
      \hline
    \end{tabular}
\end{table}

\end{document}